\documentclass[twocolumn]{aastex63}

\usepackage{amsmath}
\usepackage{natbib}
\usepackage{multirow} 
\usepackage{anyfontsize}
\usepackage{indentfirst}
\usepackage{comment}
\usepackage{booktabs}
\usepackage{bigints}
\usepackage{mathrsfs,amsmath} 
\usepackage{makecell}
\usepackage{graphicx}

\usepackage[normalem]{ulem}
\useunder{\uline}{\ul}{}
\usepackage{hyperref}
\usepackage{lineno}

\hypersetup{linkcolor=magenta,citecolor=cyan,filecolor=yellow,urlcolor=blue}

\received{ddmmyyyy}
\revised{\today}
\accepted{ddmmyyyy}
\published{ddmmyyyy}

\shorttitle{$E_{\rm p}$-Flux lag}
\shortauthors{Li.}

\begin{document}

\title{Three Subclasses of the Intensity-tracking Pattern in Gamma-Ray Burst Spectral Evolution}

\email{liang.li@icranet.org}


\author[0000-0002-1343-3089]{Liang Li}

\affiliation{Institute of Fundamental Physics and Quantum Technology, Ningbo University, Ningbo, Zhejiang 315211, People's Republic of China}

\affiliation{School of Physical Science and Technology, Ningbo University, Ningbo, Zhejiang 315211, People's Republic of China}

\begin{abstract}

The properties of the spectral evolution during the prompt emission phase of gamma-ray bursts (GRBs), which are closely related to the radiation mechanism (synchrotron or photosphere), are still a subject of debate. Two spectral evolution patterns (``hard-to-soft'' and ``intensity-tracking'') have been commonly observed in GRB prompt emission spectra. Here we present a well-defined sample of 20 single-pulse GRBs detected by \emph{Fermi} whose prompt emission spectra exhibit the intensity-tracking pattern. By performing a time-resolved spectral analysis, we derive $E_{\rm p}$ and the energy flux $F$ from the same time bins and introduce a matched-bin lag, $t_{\rm lag}^{\rm F} \equiv t_{\rm p}(E_{\rm p})-t_{\rm p}(F)$, where $t_{\rm p}$ denotes the time at which each quantity reaches its maximum. We find that the intensity-tracking pattern subdivides into three distinct subclasses: Type I (5/20), with aligned $E_{\rm p}$ and flux peaks; Type II  (13/20), with $E_{\rm p}$ peaking before the flux; and Type III (2/20), with $E_{\rm p}$ peaking after the flux. The early-peaking Type~II subclass dominates the sample. The subclasses also exhibit systematic differences in their spectral and temporal properties. Type II bursts are systematically harder than Type I, show broader flux pulses, and more often display asymmetric rising and decaying $E_{\rm p}$-$F$ branches. Type I is consistent with tightly coupled spectral and power evolution, whereas Type II is more naturally explained by nonthermal or hybrid prompt-emission scenarios in which spectral hardening precedes the peak radiative output. Type III appears to form a rare positive-lag tail whose physical origin remains uncertain.

\end{abstract}


\section{Introduction}

Gamma-ray bursts (GRBs) are the most energetic transient events in the universe. The prompt-emission spectra, observed in the keV–MeV band, are commonly described by the empirical Band function, a smoothly broken power law characterized by the low- and high-energy photon spectral indices ($\alpha$ and $\beta$) and the peak energy $E_{\rm p}$ of the $\nu F_{\nu}$ spectrum \citep{Band1993}. For typical prompt emission spectra, $E_{\rm p}$ is of order a few hundred keV, where $\alpha \approx -0.8$ and $\beta\approx -2.5$ \citep[e.g.,][]{Li2021b}. It has been recognized since the earliest temporal and spectral studies that GRB spectra evolve significantly both within individual pulses and across the burst as a whole \citep{Golenetskii1983,Norris1986,Bhat1994,Kargatis1994,Ford1995,Liang1996,Crider1997,Ryde1999,Kaneko2006,Peng2009,Lu2012,Oganesyan2018,Acuner2018,Yu2019,Li2021b}. The observed spectrum is therefore not a static quantity but a time-dependent manifestation of the underlying dissipation and radiation processes \citep[e.g.,][]{Golenetskii1983, Norris1986, Ford1995, Kaneko2006, Li2019a, Li2019b, Li2021b}. In this context, the temporal evolution of $E_{\rm p}$, and to a lesser extent that of $\alpha$, has long been considered one of the most important observational diagnostics of prompt-emission physics.

A broad phenomenological picture has emerged from decades of time-resolved spectroscopy. Two dominant patterns of $E_{\rm p}$ evolution are commonly identified, namely, the ``hard-to-soft'' pattern, in which $E_{\rm p}$ decreases monotonically regardless of the flux variation \citep[e.g.,][]{Norris1986,Bhat1994,Band1997}, and the ``intensity-tracking'' pattern, often also referred to as flux-tracking, in which $E_{\rm p}$ broadly follows the burst intensity \citep[e.g.,][]{Yu2019,Li2019b,Ryde2019,Li2021b}. With the advent of the Gamma-ray Burst Monitor (GBM, \citealt{Meegan2009}) onboard the \textit{Fermi Gamma-ray Space Telescope}, coupled with advancements in Bayesian time-resolved spectroscopy, it has become increasingly clear that the intensity-tracking behavior is both common and diverse, particularly in well-defined pulses. In several bursts, both $E_{\rm p}$ and $\alpha$ evolves in concert with the intensity, producing the so-called double-tracking behavior \citep[e.g., GRB 131231A,][]{Li2019b}, which provides a particularly direct link between spectral evolution and the underlying radiation mechanism. Large-sample analyses further indicate that joint tracking of $\alpha$ and $E_{\rm p}$ occurs in a substantial fraction of pulses, while also revealing systematic spectral differences from pulse to pulse within the same burst \citep{Li2021b}. Similar behavior has since been reported in several additional events \citep{Duan2019,Chen2021,Gupta2021,Gupta2022,Chen2022,Ror2023,Caballero-Garca2023}. Nevertheless, most previous work has still treated intensity-tracking as a single class and has focused mainly on distinguishing it from hard-to-soft evolution. Much less attention has been paid to whether, within intensity-tracking bursts themselves, the peak of $E_{\rm p}(t)$ is always contemporaneous with the pulse maximum, or whether finer temporal subclasses with distinct physical implications may exist. 

It has also become clear that reliable interpretation of spectral evolution requires genuinely time-resolved analysis. Time-integrated spectra are, by construction, averages over evolving emission episodes and can therefore conflate photons produced under different physical conditions \citep{Kaneko2006,Yu2016,Yu2019}. The problem is particularly severe for bursts with overlapping pulses, because unresolved pulse superposition can distort both the observed spectral trends and the inferred physical correlations. Indeed, simulation studies have suggested that at least some apparent intensity-tracking behavior may arise from overlapping hard-to-soft pulses \citep{Lu2012}. For this reason, single-pulse GRBs provide the cleanest laboratory for studying prompt spectral evolution. They minimize contamination from pulse overlap and allow one to compare the temporal behavior of spectral and intensity quantities in a more controlled manner.

However, even within the class of single-pulse bursts, the commonly used identification of intensity-tracking remains largely trend-based. In most previous studies, one asks whether $E_{\rm p}(t)$ generally rises and falls together with the light curve, but the relative timing of their peaks has received comparatively little attention. This issue is subtle but physically important. When the spectral evolution exhibits a well-defined hump-like structure, $E_{\rm p}(t)$ has a measurable peak time, and one may ask whether this peak coincides with, precedes, or follows the peak of the burst intensity. Addressing this question in a robust way is not straightforward, because the result may depend on temporal binning, spectral significance, model selection, and even the choice of the reference intensity curve. In particular, comparing $E_{\rm p}(t)$ with a counts light curve introduces additional ambiguity, since the two quantities are not necessarily defined on the same temporal grid.

This consideration motivates the primary methodological change adopted in the present work. Rather than using the peak time of the counts light curve as the timing reference, we compare the peak time of $E_{\rm p}$ with the peak time of the energy flux $F$. The reason is that both $E_{\rm p}$ and $F$ are derived from the same time-resolved spectral fits performed on the same Bayesian-block (BBlocks; \citealt{Scargle2013}) time bins. Consequently, the corresponding time series, $E_{\rm p}(t)$ and $F(t)$ are strictly matched bin by bin, eliminating the temporal-sampling mismatch that is unavoidable when $E_{\rm p}$ is compared with an independently binned counts light curve. This allows us to define a self-consistent lag quantity, $t_{\rm lag}^{\rm F}$ (Eq.\ref{eg:t_lag}), which measures the relative timing between spectral hardening and radiative power within a unified time-resolved spectroscopic framework.

The physical motivation for quantifying this lag is compelling. If the prompt emission is dominated by a photosphere emission component, one may expect relatively tight coupling between luminosity and temperature, and hence a tendency toward nearly exact tracking between the $E_{\rm p}$ peak and the intensity peak. In contrast, for optically thin synchrotron or magnetic-dissipation models, particularly the Internal-Collision-induced MAgnetic Reconnection and Turbulence (ICMART)-like scenarios \citep{ZhangYan2011}, the characteristic photon energy and the total radiated power need not peak simultaneously, since they depend differently on the magnetic field strength, particle acceleration efficiency, radiative cooling, and the energy-dissipation history \citep{ZhangYan2011,Uhm2014NP,Uhm2018,Shao2022}. In this sense, the relative timing of $t_{\rm p}(E_{\rm p})$ and $t_{\rm p}(F)$ may therefore serve as a new observational discriminator among prompt-emission models. A complementary diagnostic is provided by the low-energy photon index $\alpha$, because a systematically harder $\alpha$ distribution in the aligned subclass would be more naturally consistent with photosphere emission, whereas similar $\alpha$ distributions across subclasses would favor a broader nonthermal interpretation.

In this paper, we construct a sample of single-pulse \textit{Fermi}/GBM bursts exhibiting intensity-tracking spectral evolution and revisit this class from the perspective of peak-time ordering. We employ BBlocks algorithm to define temporal bins for time-resolved spectroscopy and fit each bin with either the Band function or a cutoff power-law (CPL) model. For spectra best described by a cutoff power law, the peak energy is calculated through $E_{\rm p}=(2+\alpha)E_{\rm c}$. This procedure yields one $E_{\rm p}$ value and one energy-flux measurement per spectral bin, allowing a direct comparison between $t_{\rm p}(E_{\rm p})$ and $t_{\rm p}(F)$. We demonstrate that the traditionally defined intensity-tracking pattern is not monolithic. It encompasses an aligned subclass in which the $E_{\rm p}$ peak is consistent with the flux peak, an early-peaking subclass in which the $E_{\rm p}$ peak precedes the flux peak, and a rarer late-peaking subclass in which the $E_{\rm p}$ peak lags behind the flux peak. For completeness, we also compare $E_{\rm p}$ with the counts light-curve peak, but this counts-based comparison is treated only as a reference rather than the formal classification criterion.

This paper is organized as follows. Section \ref{sec:Methodology} describe the data analysis procedure, including the sample construction, BBlocks algorithm to rebin the light curve, time-resolved spectral fitting, and the definition of peak-time measurements and the flux-based lag. Section \ref{sec:Result} present the observational results, including the final sample, representative bursts, the distribution of peak-time offsets, and inter-subclass differences in spectral and temporal properties. Section \ref{sec:Dis} discusses the robustness of the classification against potential systematic effects and explores the implications of our findings for photosphere and nonthermal prompt-emission scenarios. Our conclusions are summarized in Section \ref{sec:Con}. Throughout this paper, all parameter uncertainties are quoted at the $68\%$ (1$\sigma$) credible interval unless otherwise stated.

\section{Methodology} \label{sec:Methodology}

\subsection{Data reduction and Sample Selection} \label{sec:sample}

We analyze the prompt-emission data recorded by the Fermi-GBM, employing the time-tagged event (TTE) data for both temporal and spectral analyses. The GBM comprises 12 NaI detectors and 2 BGO detectors, spanning a broad energy range from a few keV to several tens of MeV \citep{Meegan2009}. For each burst, we select the NaI detectors with the most favorable viewing angles and the brightest BGO detector follow the standard GBM practice. Background is estimated from source-free time intervals prior to and following each burst, and the corresponding detector response matrices are generated for spectral analysis. Unless otherwise noted, we adopt the standard GBM energy ranges used in previous catalog studies, namely, 10-30 keV and 40-9000 keV for the NaI detectors, avoiding the K-edge region, and 30 keV-30 MeV for the BGO detector \citep{Goldstein2012,Gruber2014,Yu2016,Yu2019}.

Our procedure includes the following main steps.

\begin{itemize}

\item Preliminary burst selection. To quantify the relative timing between the spectral-evolution peak and the intensity peak, pulse overlap must be minimized as much as possible. To avoid overlapping effects between pulses within a burst, we pay special attention to single-pulse bursts. We visually inspect the GBM TTE light curves of all {\it Fermi}/GBM bursts from July 2008 to December 2022, and identify more than 300 well-defined single-pulse bursts. These bursts form our initial sample, which is further screened below.  

\item Light-curve binning. To investigate the spectral evolution inside a burst, a refined time-resolved spectral analysis is required. For each candidate burst in our target sample, we apply the BBlocks algorithm to refine the TTE lightcurve of the brightest NaI detector in order to define the time bins for the time-resolved spectral analysis. The algorithm creates multiple-time events over the entire emission period, and a refined spectral analysis is therefore performed on each timing event individually. The statistical significance ($S$; \citealt{Vianello2018a, Li1983}) is computed for each individual time bin. To ensure well-determined spectral fits and tightly constrained spectral parameters, we select only bursts with at least five time bins, each satisfying $S>15$. This criterion guarantees that the peak energy $E_{\rm p}$ and the corresponding energy flux $F$ are derived from identical temporal bins.

\item Spectral fitting. We perform time-resolved spectral analysis for each individual time bin within a Bayesian framework using {\tt the Multi-Mission Maximum Likelihood Framework} ({\tt 3ML}; \citealt{Vianello2015}) and standard forward-folding procedures, following the standard practices of \citep{Li2019b,Li2019c,Yu2019,Burgess2019,Li2020,Li2021a,Li2021b}. Best-fit spectral parameters are determined via a Markov Chain Monte Carlo (MCMC) sampler. Each spectrum is first fit with the Band function \citep{Band1993}. If the high-energy power-law index is unconstrained, the spectrum is refit with a CPL model. For Band fits, $E_{\rm p}$ is obtained directly from the best-fit parameters. For CPL fits, the spectral peak energy is given by $E_{\rm p}=(2+\alpha)E_{\rm c}$, where $\alpha$ is the low-energy photon index and $E_{\rm c}$ is the cutoff energy. The corresponding energy flux $F$ is derived from the same best-fit model over the same time bin. Consequently, the full burst duration is represented by two rigorously matched time series, $E_{\rm p}(t)$ and $F(t)$, both defined on the same Bayesian-block time bins. 

\item Spectral evolution pattern. From the time-resolved spectral results, we examine the $E_{\rm p}(t)$ evolution of each burst and retain only those displaying an intensity-tracking pattern as our final sample. After applying all selection criteria, the final sample comprises 20 bursts. The flux-based timing properties of each burst are summarized in Table~\ref{tab:burst-summary}. 
	
\end{itemize}

\subsection{Peak-time Measurement, Lag Definition, and Classification Criteria}

We quantify the relative timing between the spectral evolution peak and the intensity peak as follows. Since both $E_{\rm p}$ and the energy flux $F$ are derived from the same BBlocks time bins, their peak times are defined directly from the time-resolved spectral sequence. If the maximum $E_{\rm p}$ occurs in the time bin $[t_{{\rm start}},\,t_{{\rm stop}}]$, then the peak time of the spectral evolution is defined as $t_{\rm p}(E_{\rm p}) = [t_{\rm start}+t_{\rm stop})/2]E_{\rm p}$, with timing uncertainty $\sigma_{t_{\rm p}} =[(t_{\rm stop}- t_{\rm start})/2]E_{\rm p}$. The peak time of the energy flux is defined analogously, $t_{\rm p}(F) =[t_{\rm start} + t_{\rm stop}/2]F$, and $\sigma_{t_{\rm p}(F)} =[(t_{\rm stop}-t_{\rm start})/2]F$. By construction, the two peak times are measured on the same temporal grid and carry directly comparable uncertainties.

Based on the energy-flux peak time\footnote{For completeness, we also reconstruct the photon-counts light curve for each burst and compare its peak time with that of $E_{\rm p}$, the counts-based comparison serves only as a reference and cross-check (see the Appendix).} derived from the same spectral bins as $E_{\rm p}$, we define the spectral-intensity lag as
\begin{equation}
t_{\rm lag}^{\rm F} \equiv t_{\rm p}(E_{\rm p}) - t_{\rm p}(F).
\label{eg:t_lag}
\end{equation}
A negative value indicates that the $E_{\rm p}$ peak precedes the energy-flux peak, a value consistent with zero implies that the two peaks are aligned within the timing uncertainty, and a positive value indicates that the $E_{\rm p}$ peak follows the energy-flux peak. The corresponding uncertainty is defined as
\begin{equation}
\sigma_{t_{\rm lag}^{\rm F}} = \left[ \sigma_{t_{\rm p}(E_{\rm p})}^{2}+\sigma_{t_{\rm p}(F)}^{2} \right]^{1/2}.
\end{equation}

We classify the sample according to the sign and significance of $t_{\rm lag}^{\rm F}$. 
\begin{itemize}

\item Type~I (aligned-tracking): $|t_{\rm lag}^{\rm F}| \leq \sigma_{t_{\rm lag}^{\rm F}}$, indicating that the $E_{\rm p}(t)$ peak is consistent with the energy-flux peak within the timing uncertainty. Five of the 20 bursts belong to this subclass.

\item Type~II (early-peaking tracking): $t_{\rm lag}^{\rm F}< -\sigma_{t_{\rm lag}^{\rm F}}$, indicating that the $E_{\rm p}(t)$ peak precedes the energy-flux peak, with negative values of $t_{\rm lag}^{\rm F}$. This is the dominant subclass, comprising 13 bursts.

\item Type~III (late-peaking tracking): $t_{\rm lag}^{\rm F}> \sigma_{t_{\rm lag}^{\rm F}}$, indicating that the $E_{\rm p}(t)$ peak occurs later than the energy-flux peak, with a positive value of $t_{\rm lag}^{\rm F}$. This subclass contains two bursts. 

\end{itemize}
This classification naturally separates the intensity-tracking bursts into temporally aligned, early-peaking, and late-peaking subclasses. 

\section{Results} \label{sec:Result}

\subsection{Global Properties of the Sample}

Using the time-resolved spectral results based on the 3ML analysis, we construct a unified bin-level database for the 20 single-pulse GRBs in our final sample, retaining only bins with $S>15$. For each burst, we measured the peak time of the energy flux, $t_{\rm p}(F)$, the peak time of the spectral peak energy, $t_{\rm p}(E_{\rm p})$, and the flux-based lag $t_{\rm lag}^{\rm F}$. We further derived the normalized lag $\hat{t}_{\rm lag}=t_{\rm lag}^{\rm F}/{\rm FWHM}_{F}$, the classification probabilities $(P_{\rm I},P_{\rm II},P_{\rm III})$ from a Gaussian approximation, summary statistics of the low-energy photon index $\alpha$, and a set of flux-pulse morphology parameters. The burst-level properties are summarized in Table~\ref{tab:burst-summary}, and the pulse morphology and correlation quantities are listed in Table~\ref{tab:morph-summary}.

\begin{figure*}
\centering
\includegraphics[width=0.78\textwidth]{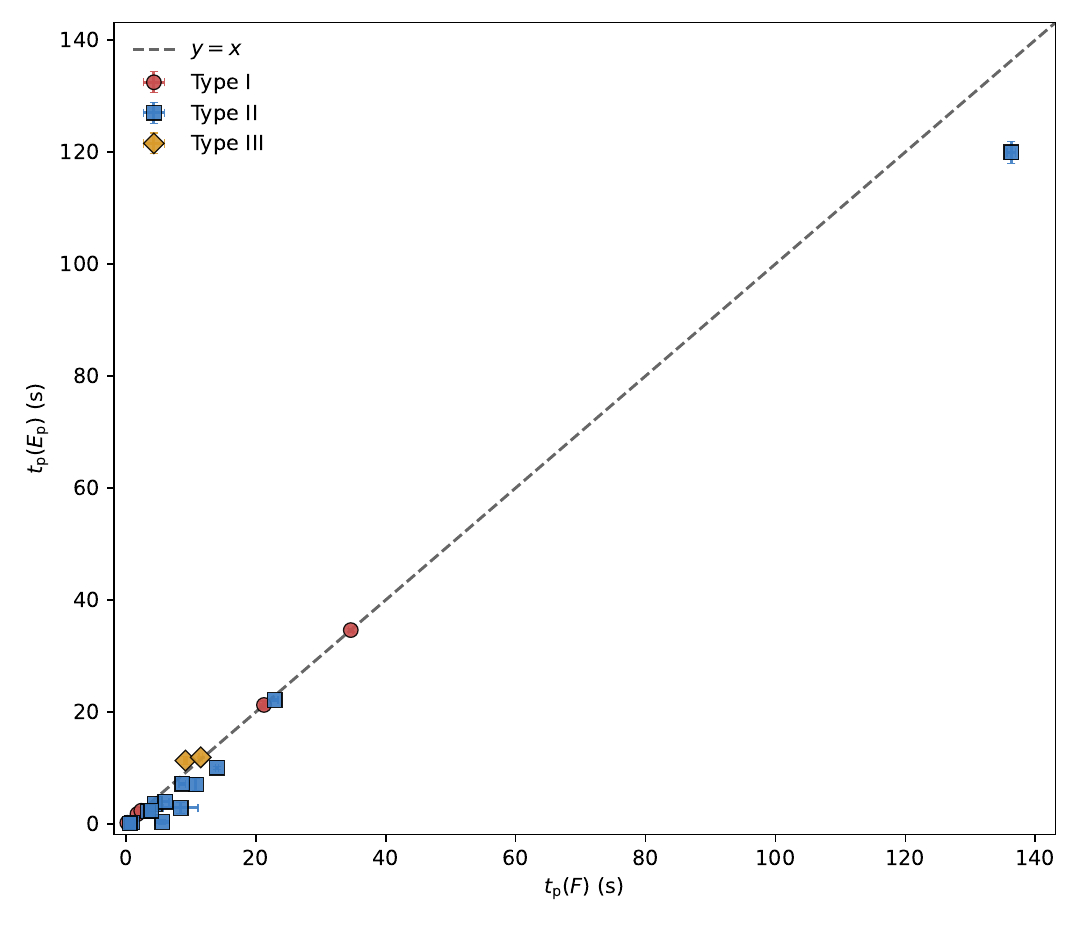}
\caption{Comparison between the peak times of the spectral peak energy and the energy flux. The horizontal axis shows $t_{\rm p}(F)$, the vertical axis shows $t_{\rm p}(E_{\rm p})$, and the dashed line marks $t_{\rm p}(E_{\rm p})=t_{\rm p}(F)$. Type~I bursts cluster around the one-to-one relation, Type~II bursts lie below it, and the two Type~III bursts lie above it.}
\label{fig:tp-scatter}
\end{figure*}

The peak-time relation between $t_{\rm p}(E_{\rm p})$ and $t_{\rm p}(F)$ is shown in Figure~\ref{fig:tp-scatter}. The bursts separate clearnly around the $y=x$ diagonal into three subclasses. Five bursts satisfy $|t_{\rm lag}^{\rm F}|\leq \sigma_{t_{\rm lag}^{\rm F}}$ (Type~I), 13 bursts having negative lags are classified as Type~II, and two bursts having positive lags are classified as Type~III. The flux-tracking sample is thus dominated by early-peaking events (65\%), while the aligned and late-peaking subclasses account for 25\% and 10\% of the sample, respectively. These fractions demonstrate that the traditional intensity-tracking category is not internally homogeneous once the peak ordering is resolved on the spectrally matched $E_{\rm p}(t)$ and $F(t)$ time series.

\begin{deluxetable*}{ccccccccccccccccc}
\tablewidth{0pt}
\tabletypesize{\scriptsize}
\tablecaption{Burst-level Summary for the Flux-based Classification\label{tab:burst-summary}}
\tablehead{
\colhead{GRB} & \colhead{$z$} & \colhead{$T_{90}$} & \colhead{Model} & \colhead{$N_{\rm bin}$} & \colhead{$t_{\rm p}(F)$} & \colhead{$t_{\rm p}(E_{\rm p})$} & \colhead{$t_{\rm lag}^{\rm F}$} & \colhead{$Z_{\rm lag}$} & \colhead{$\hat{t}_{\rm lag}$} & \colhead{$P_{\rm I}$} & \colhead{$P_{\rm II}$} & \colhead{$P_{\rm III}$} & \colhead{$\bar{\alpha}_{\rm w}$} & \colhead{$\alpha_{\rm peak}^{F}$} & \colhead{$f_{-2/3}$} & \colhead{Class} \\
 & & (s) & & & (s) & (s) & (s) & & & & & & & & &
}
\startdata
090620400 & \nodata & 13.57$\pm$0.72 & CPL & 7 & 4.44$\pm$0.36 & 3.58$\pm$0.49 & -0.85$\pm$0.61 & -1.40 & -0.32 & 0.34 & 0.65 & 0.01 & -0.21 & -0.05 & 0.86 & II \\
090719063 & \nodata & 11.39$\pm$0.47 & CPL & 11 & 5.61$\pm$1.17 & 0.36$\pm$0.36 & -5.25$\pm$1.22 & -4.30 & -0.72 & 0.00 & 1.00 & 0.00 & -0.40 & -0.41 & 0.64 & II \\
090804940 & \nodata & 5.57$\pm$0.36 & CPL & 8 & 1.79$\pm$0.51 & 1.79$\pm$0.51 & 0.00$\pm$0.72 & 0.00 & 0.00 & 0.68 & 0.16 & 0.16 & -0.36 & -0.37 & 1.00 & I \\
090820027 & \nodata & 12.42$\pm$0.18 & CPL & 19 & 34.63$\pm$0.34 & 34.63$\pm$0.34 & 0.00$\pm$0.48 & 0.00 & 0.00 & 0.68 & 0.16 & 0.16 & -0.70 & -0.69 & 0.32 & I \\
100122616 & \nodata & 22.53$\pm$2.77 & Band & 6 & 21.26$\pm$0.71 & 21.26$\pm$0.71 & 0.00$\pm$1.00 & 0.00 & 0.00 & 0.68 & 0.16 & 0.16 & -1.19 & -1.34 & 0.17 & I \\
100528075 & \nodata & 22.46$\pm$0.75 & CPL & 8 & 8.43$\pm$2.66 & 2.89$\pm$1.00 & -5.54$\pm$2.84 & -1.95 & -0.54 & 0.17 & 0.83 & 0.00 & -1.02 & -1.00 & 0.00 & II \\
100707032 & \nodata & 81.79$\pm$1.22 & CPL & 13 & 0.95$\pm$0.24 & 0.27$\pm$0.09 & -0.68$\pm$0.26 & -2.60 & -0.24 & 0.05 & 0.95 & 0.00 & -0.06 & 0.38 & 0.69 & II \\
120426090 & \nodata & 2.69$\pm$0.09 & CPL & 12 & 0.62$\pm$0.11 & 0.11$\pm$0.07 & -0.51$\pm$0.13 & -3.76 & -0.28 & 0.00 & 1.00 & 0.00 & -0.52 & -0.53 & 0.67 & II \\
120919309 & \nodata & 21.25$\pm$1.81 & CPL & 6 & 3.43$\pm$0.76 & 2.27$\pm$0.41 & -1.16$\pm$0.86 & -1.35 & -0.34 & 0.35 & 0.64 & 0.01 & -0.78 & -0.72 & 0.00 & II \\
130614997 & \nodata & 9.28$\pm$1.97 & CPL & 5 & 0.23$\pm$0.23 & 0.23$\pm$0.23 & 0.00$\pm$0.33 & 0.00 & 0.00 & 0.68 & 0.16 & 0.16 & -1.36 & -1.39 & 0.00 & I \\
131231198 & 0.642 & 31.23$\pm$0.57 & CPL & 30 & 22.91$\pm$0.50 & 22.16$\pm$0.25 & -0.75$\pm$0.56 & -1.34 & -0.14 & 0.36 & 0.63 & 0.01 & -1.05 & -0.79 & 0.00 & II \\
141028455 & 2.330 & 31.49$\pm$2.43 & CPL & 11 & 10.79$\pm$0.12 & 7.06$\pm$1.11 & -3.73$\pm$1.11 & -3.35 & -0.63 & 0.01 & 0.99 & 0.00 & -0.88 & -0.53 & 0.18 & II \\
150213001 & \nodata & 4.10$\pm$0.09 & CPL & 16 & 2.38$\pm$0.15 & 2.38$\pm$0.15 & 0.00$\pm$0.21 & 0.00 & 0.00 & 0.68 & 0.16 & 0.16 & -1.27 & -0.96 & 0.00 & I \\
150902733 & \nodata & 13.57$\pm$0.36 & CPL & 15 & 9.17$\pm$0.24 & 11.33$\pm$0.72 & 2.16$\pm$0.76 & 2.84 & 0.73 & 0.03 & 0.00 & 0.97 & -0.60 & -0.42 & 0.67 & III \\
160509374 & 1.170 & 369.67$\pm$0.81 & Band & 23 & 14.03$\pm$0.22 & 10.05$\pm$0.25 & -3.98$\pm$0.33 & -11.99 & -0.47 & 0.00 & 1.00 & 0.00 & -0.82 & -0.69 & 0.13 & II \\
160530667 & \nodata & 9.02$\pm$0.18 & Band & 39 & 6.08$\pm$0.58 & 3.98$\pm$0.10 & -2.10$\pm$0.59 & -3.59 & -0.53 & 0.00 & 1.00 & 0.00 & -0.67 & -0.61 & 0.49 & II \\
160821857 & \nodata & 43.01$\pm$0.72 & Band & 35 & 136.30$\pm$0.67 & 119.93$\pm$1.96 & -16.37$\pm$2.07 & -7.92 & -2.48 & 0.00 & 1.00 & 0.00 & -0.96 & -0.92 & 0.00 & II \\
160910722 & \nodata & 24.32$\pm$0.36 & Band & 11 & 8.68$\pm$0.48 & 7.20$\pm$0.21 & -1.48$\pm$0.52 & -2.84 & -0.51 & 0.03 & 0.97 & 0.00 & -0.39 & -0.24 & 1.00 & II \\
180305393 & \nodata & 13.06$\pm$0.81 & CPL & 10 & 3.92$\pm$0.47 & 2.37$\pm$0.41 & -1.54$\pm$0.62 & -2.48 & -0.46 & 0.07 & 0.93 & 0.00 & -0.28 & -0.20 & 1.00 & II \\
180728728 & 0.117 & 6.40$\pm$0.36 & CPL & 18 & 11.53$\pm$0.26 & 11.91$\pm$0.12 & 0.38$\pm$0.29 & 1.33 & 0.16 & 0.36 & 0.01 & 0.63 & -1.65 & -1.61 & 0.00 & III \\
\enddata
\tablecomments{All quantities are derived from the same time-resolved spectral bins. The lag probabilities are computed assuming a Gaussian approximation for $t_{\rm lag}^{\rm F}$ with width $\sigma_{t_{\rm lag}^{\rm F}}$, and $f_{-2/3}$ denotes the fraction of bins with $\alpha>-2/3$.}
\end{deluxetable*}

\subsection{Distribution of Peak-time Offsets}

\begin{figure*}
\centering
\includegraphics[width=0.95\textwidth]{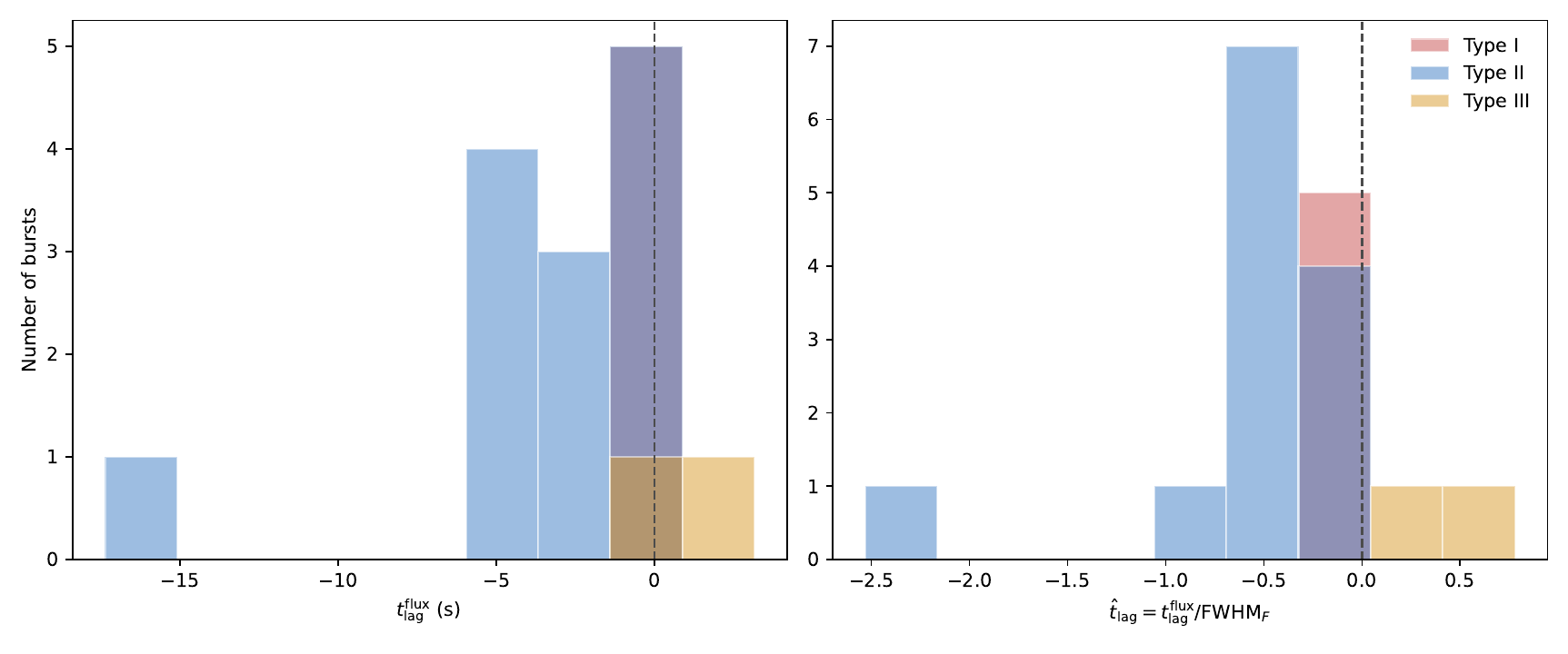}
\caption{Distributions of the peak-time lag. Left: the absolute lag $t_{\rm lag}^{\rm F}$. Right: the width-normalized lag $\hat{t}_{\rm lag}$. The early-peaking Type~II events dominate both distributions, while the two Type~III bursts form a positive-lag tail.}
\label{fig:lag-dist}
\end{figure*}

The subclass separation is particularly clear in the lag distribution shown in Figure~\ref{fig:lag-dist}. The full sample has a median lag of $-0.80$~s, with subclass medians of $0.00$~s for Type~I, $-1.54$~s for Type~II, and $1.27$~s for Type~III. This ordering persists after normalizing by the flux-pulse width, yielding median values of $\hat{t}_{\rm lag}=0.00$, $-0.46$, and $0.45$, respectively. The consistency of this ordering in the width-normalized distribution indicates that the predominance of Type~II bursts cannot be attributed to a small number of exceptionally long pulses.

Notable Type~II events include GRB~160509374, GRB~160821857, GRB~090719063, GRB~120426090, and GRB~160530667, all with $P_{\rm II}>0.99$, their negative lags remain substantial following normalization by ${\rm FWHM}_{F}$. At the other extreme, GRB~150902733 represents the most unambiguous Type~III burst, with $P_{\rm III}=0.97$, while GRB~180728728 occupies the same positive-lag regime with lower significance. The aligned Type I subclass is correspondingly well-concentrated: lags for all five Type~I bursts are consistent with zero within the bin-scale timing uncertainty.

\subsection{Temporal Properties of the Three Types}

\begin{figure*}
\includegraphics[width=0.5\hsize,clip]{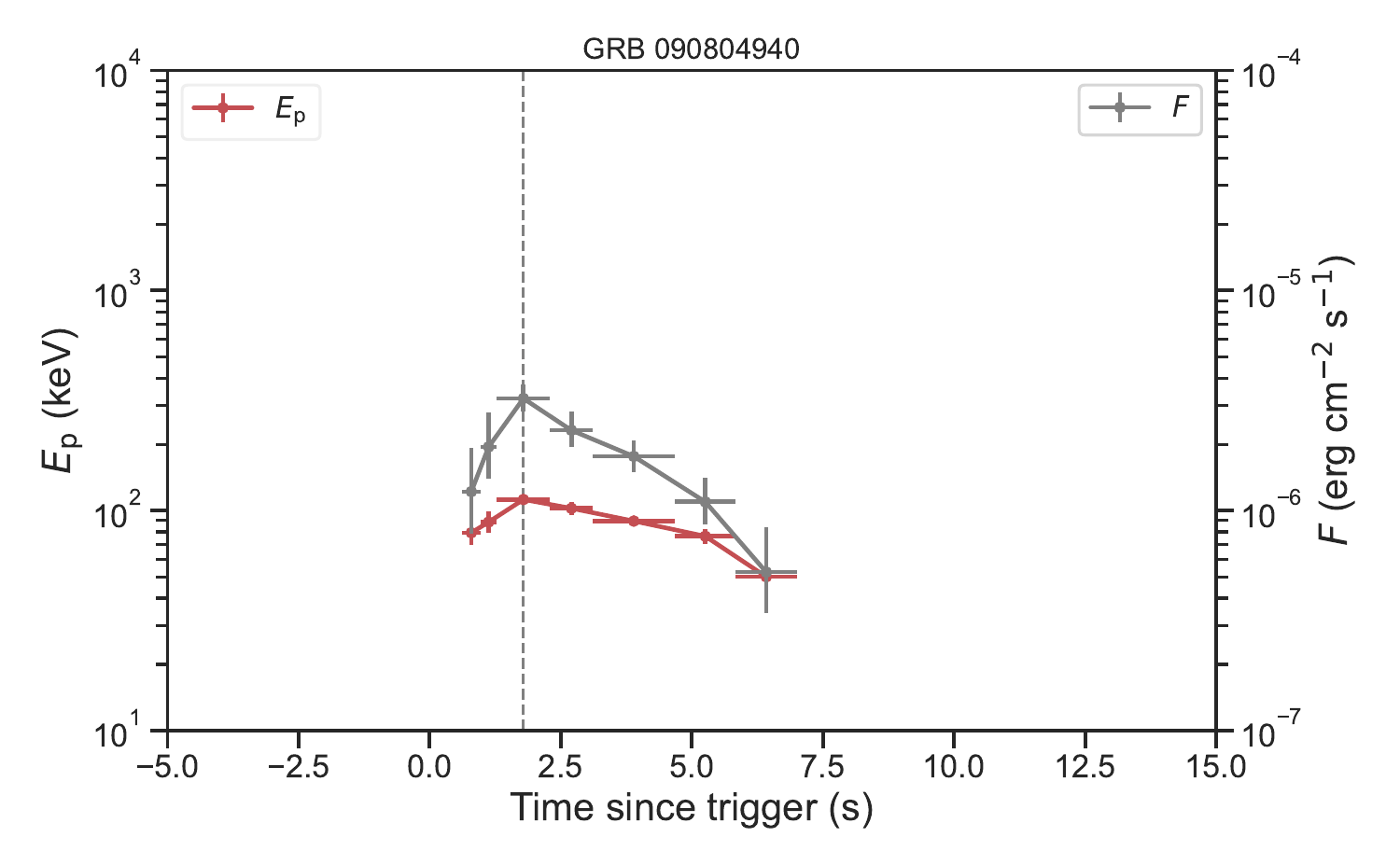}
\includegraphics[width=0.5\hsize,clip]{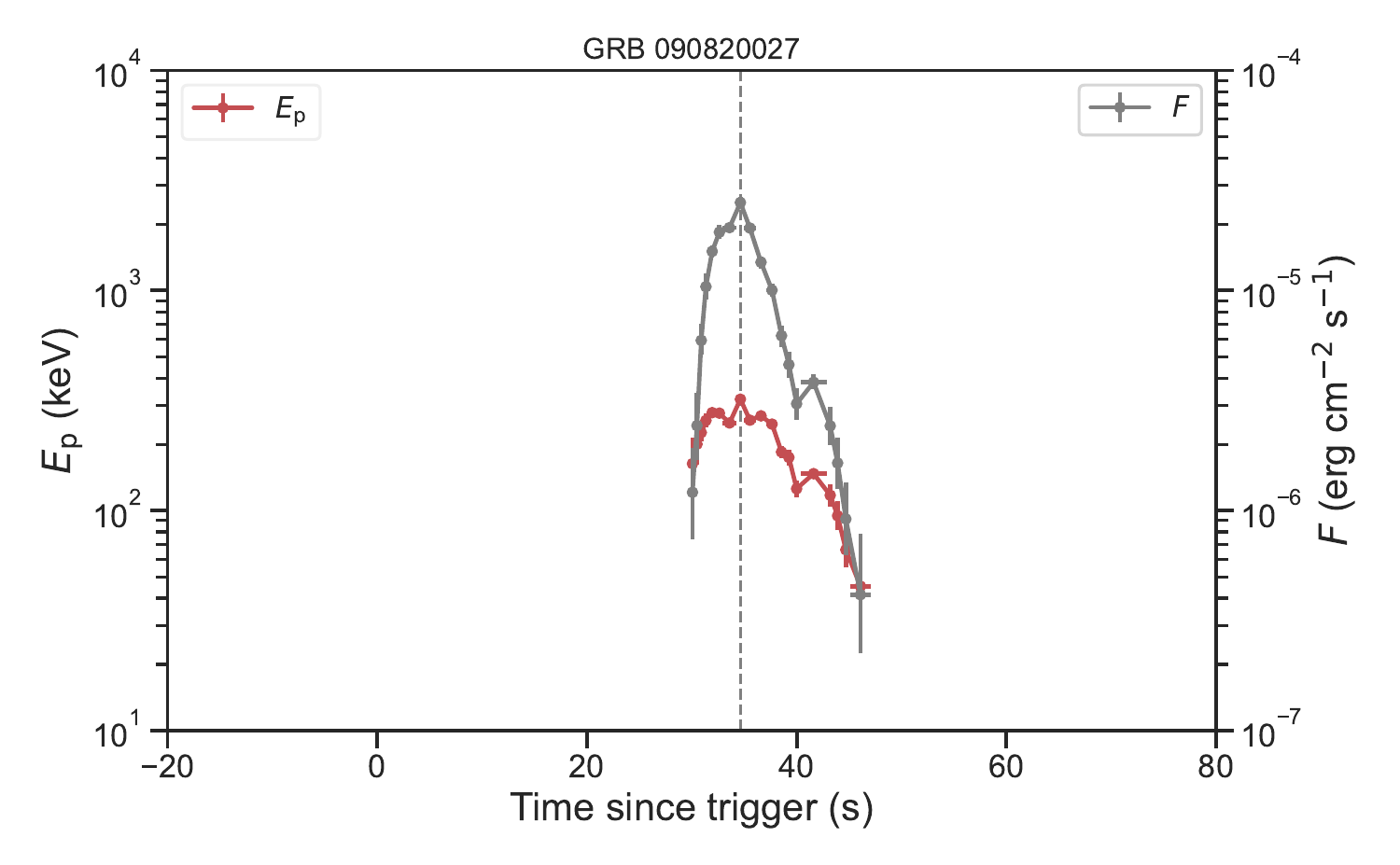}
\includegraphics[width=0.5\hsize,clip]{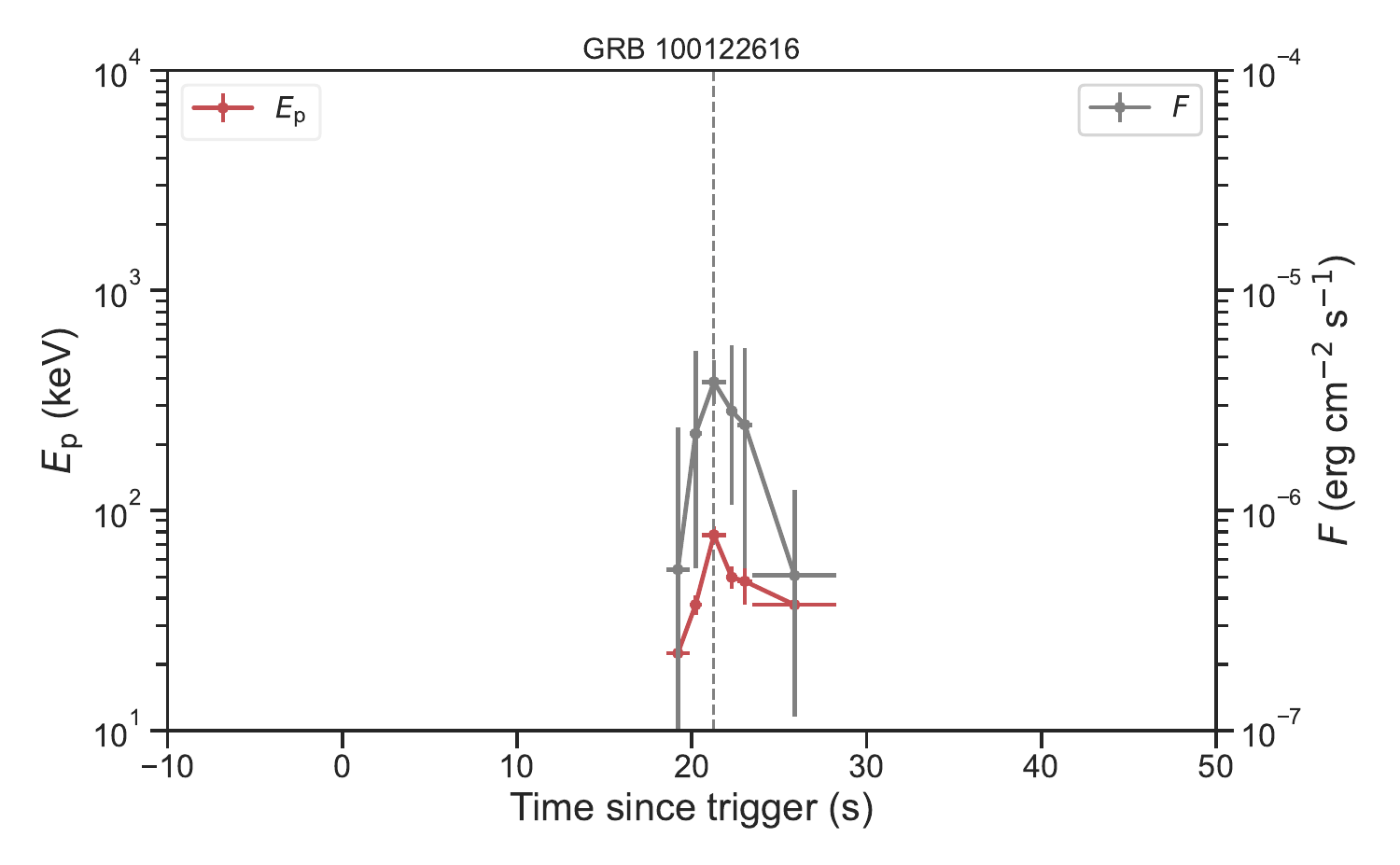}
\includegraphics[width=0.5\hsize,clip]{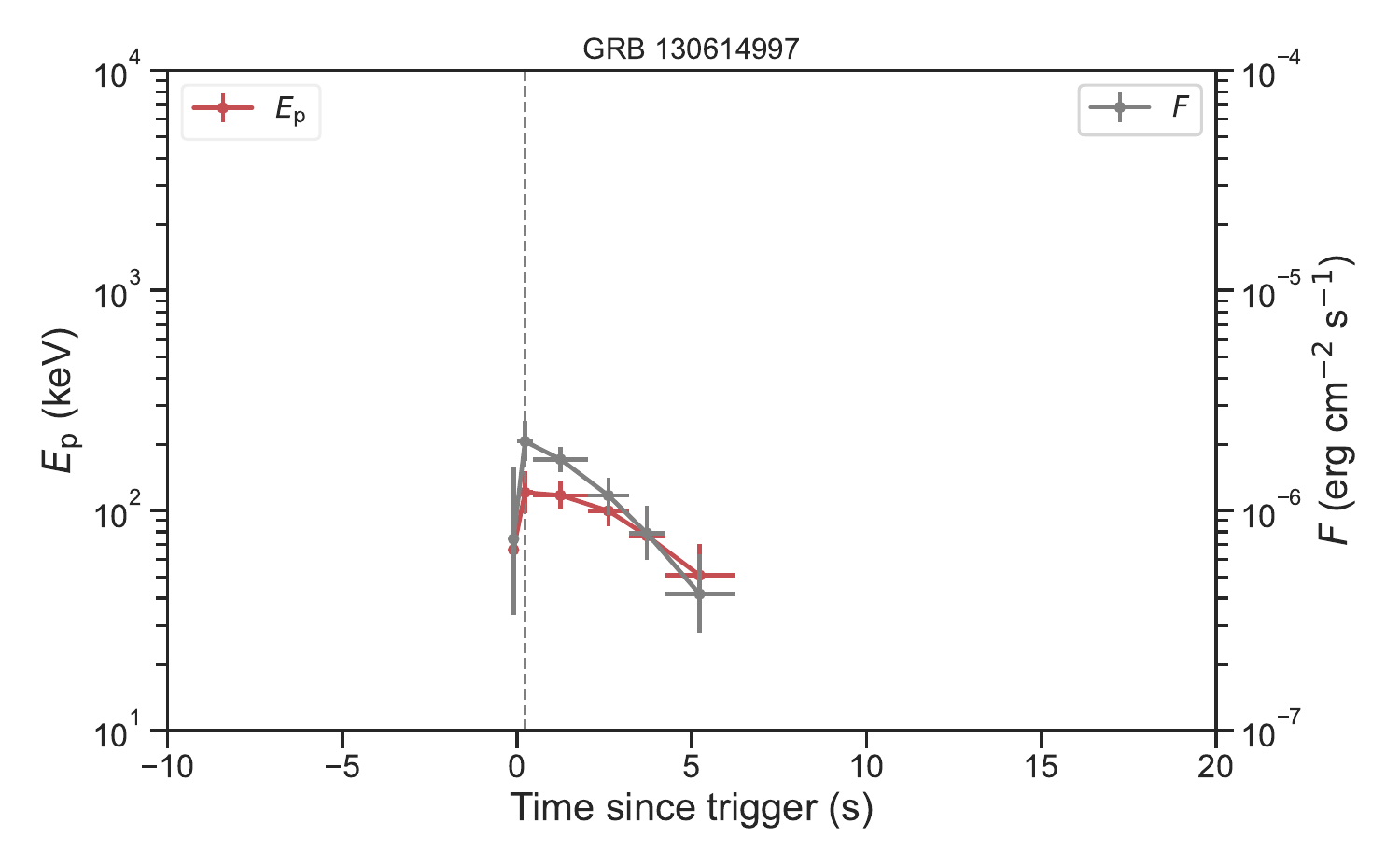}
\includegraphics[width=0.5\hsize,clip]{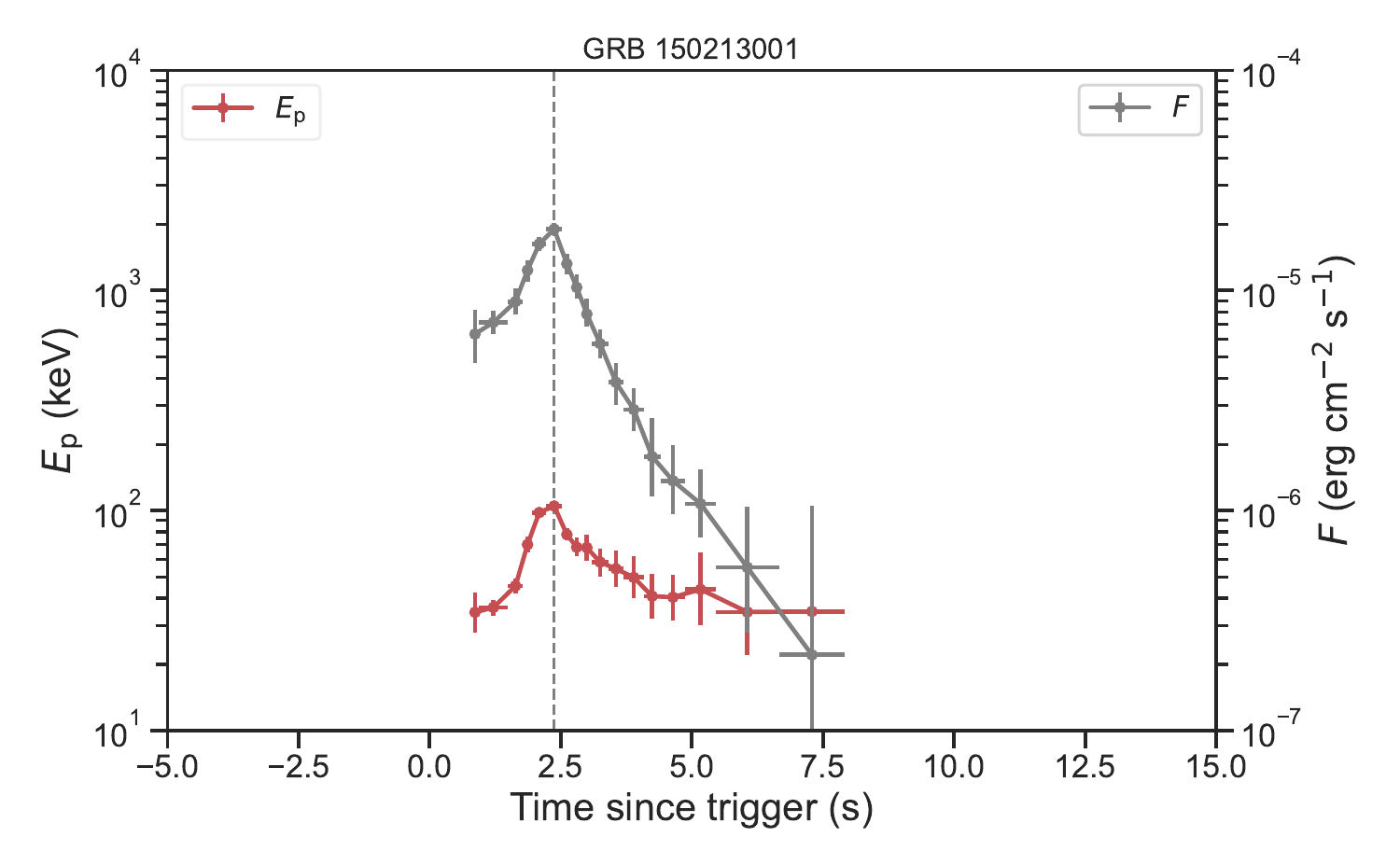}
\caption{All Type~I bursts in the sample. In each panel, the red points show $E_{\rm p}(t)$ and the grey points show $F(t)$ measured from the same time-resolved spectral bins. The peak times are aligned within the timing uncertainty.}
\label{fig:typeI-all}
\end{figure*}

\begin{figure*}
\includegraphics[width=0.5\hsize,clip]{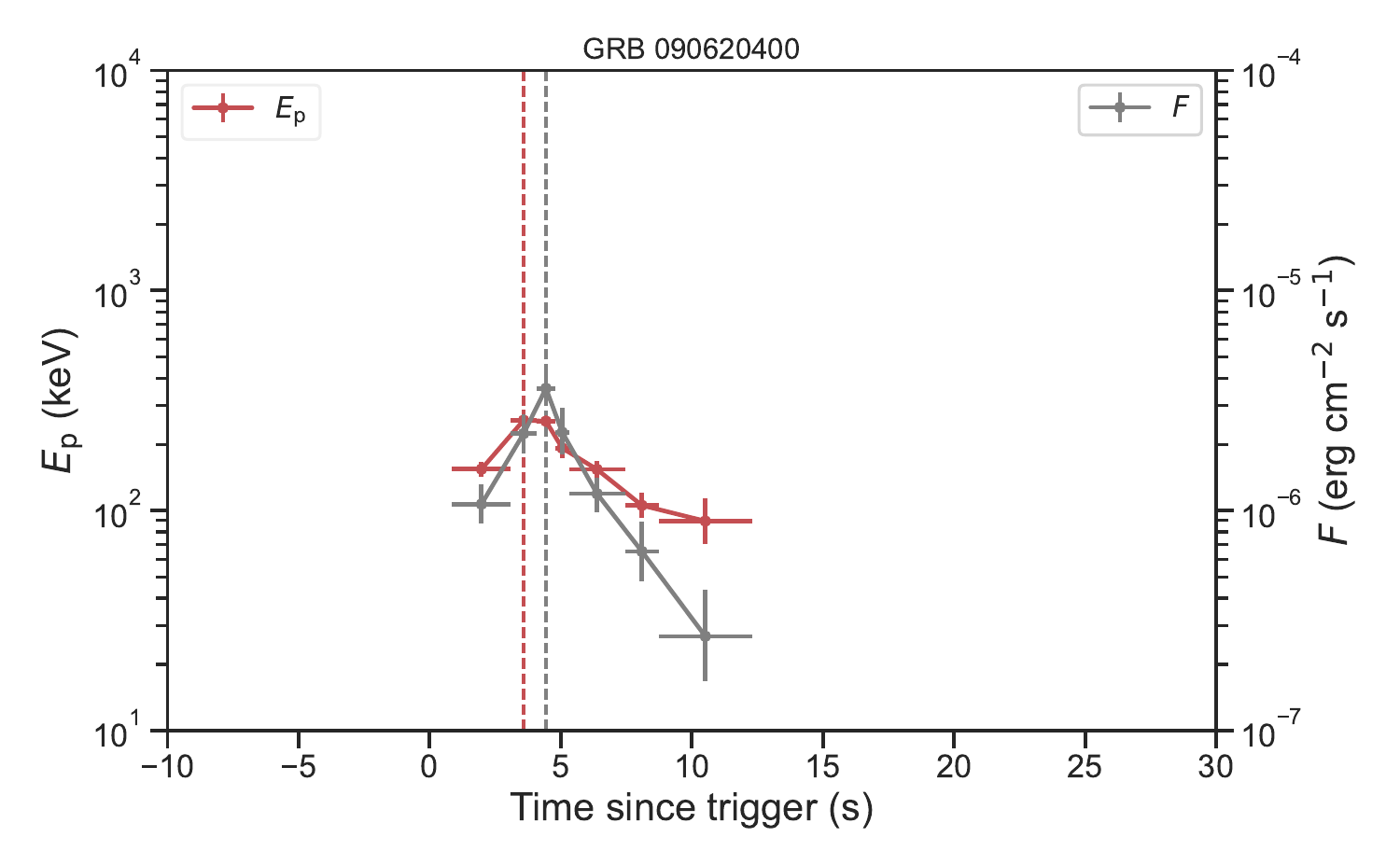}
\includegraphics[width=0.5\hsize,clip]{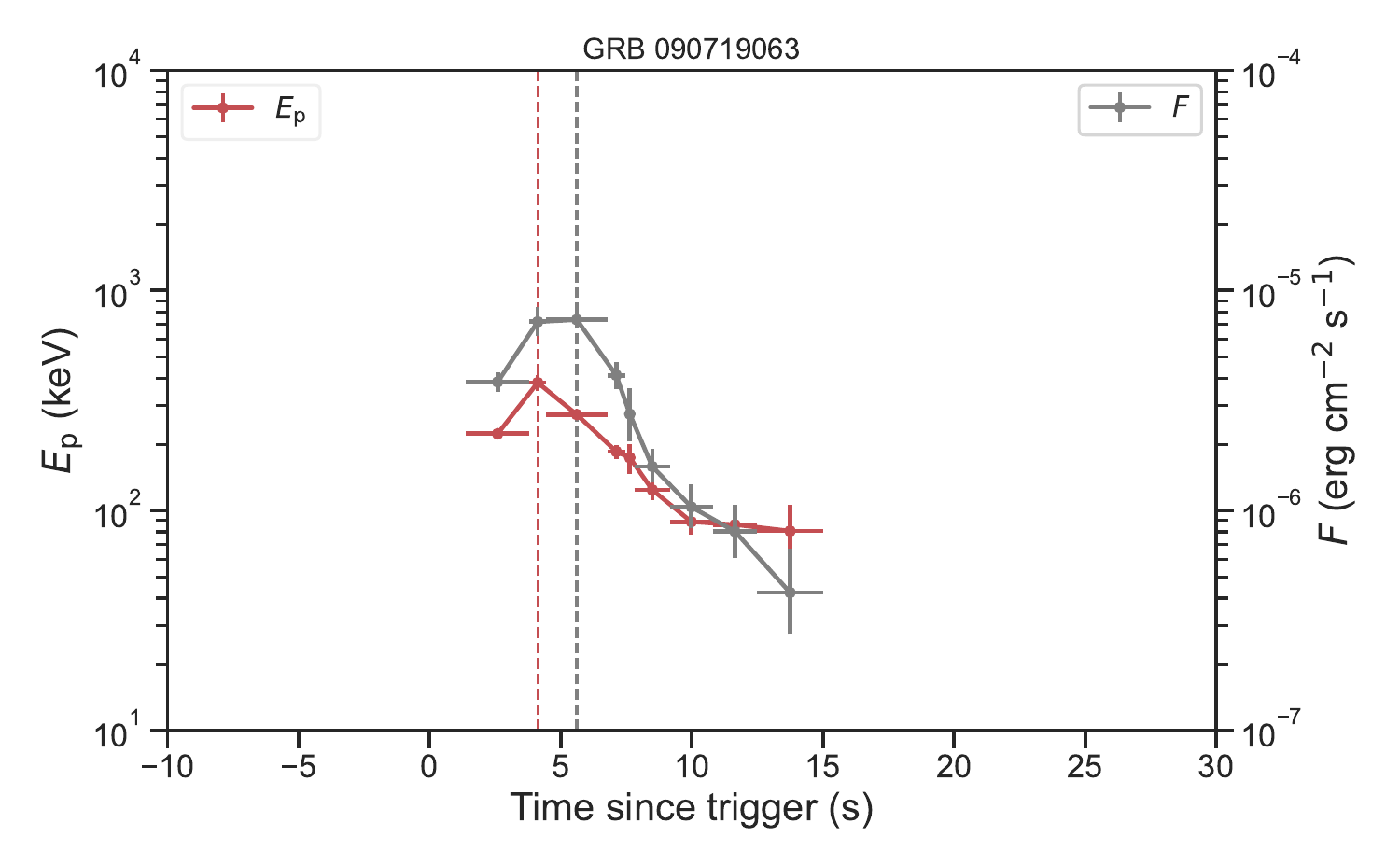}
\includegraphics[width=0.5\hsize,clip]{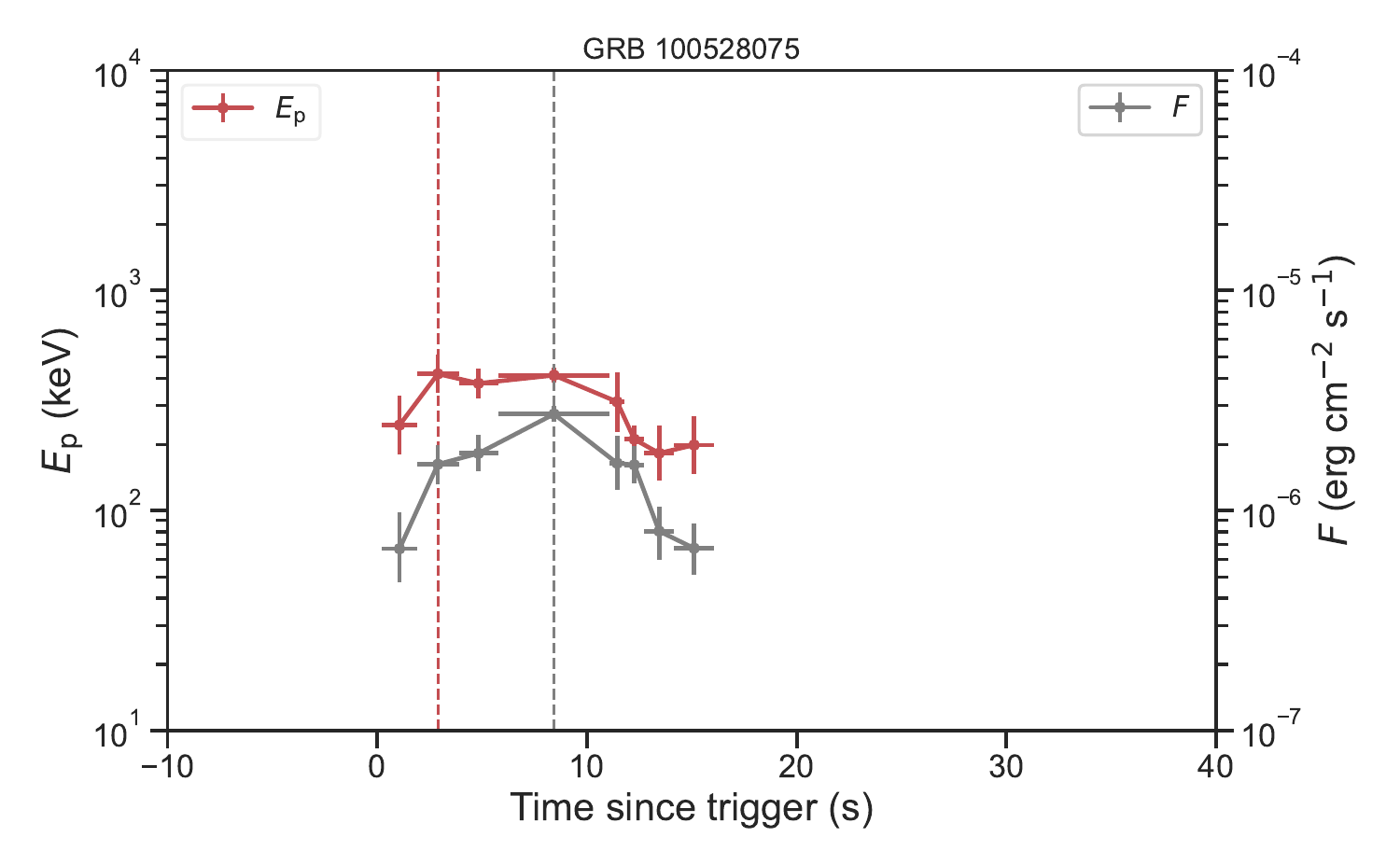}
\includegraphics[width=0.5\hsize,clip]{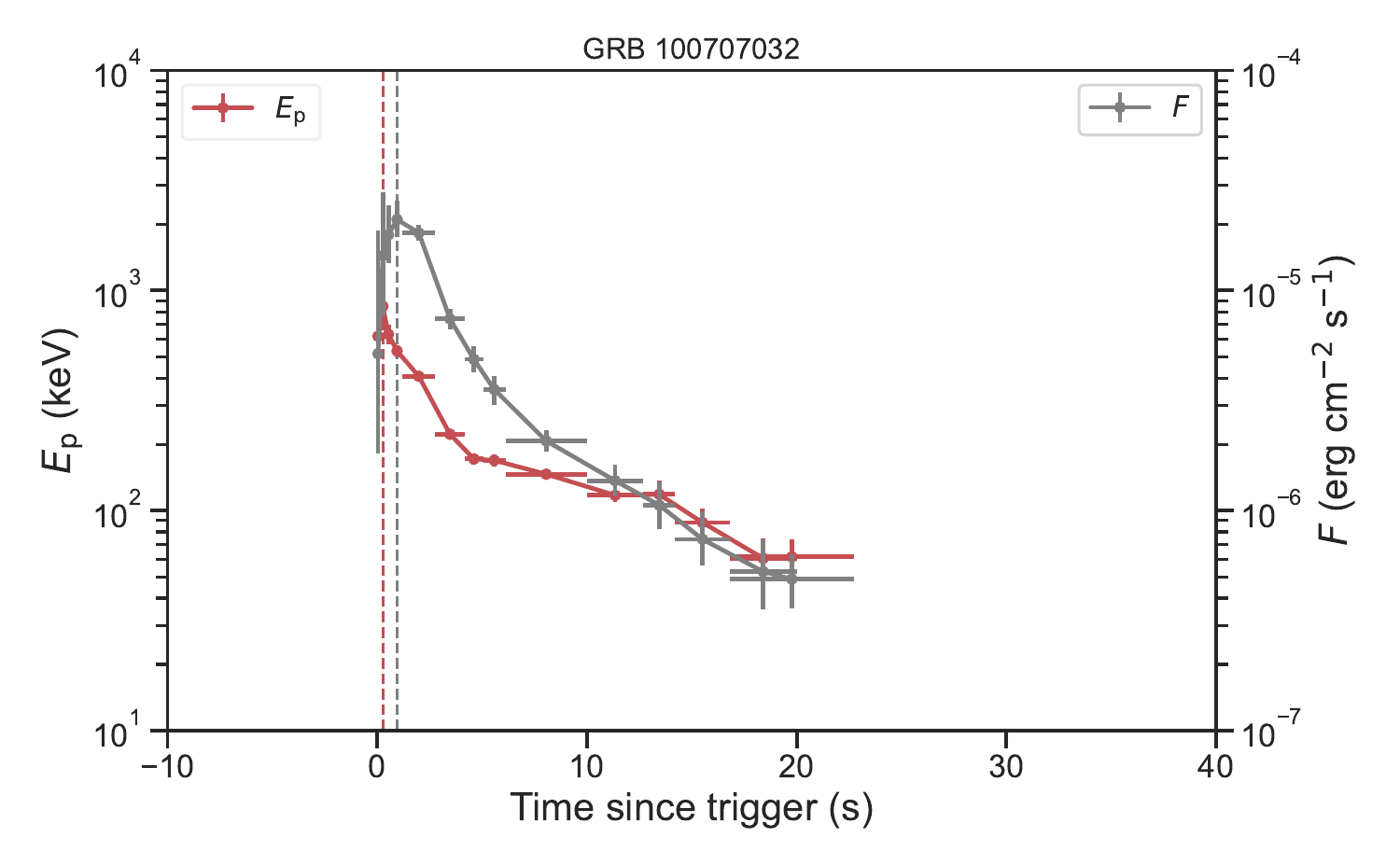}
\includegraphics[width=0.5\hsize,clip]{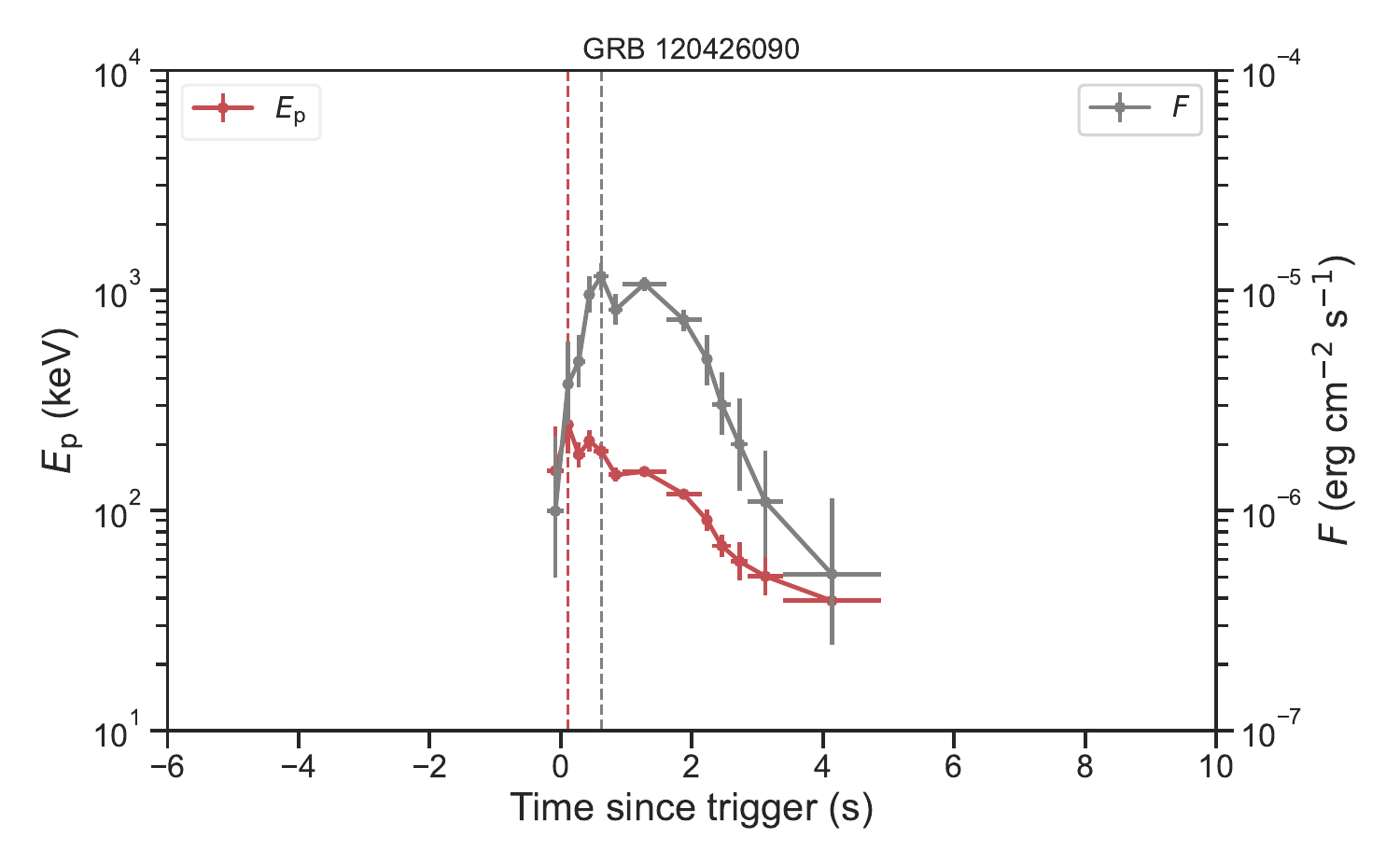}
\includegraphics[width=0.5\hsize,clip]{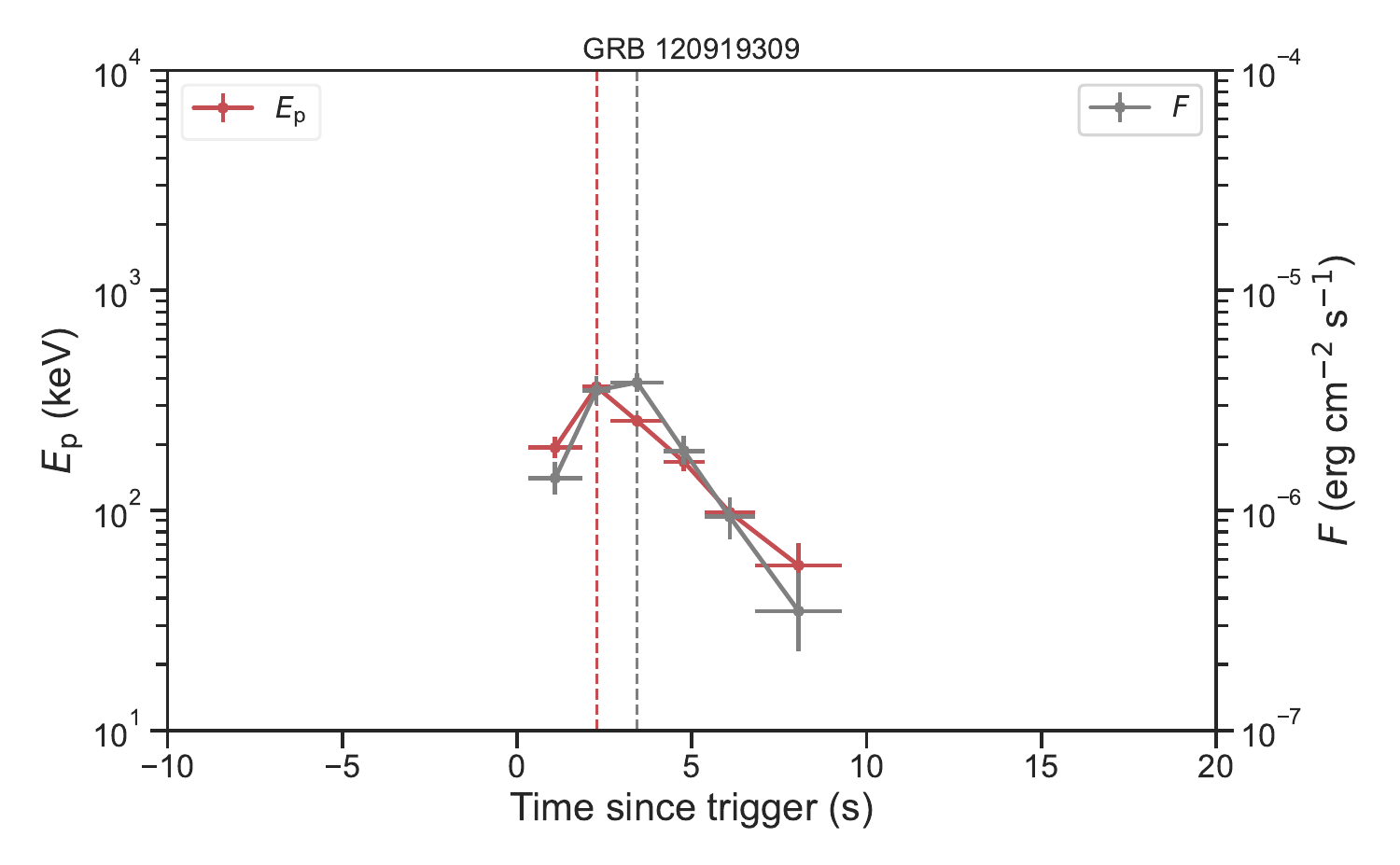}
\caption{Same as Fig. \ref{fig:typeI-all} but for all the Type~II bursts.}
\label{fig:typeII-all}
\end{figure*}

\begin{figure*}
\includegraphics[width=0.5\hsize,clip]{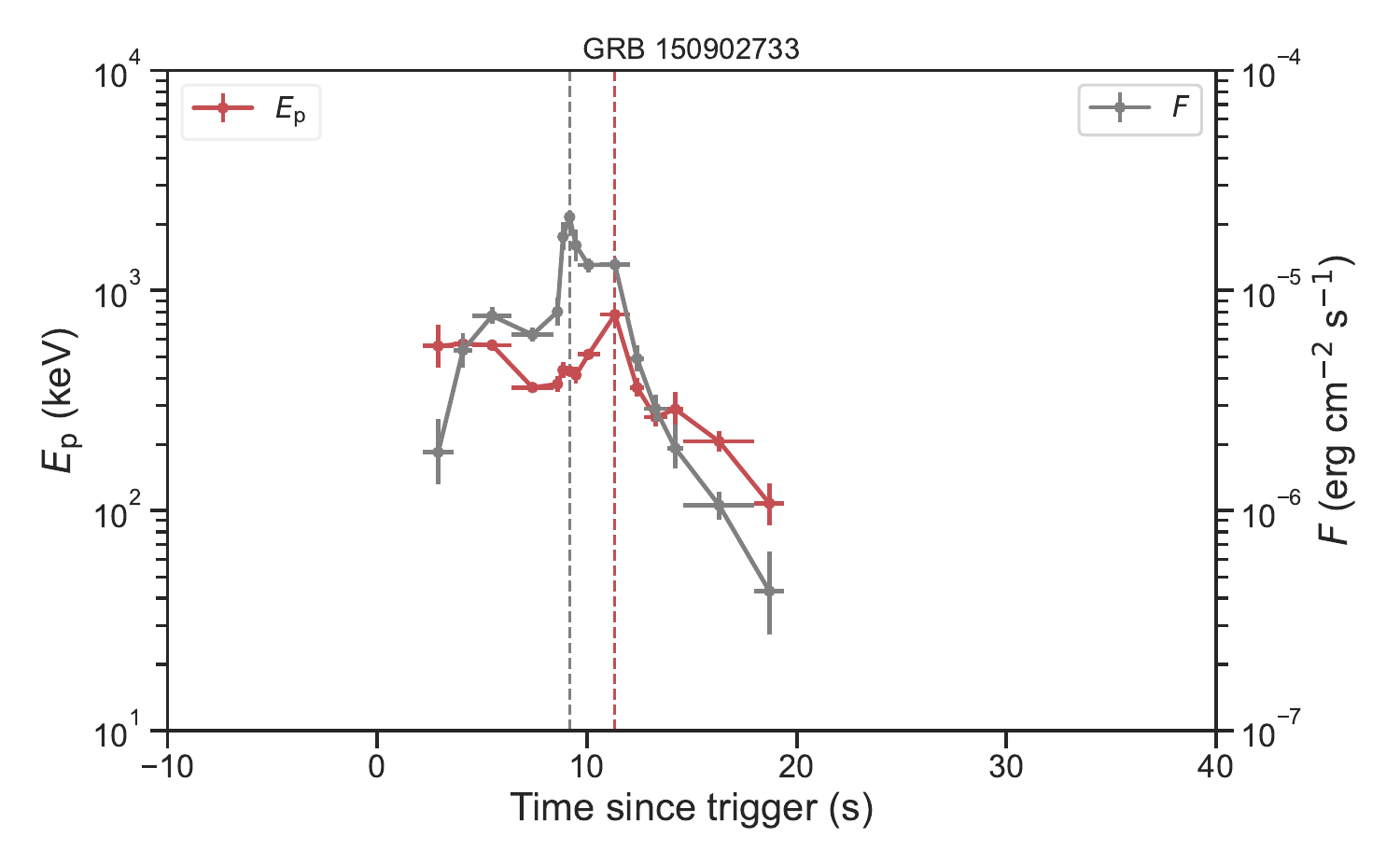}
\includegraphics[width=0.5\hsize,clip]{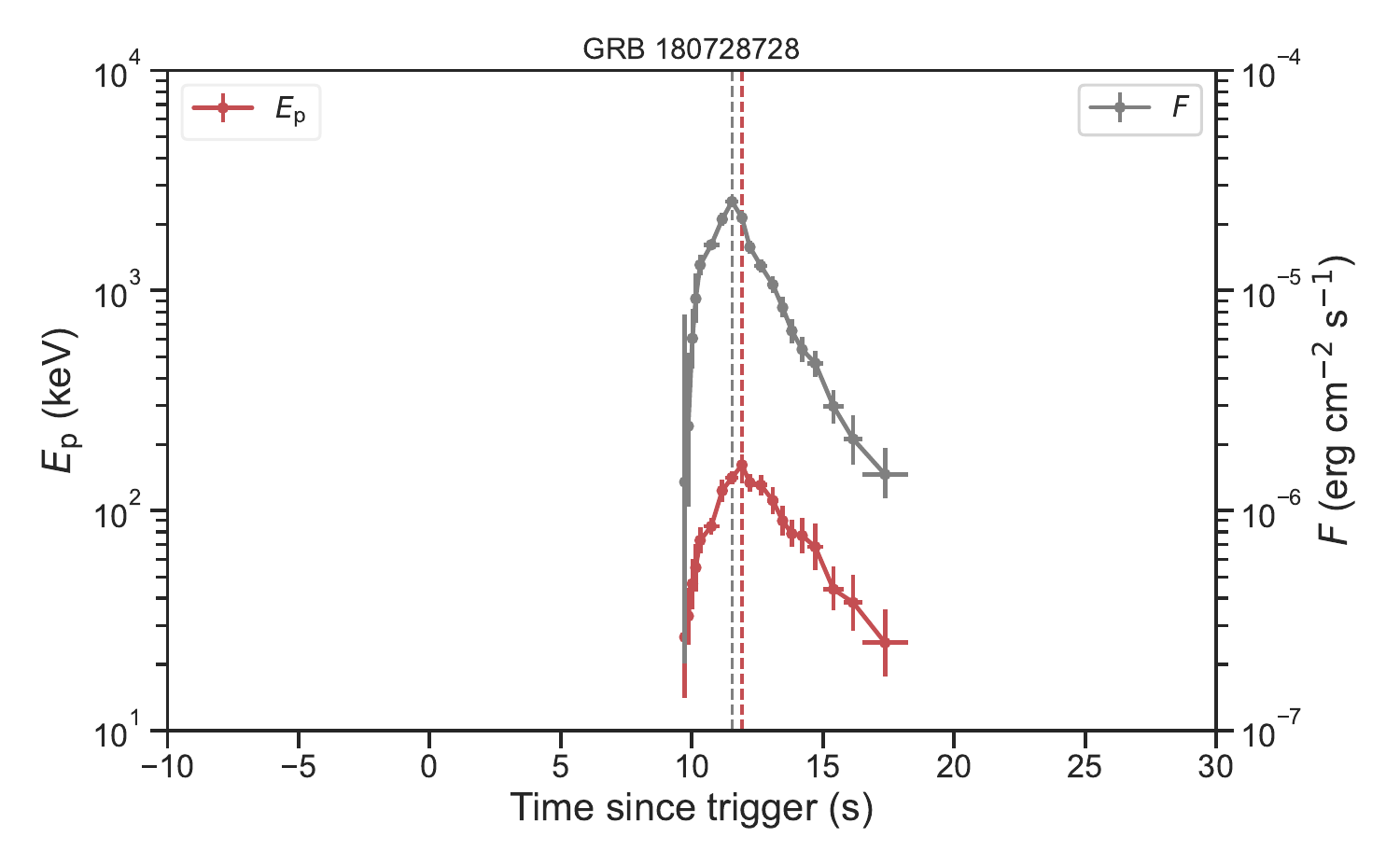}
\caption{Same as Fig. \ref{fig:typeI-all} but for all the Type~III bursts.}
\label{fig:typeIII-all}
\end{figure*}

The full set of matched $E_{\rm p}(t)$ and $F(t)$ sequences is shown in Figures~\ref{fig:typeI-all}-\ref{fig:typeIII-all}. These figures display only the spectrally matched time series used in the flux-based classification, thereby isolating the timing relation that defines each subclasses. All five Type~I bursts are shown in Figure~\ref{fig:typeI-all}. In each case, the $E_{\rm p}$ and $F$ maxima occur in the same time bin, or in bins whose centers are consistent within the timing uncertainty. The aligned Type I class therefore remains compact in peak-time alignment even though the detailed pulse widths and off-peak spectral evolution differ among burst.

The 13 Type~II bursts are displayed in Figures~\ref{fig:typeII-all}. These events define the dominant pattern in the sample. The spectral peak energy reaches its maximum while the energy flux is still rising, such that the hardest spectrum precedes the flux maximum. The magnitude of the offset varies substantially, from relatively modest offsets in GRB~090620400 and GRB~131231198 to markedly larger lags in GRB~141028455, GRB~160509374, and GRB~160821857. The two Type~III bursts, shown in Figure~\ref{fig:typeIII-all}, exhibit the opposite ordering, with the flux maximum occurs first, with the $E_{\rm p}$ peak following at later times.

\begin{figure*}
\includegraphics[width=0.5\hsize,clip]{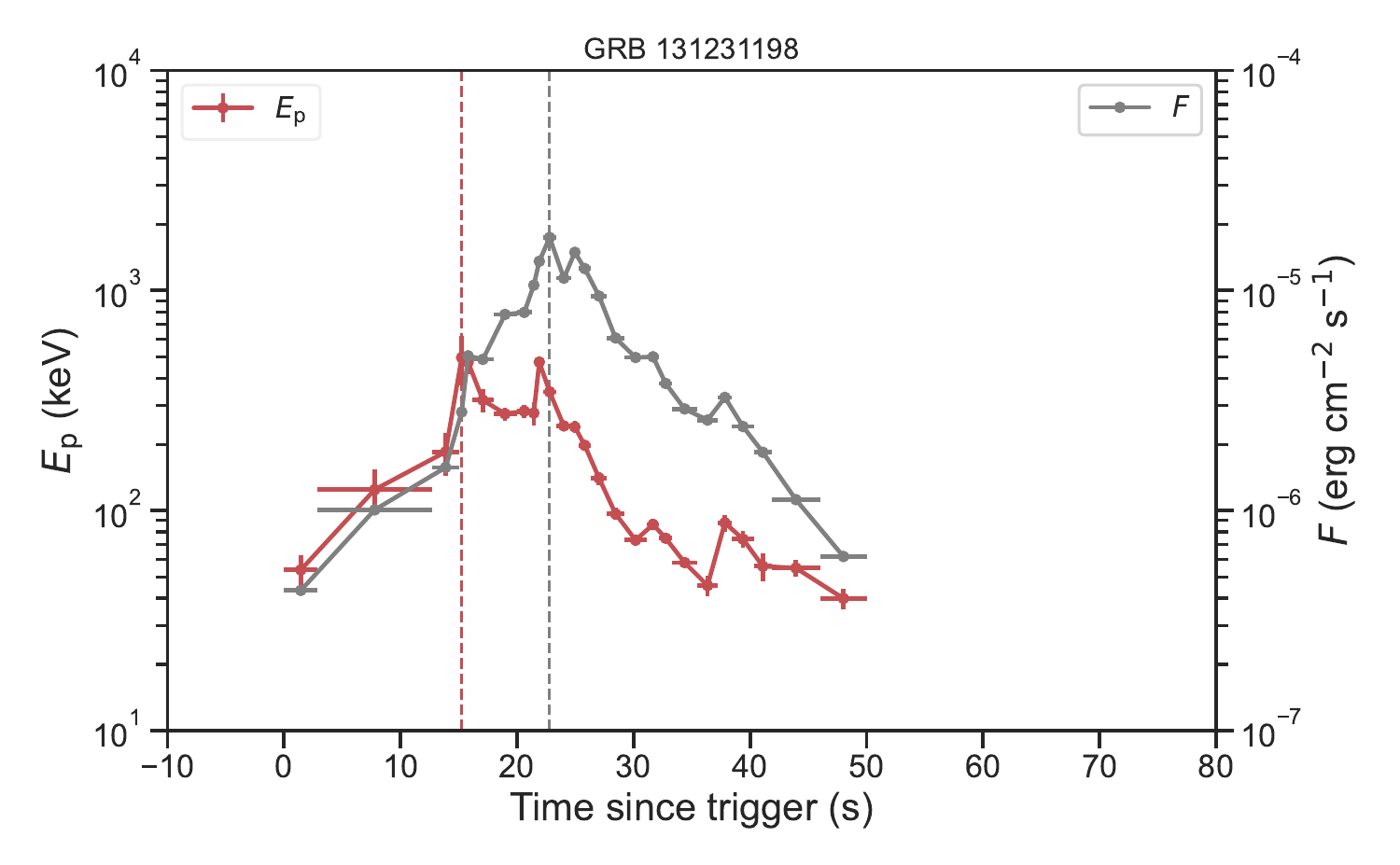}
\includegraphics[width=0.5\hsize,clip]{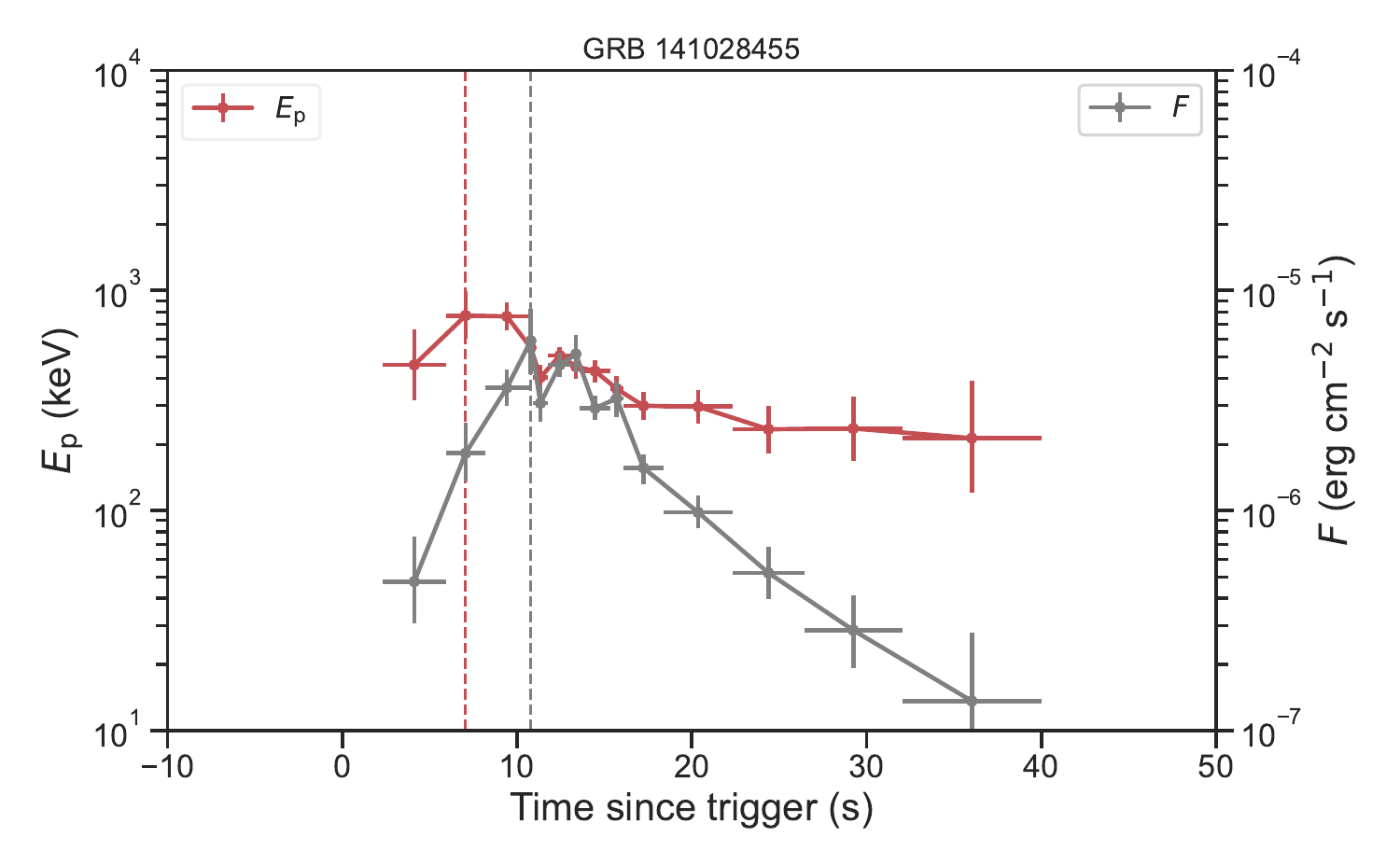}
\includegraphics[width=0.5\hsize,clip]{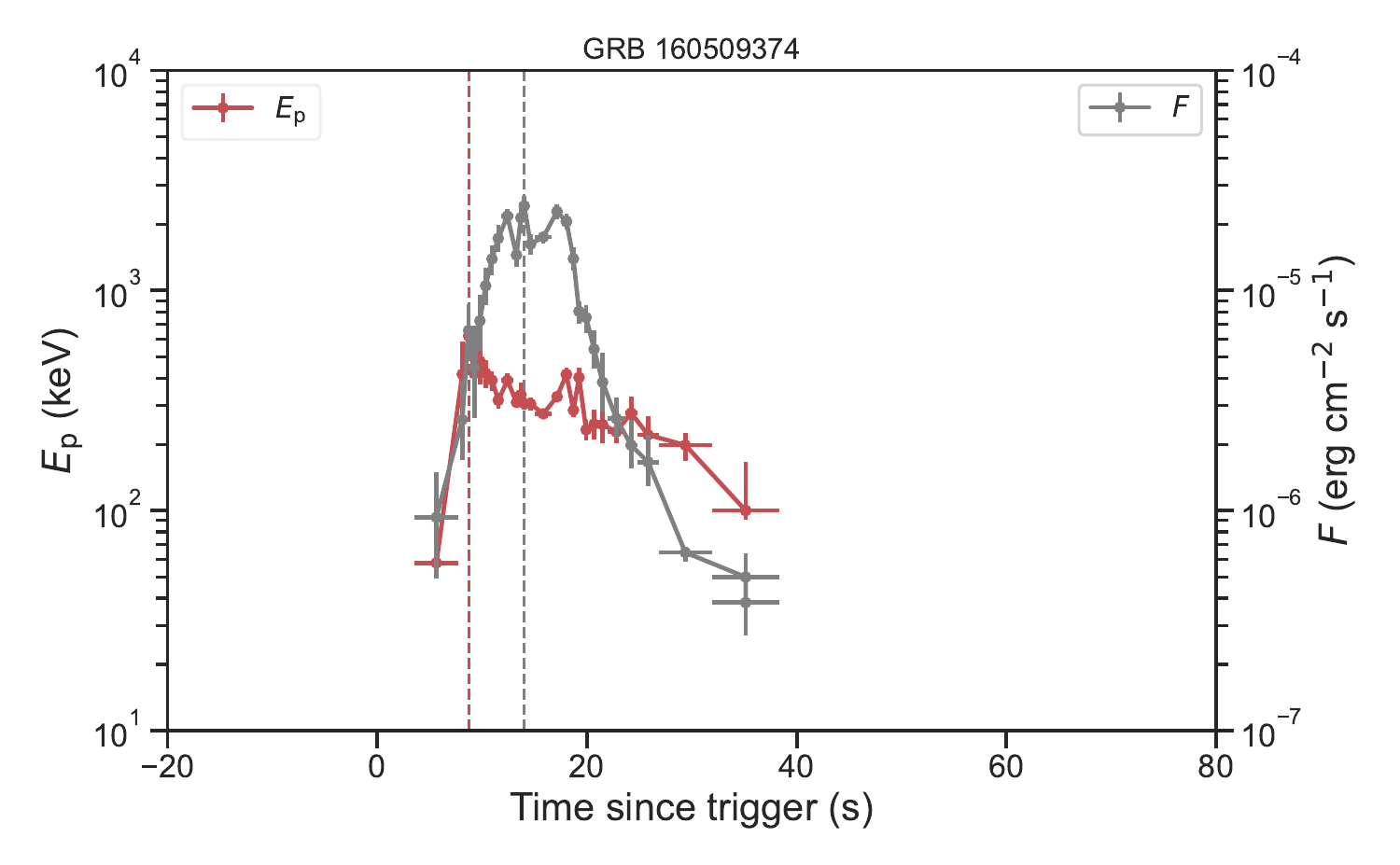}
\includegraphics[width=0.5\hsize,clip]{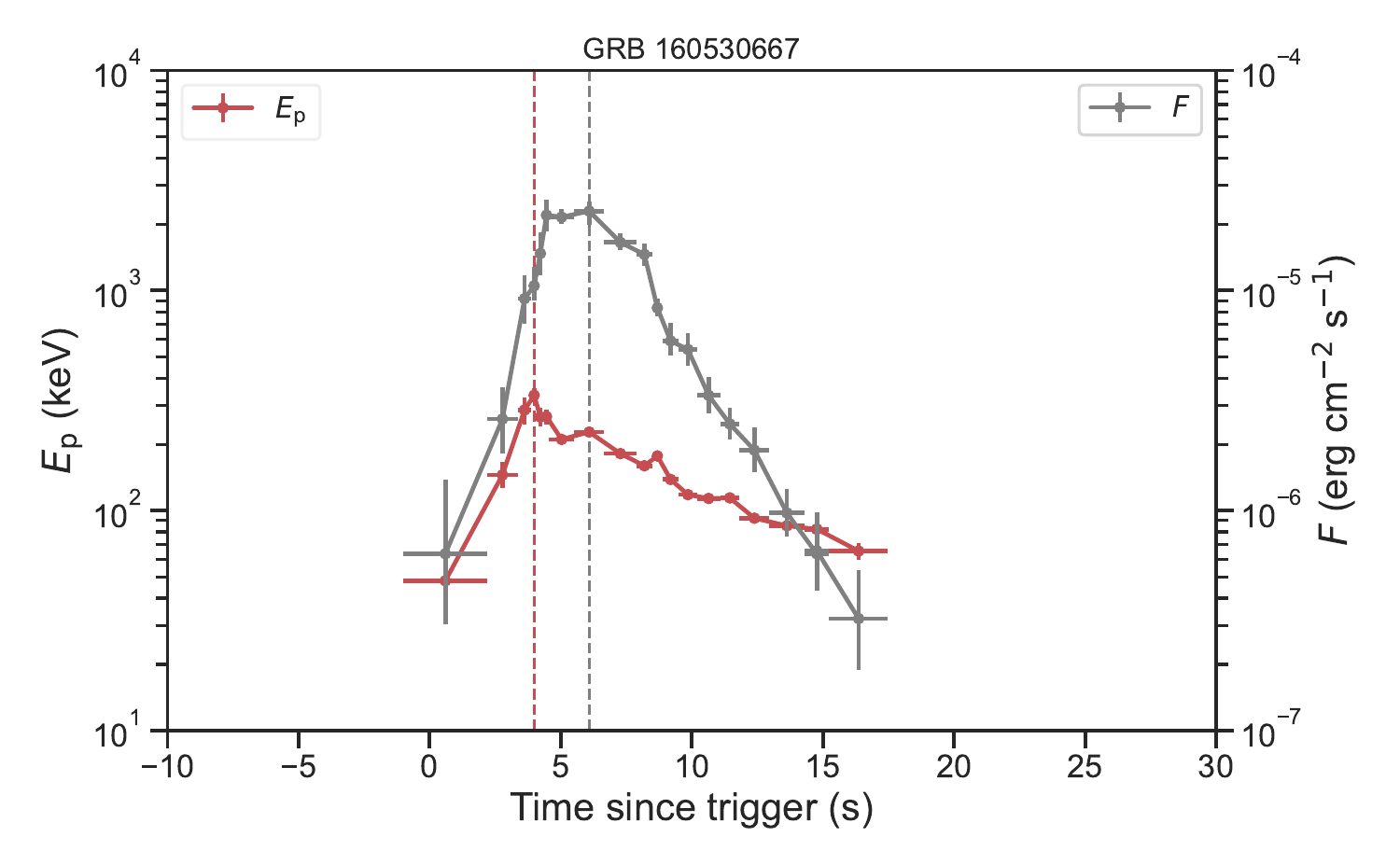}
\includegraphics[width=0.5\hsize,clip]{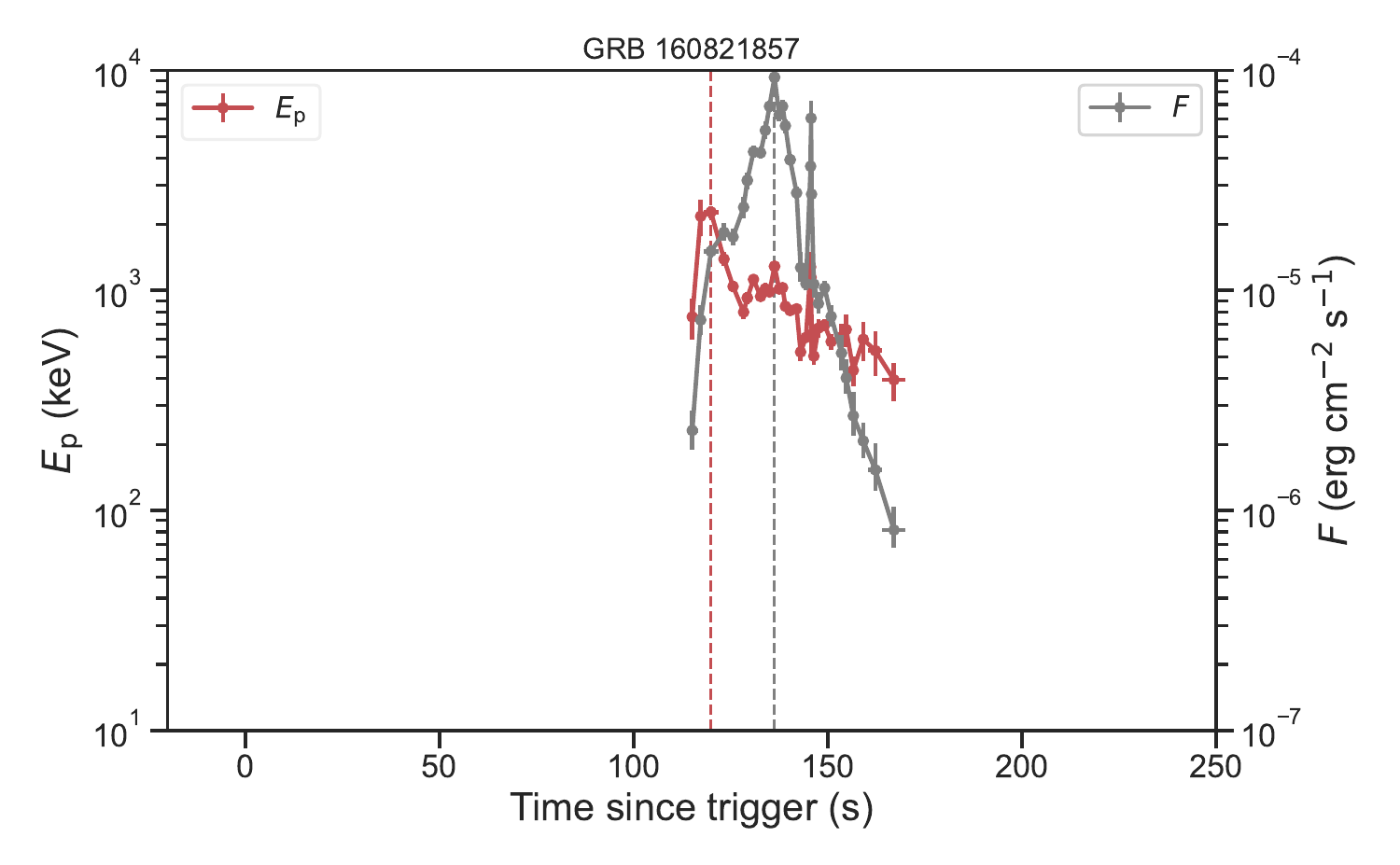}
\includegraphics[width=0.5\hsize,clip]{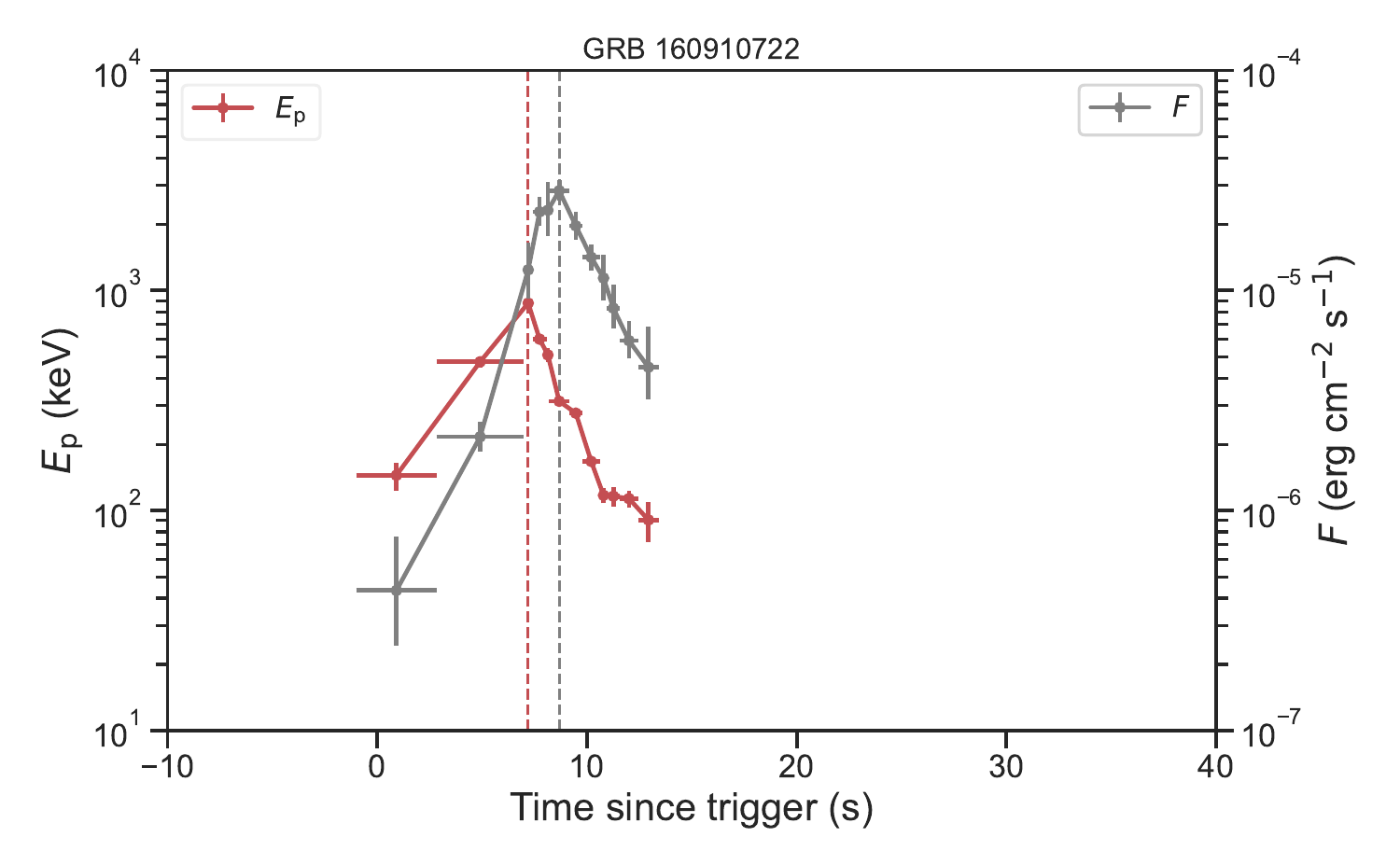}
\includegraphics[width=0.5\hsize,clip]{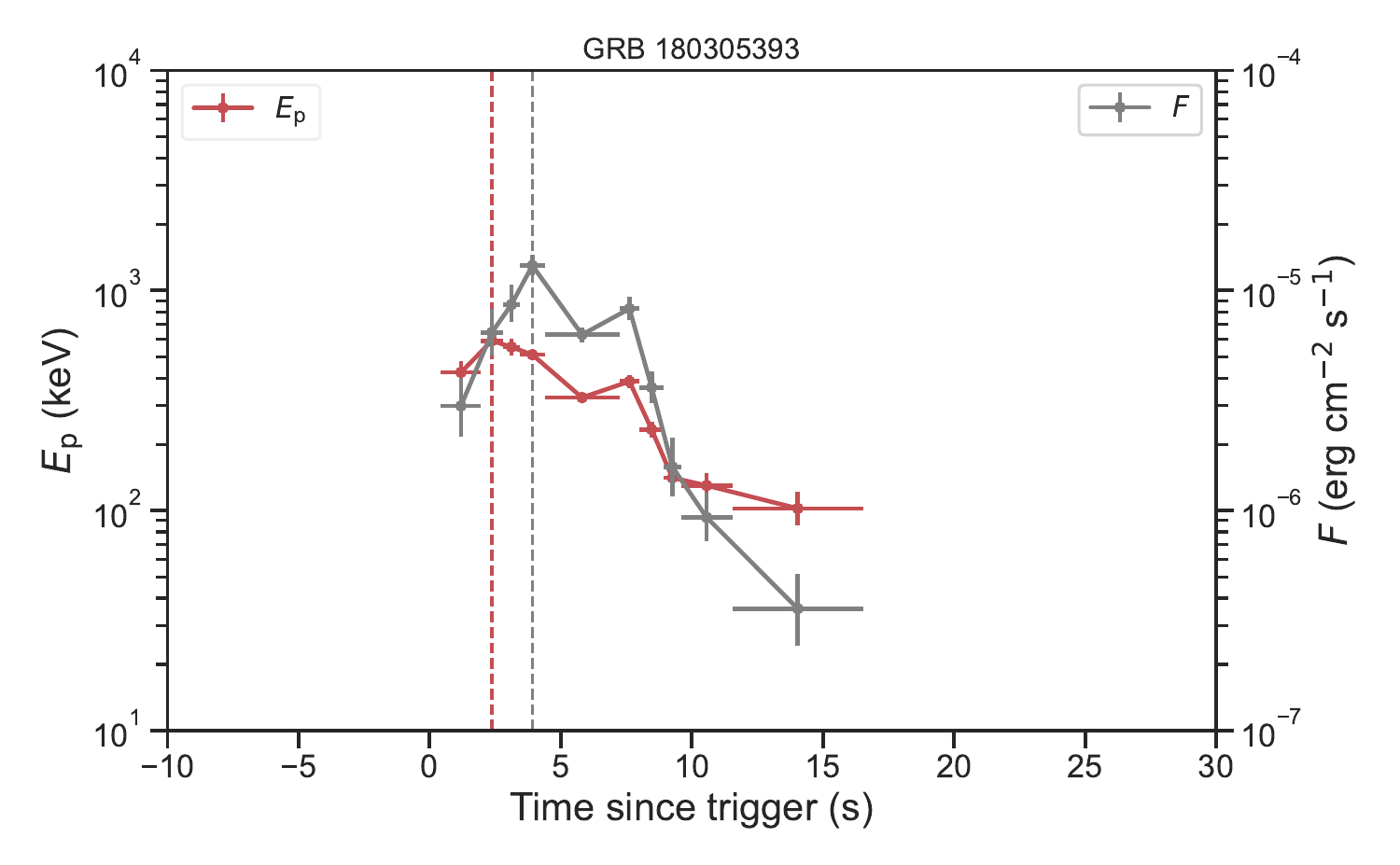}
\center{Fig. \ref{fig:typeII-all}- Continued}
\end{figure*}

\subsection{Spectral Properties of the Three Types}

\begin{figure*}
\centering
\includegraphics[width=0.95\textwidth]{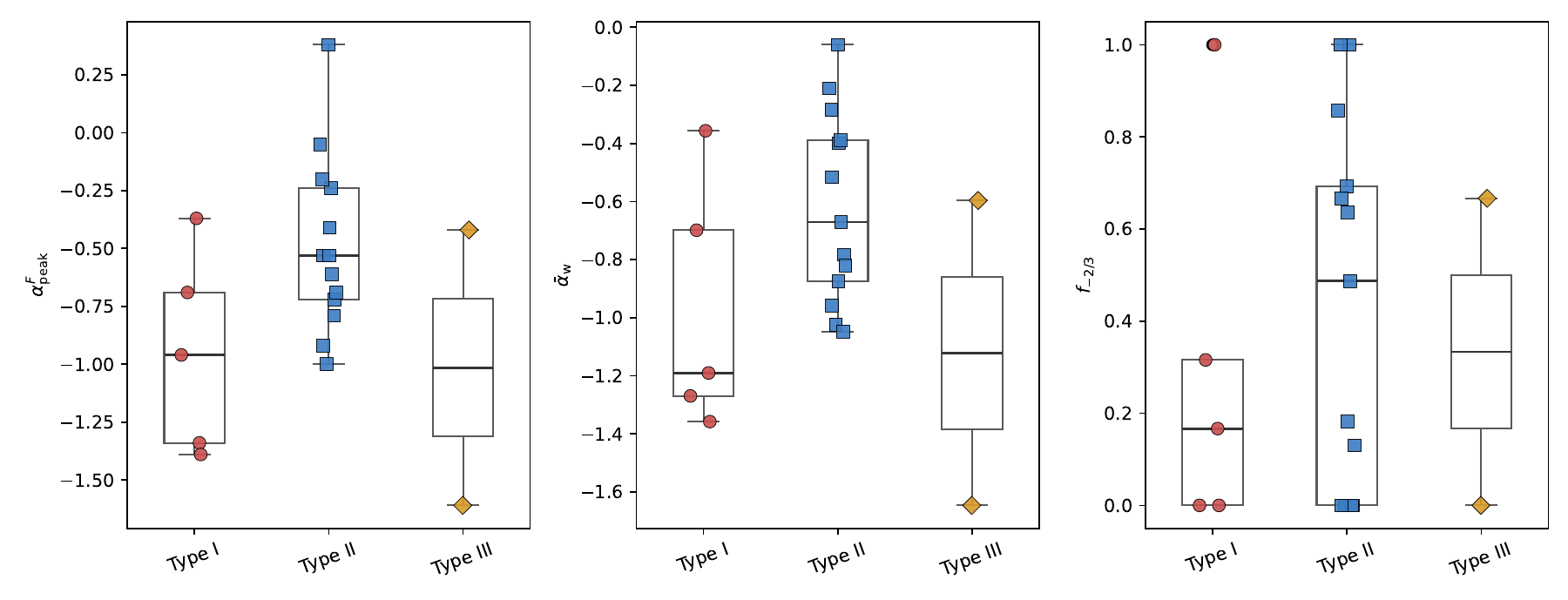}
\caption{Comparison of the burst-level $\alpha$ summaries for the three subclasses. From left to right, the panels show, the low-energy photon index at the flux peak, the inverse-variance weighted mean $\bar{\alpha}_{\rm w}$, and the hard-bin fraction $f_{-2/3}$. The scatter points correspond to individual bursts and the box plots mark the overall distribution of each subclass.}
\label{fig:alpha-type}
\end{figure*}

To characterize the spectral differences among the subclasses, we extracted three burst-level statistics from the low-energy photon index evolution: the value measured at the flux peak, $\alpha_{\rm peak}^{F}$, the inverse-variance weighted mean $\bar{\alpha}_{\rm w}$, and the hard-bin fraction $f_{-2/3}=N(\alpha>-2/3)/N_{\rm bin}$. Their distributions are shown in Figure~\ref{fig:alpha-type}. The Type~II sample is, systematically spectrally harder than the Type I sample. The median values of $\alpha_{\rm peak}^{F}$ are $-0.53$ for Type~II and $-0.96$ for Type~I, and the medians of $\bar{\alpha}_{\rm w}$ are $-0.67$ and $-1.19$, respectively. The two Type~III bursts lie between these two groups, with median values approximately $-1.02$ for $\alpha_{\rm peak}^{F}$ and $-1.12$ for $\bar{\alpha}_{\rm w}$.

The same trend is reflected in the hard-bin fraction. Type~II bursts have a median $f_{-2/3}=0.49$, compared with $0.17$ for Type~I bursts. Formally, three-sample Anderson-Darling tests yield probabilities of order $10^{-3}$ for all three summary metrics, indicating that the subclass distributions are inconsistent with a common parent population. Although the small Type~III sample ($N=2$) limits the robustness of these statistical tests, the available data indicate that early-peaking events exhibit a harder spectrum than aligned events.

\subsection{$E_{\rm p}$-$F$ relation of the Three Types}

\begin{figure*}
\centering
\includegraphics[width=0.72\textwidth]{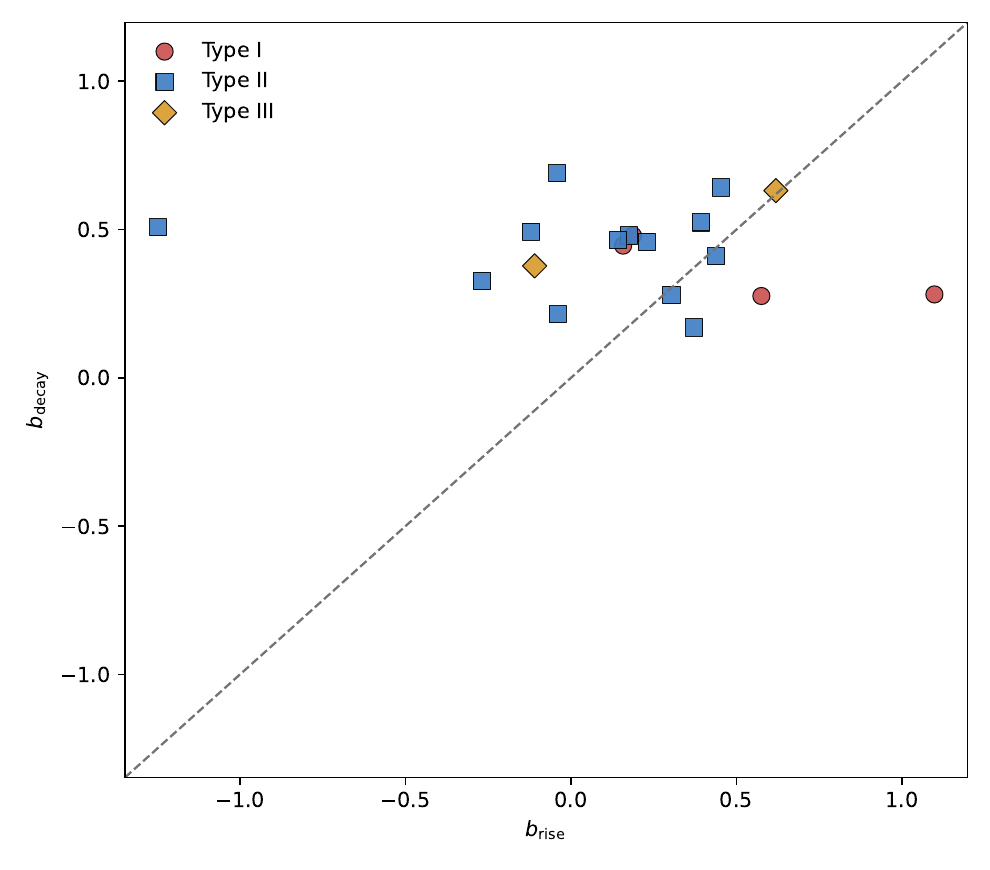}
\caption{Summary of the branch-dependent $E_{\rm p}$-$F$ slopes. Each point represents one burst, with the slope measured on the rising branch plotted against the slope measured on the decaying branch from the relation $\log E_{\rm p}=a+b\log F$. The dashed diagonal marks $b_{\rm rise}=b_{\rm decay}$. Type~I bursts tend to cluster near the diagonal, whereas many Type~II bursts lie below it (corresponding to $b_{\rm rise}<b_{\rm decay}$). The two Type~III bursts are shown for completeness.}
\label{fig:slope-summary}
\end{figure*}

For each burst, we fitted the relation $\log E_{\rm p}=a+b\log F$ separately to the rising and decaying branches defined with respect to the flux peak. The resulting slope comparison is shown in Figure~\ref{fig:slope-summary}. Type~I bursts cluster near the one-to-one line $b_{\rm rise}=b_{\rm decay}$, with a median offset $\Delta b=b_{\rm rise}-b_{\rm decay}$ close to zero ($\Delta b_{\rm med}=0.01$). In contrast, Type~II bursts preferentially lie below this line, with a median $\Delta b=-0.25$, indicating that for this subclass the $E_{\rm p}$-$F$ slope is systematically shallower on the rising branch than on the decaying branch.

This branch asymmetry is consistent with the time-domain lag itself. When the spectral maximum occurs before the flux maximum, the subsequent decay spans a substantial fraction of the dynamic range in the $E_{\rm p}$-$F$ plane. Despite this asymmetry, the overall correlation remains positive for the majority of bursts. The median Spearman coefficient for the full sample is $\rho_{E_{\rm p},F}=0.81$, demonstrating that a positive $E_{\rm p}$-$F$ correlation is common to all three subclasses and that the key discriminator is the ordering of the maxima rather than the sign of the correlation alone. 

\begin{figure*}
\centering
\includegraphics[width=0.49\textwidth]{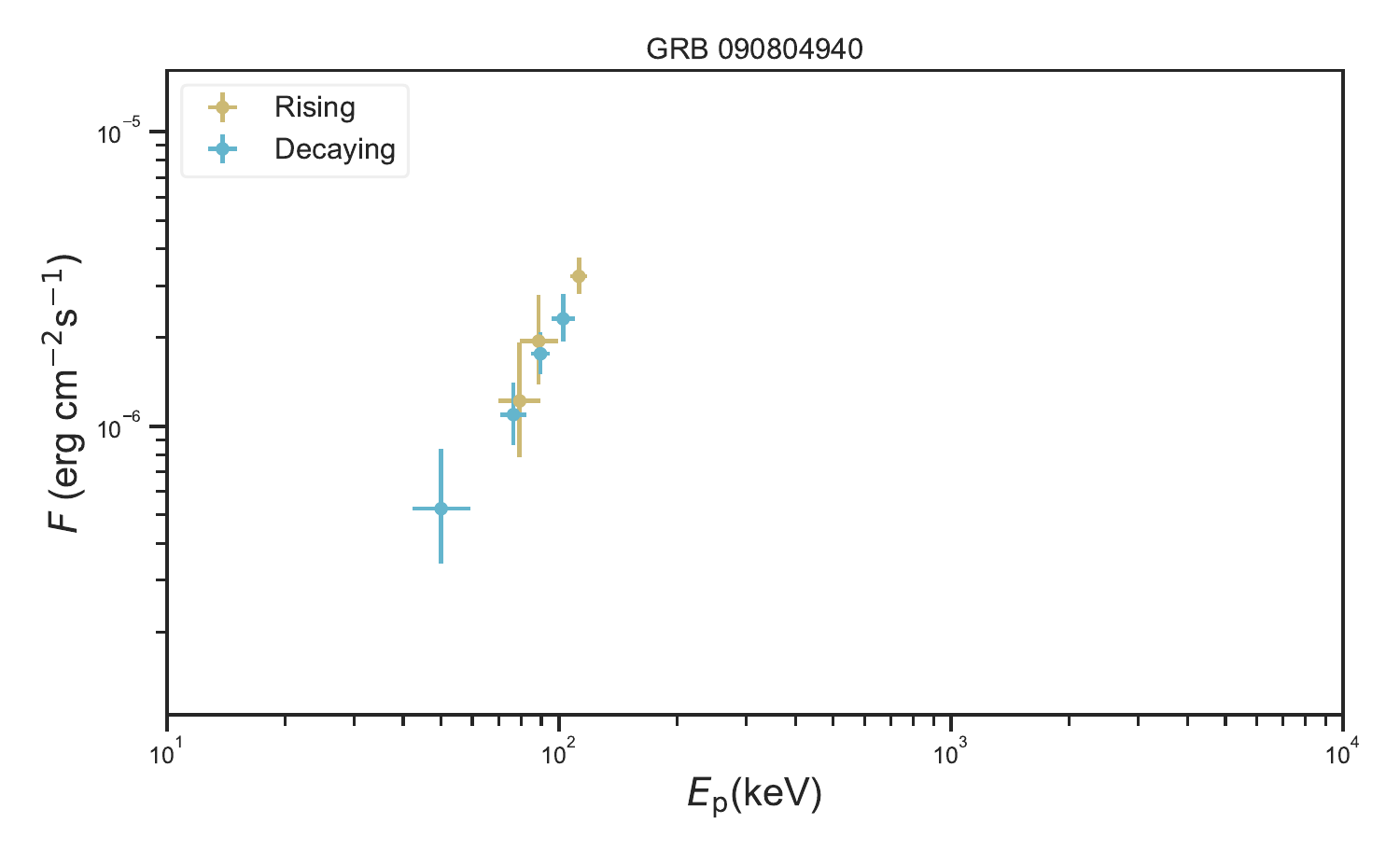}
\includegraphics[width=0.49\textwidth]{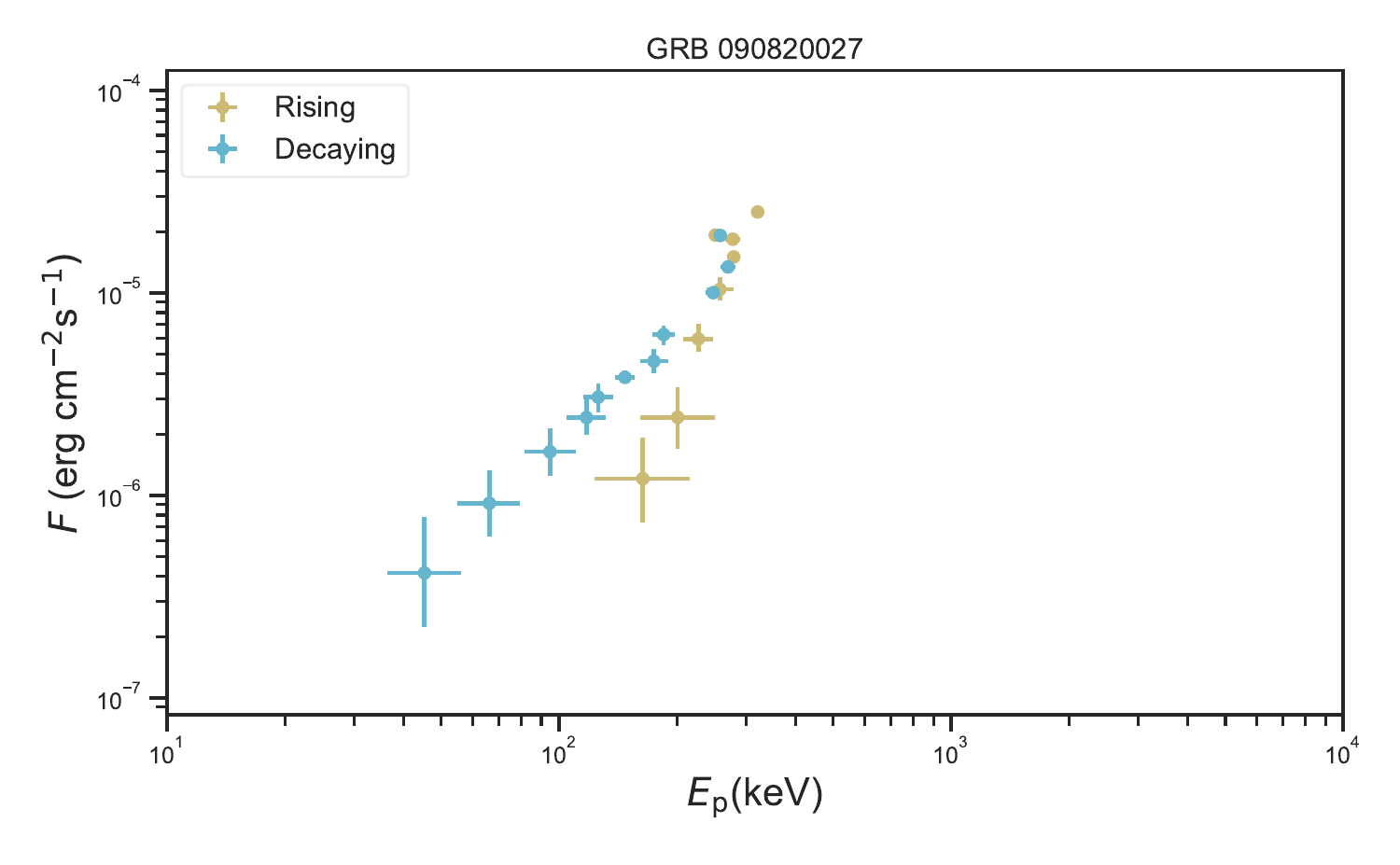}
\includegraphics[width=0.49\textwidth]{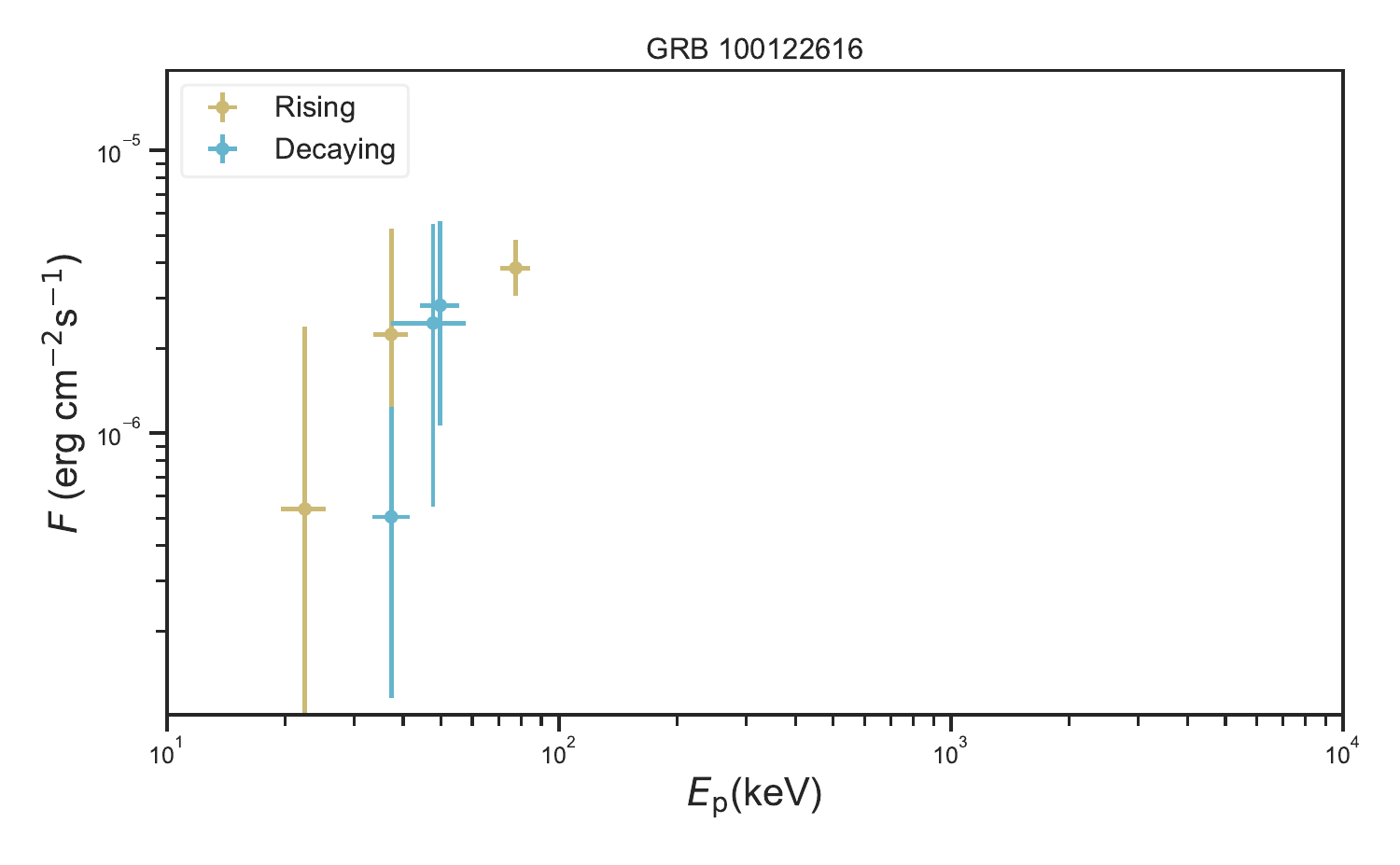}
\includegraphics[width=0.49\textwidth]{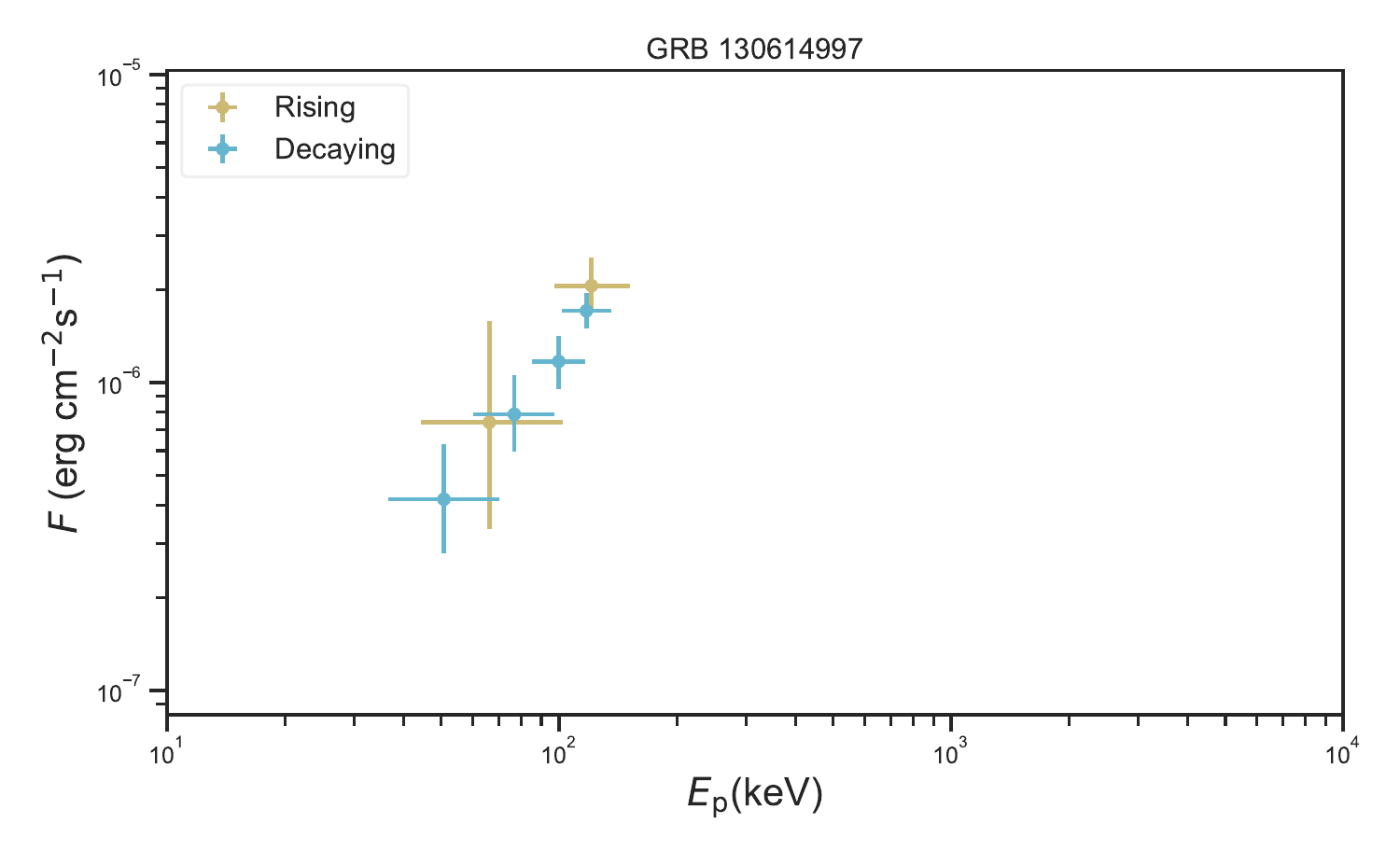}
\includegraphics[width=0.49\textwidth]{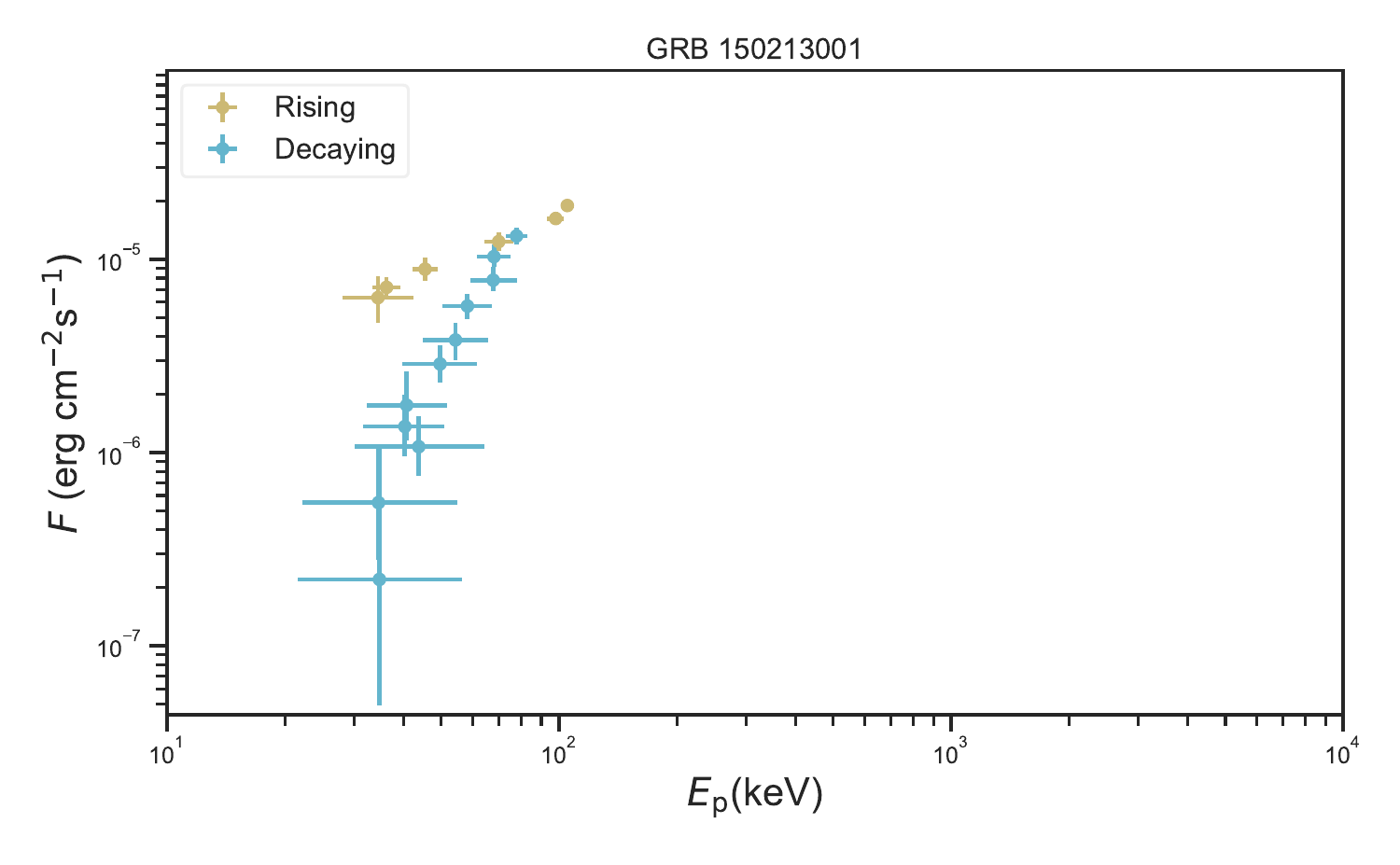}
\caption{Representative $E_{\rm p}$-$F$ relations for selected Type~I bursts. In each panel, the time-resolved spectral bins are divided into rising and decaying branches with respect to the flux peak. The two branches largely overlap in the $\log E_{\rm p}$-$\log F$ plane, consistent with the near-aligned peak ordering of this subclass.}
\label{fig:Ep-Flux-TypeI}
\end{figure*}

\begin{figure*}
\centering
\includegraphics[width=0.49\textwidth]{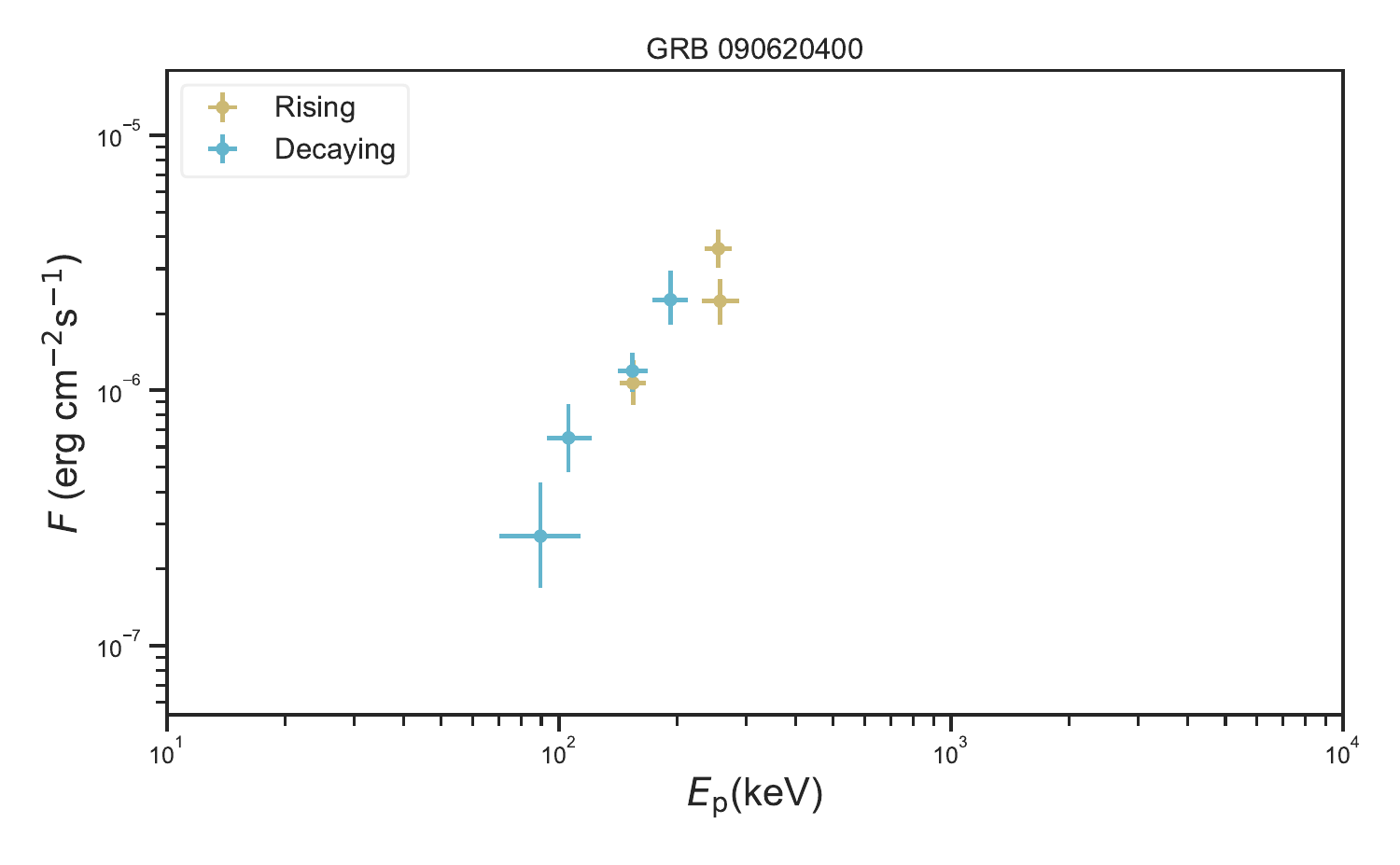}
\includegraphics[width=0.49\textwidth]{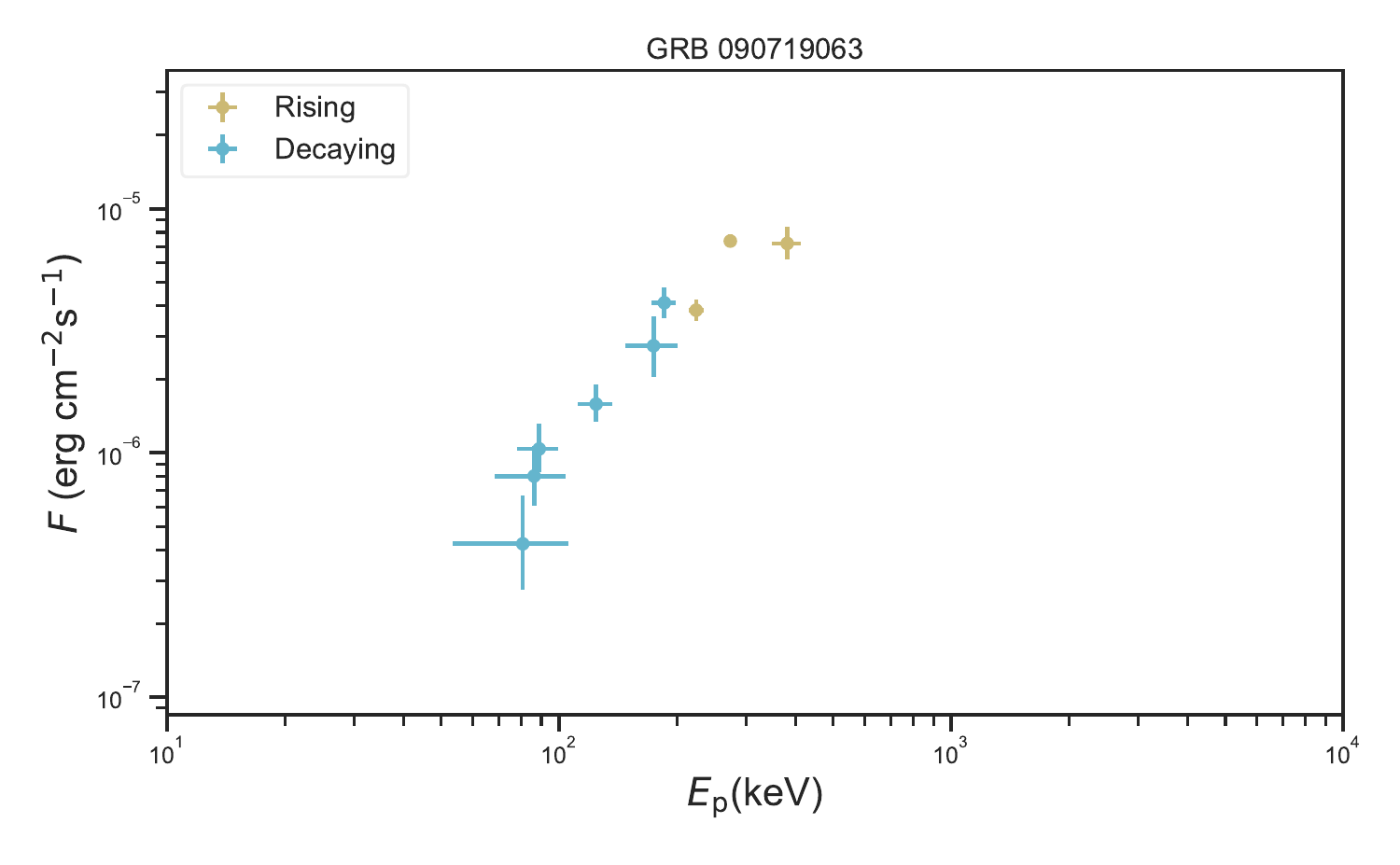}
\includegraphics[width=0.49\textwidth]{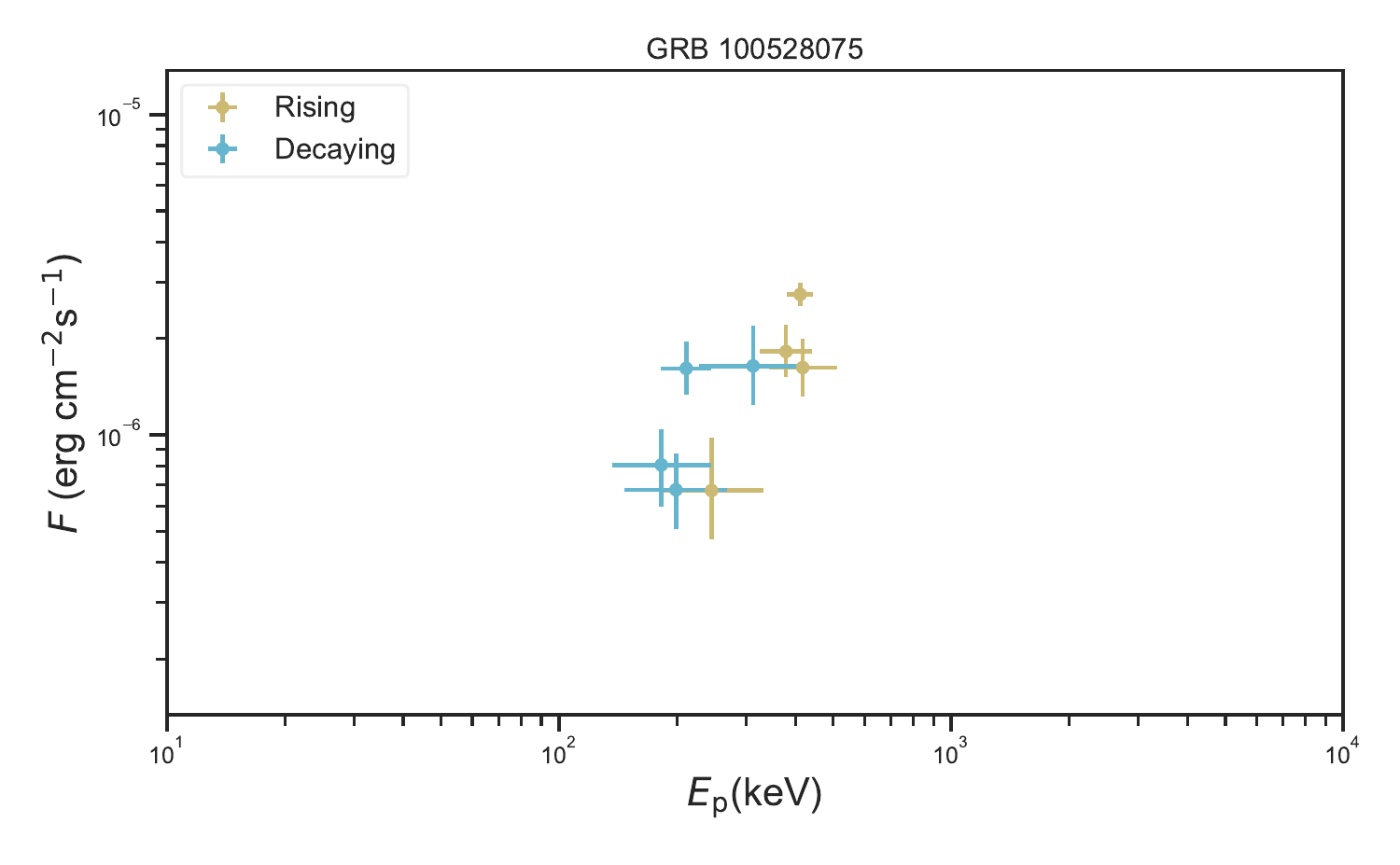}
\includegraphics[width=0.49\textwidth]{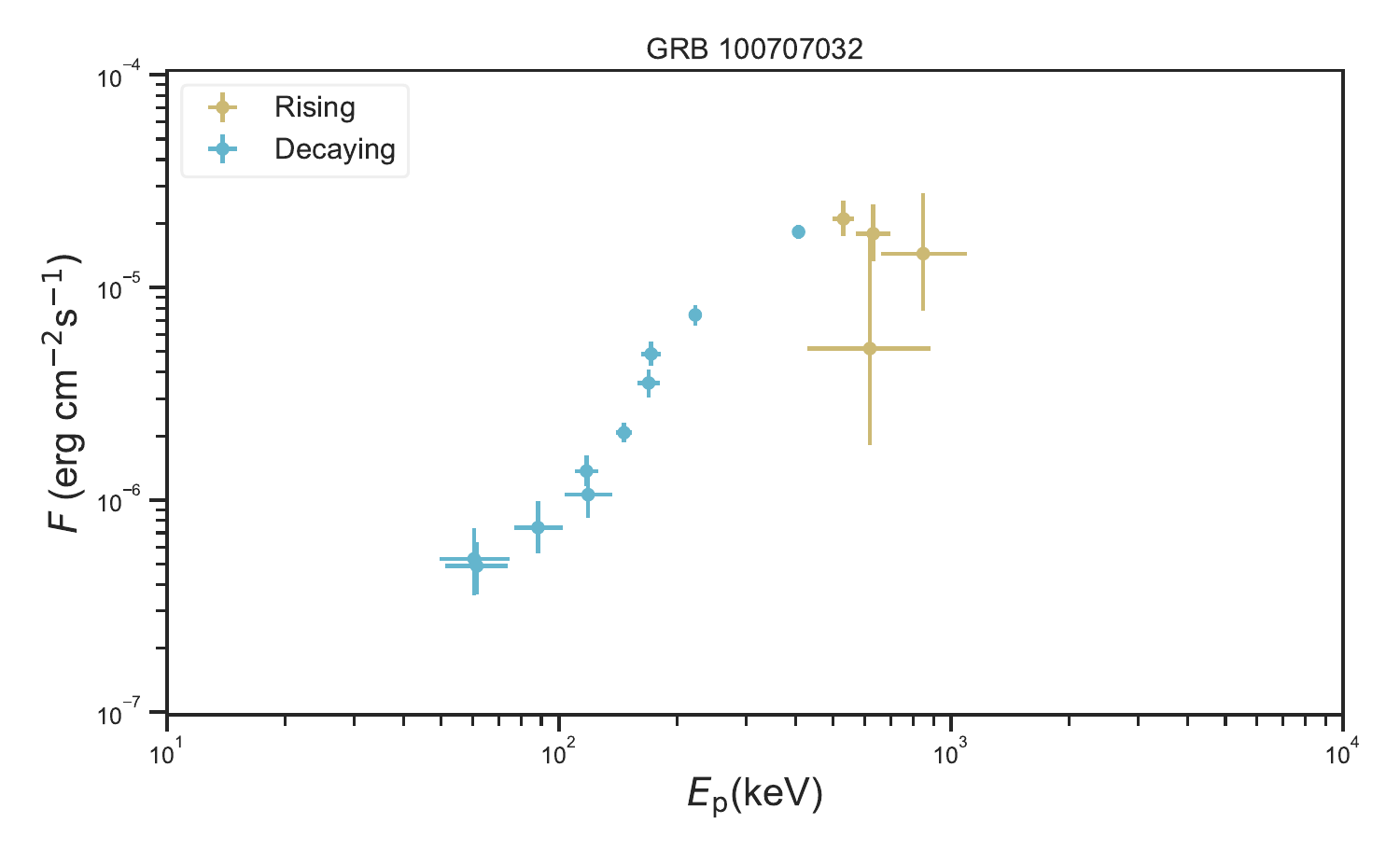}
\includegraphics[width=0.49\textwidth]{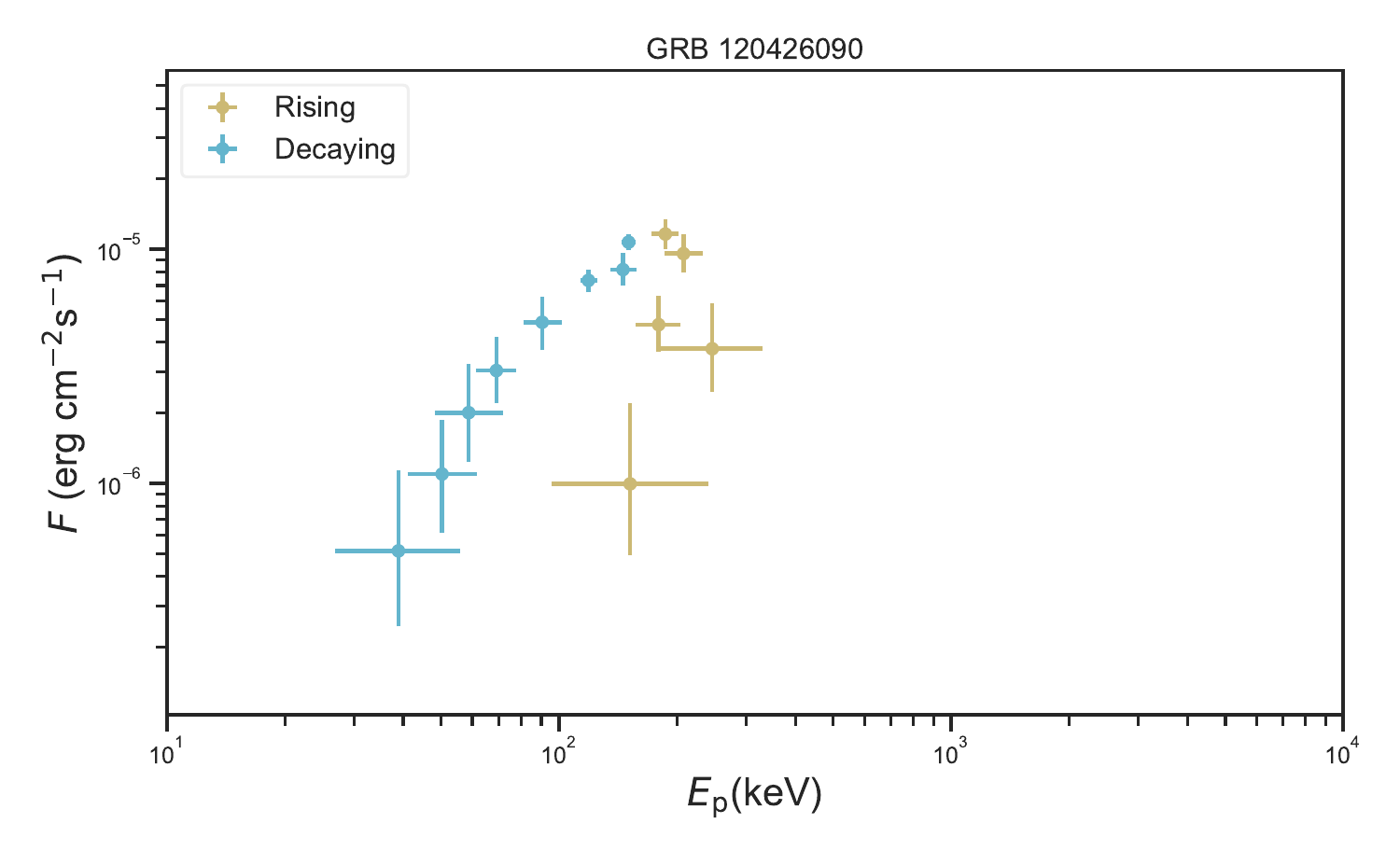}
\includegraphics[width=0.49\textwidth]{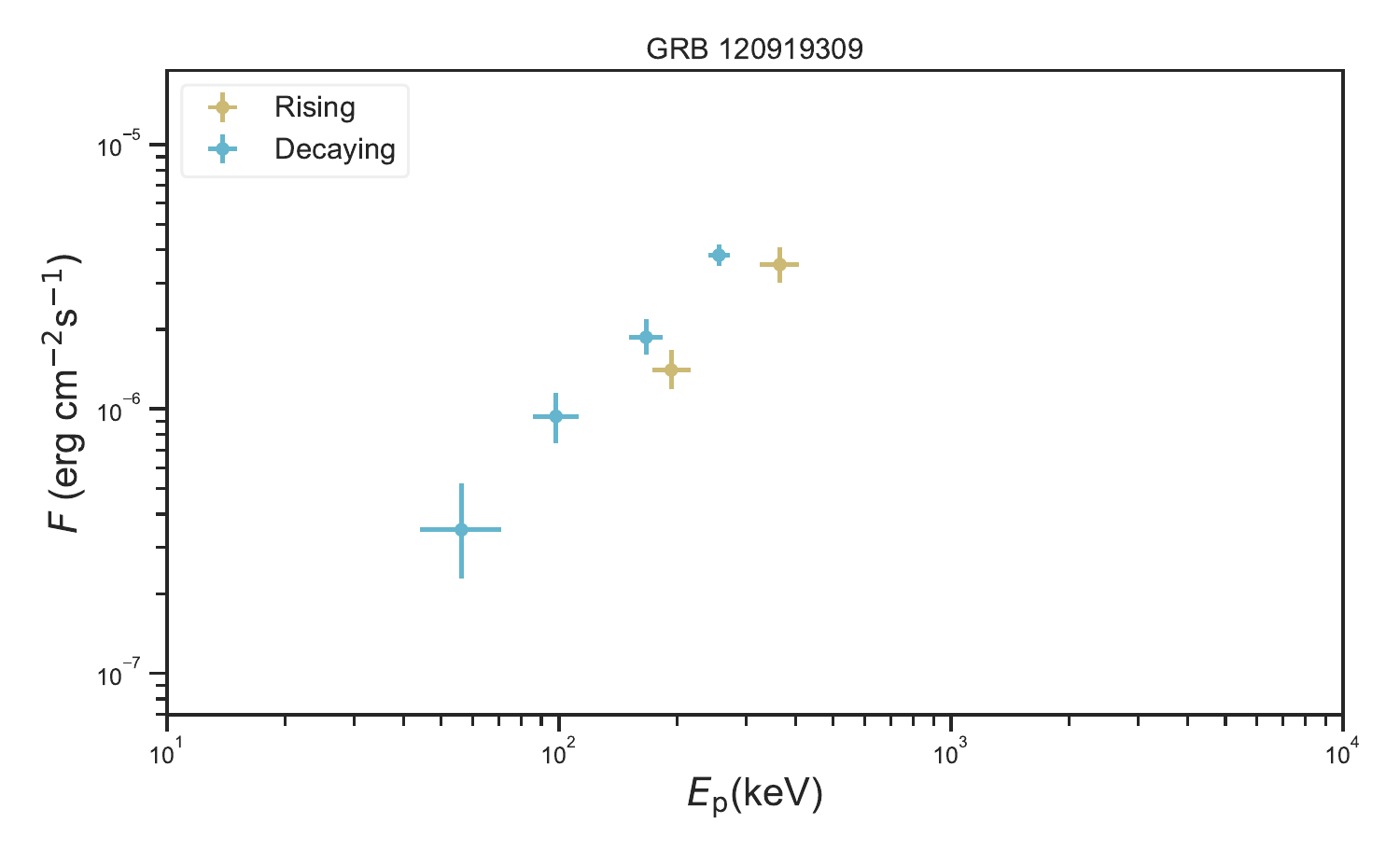}
\caption{Same as Figure \ref{fig:Ep-Flux-TypeI} but for selected Type~II bursts. Compared with Type~I, many Type~II bursts show a clearer branch separation, with the rising branch typically shallower than the decaying branch.}
\label{fig:Ep-Flux-TypeII}
\end{figure*}

\begin{figure*}
\centering
\includegraphics[width=0.49\textwidth]{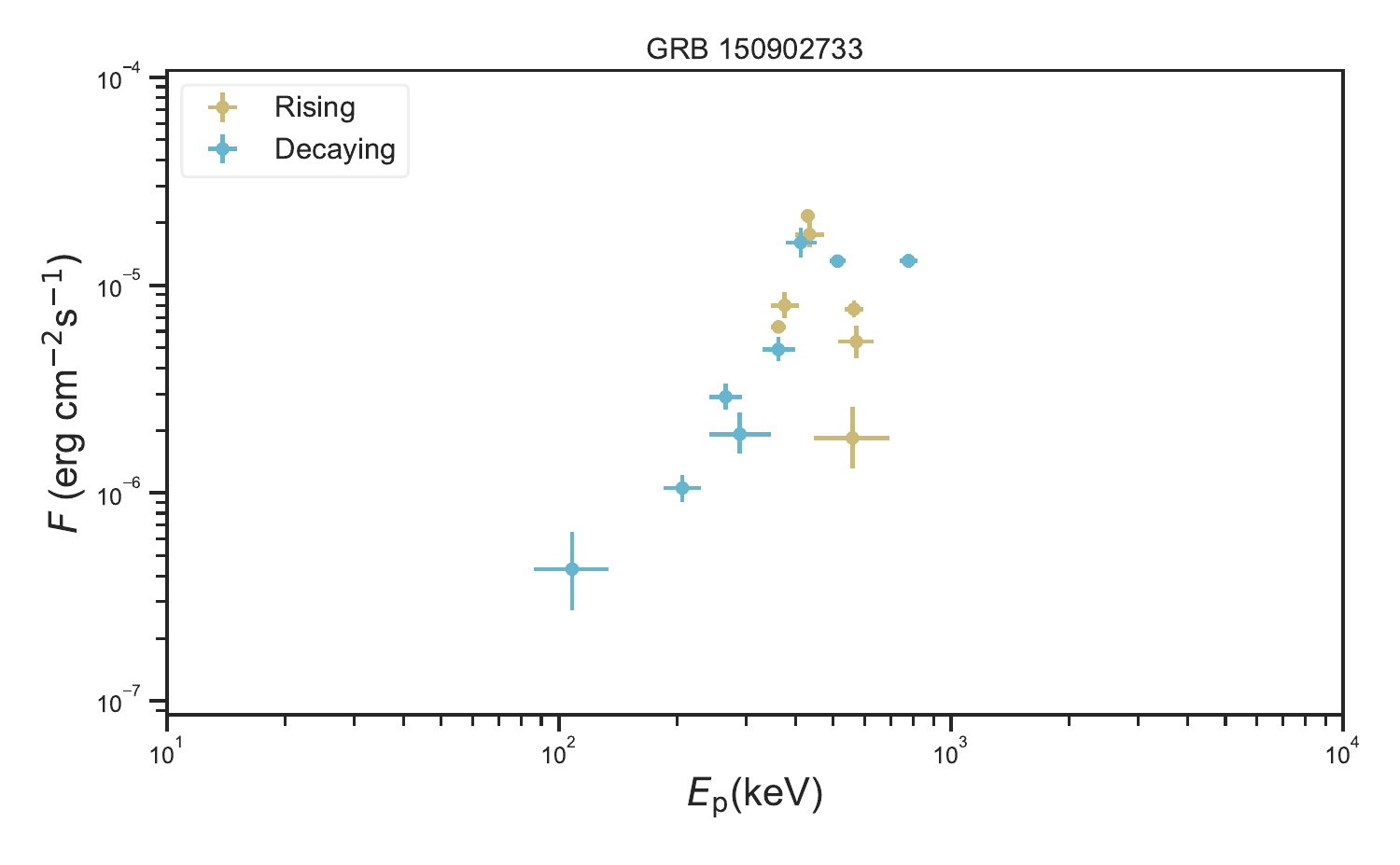}
\includegraphics[width=0.49\textwidth]{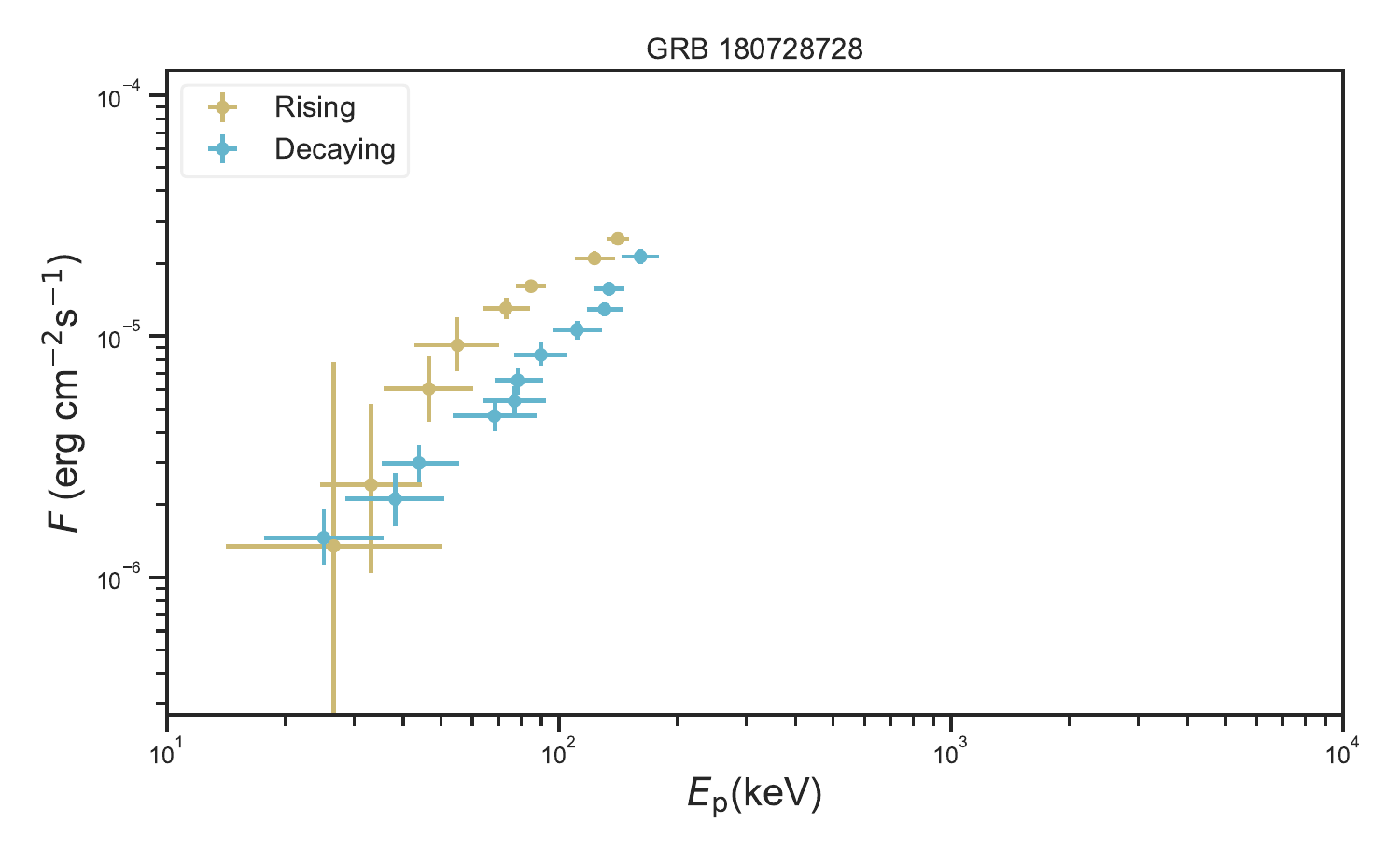}
\caption{Same as Figure \ref{fig:Ep-Flux-TypeI} but for Type III bursts. Both bursts preserve an overall positive $E_{\rm p}$-$F$ correlation, but the branch geometry does not yet define a common pattern for this rare subclass.}
\label{fig:Ep-Flux-TypeIII}
\end{figure*}

The branch separation is further illustrated by the individual $E_{\rm p}$-$F$ tracks shown in Figures~\ref{fig:Ep-Flux-TypeI}-\ref{fig:Ep-Flux-TypeIII}, where spectral bins on the rising and decaying sides of the flux pulse are plotted separately. Type~I bursts generally trace a nearly single-valued trajectory in the $\log E_{\rm p}$-$\log F$ plane, with considerable overlap between the rising and decaying branches. Many Type~II bursts, by contrast, exhibit more pronounced branch separation, with the rising branch shallower and the decaying branch steeper, consistent with their negative median $\Delta b$. The two Type~III bursts also show a positive $E_{\rm p}$-$F$ correlation overall, but the sample is too small to characterize a representative branch geometry. These tracks confirm that the subclasses are distinguished not by the sign of the $E_{\rm p}$-$F$ correlation, but by the relative geometry of the rising and decaying branches.

\begin{figure*}
\centering
\includegraphics[width=0.49\textwidth]{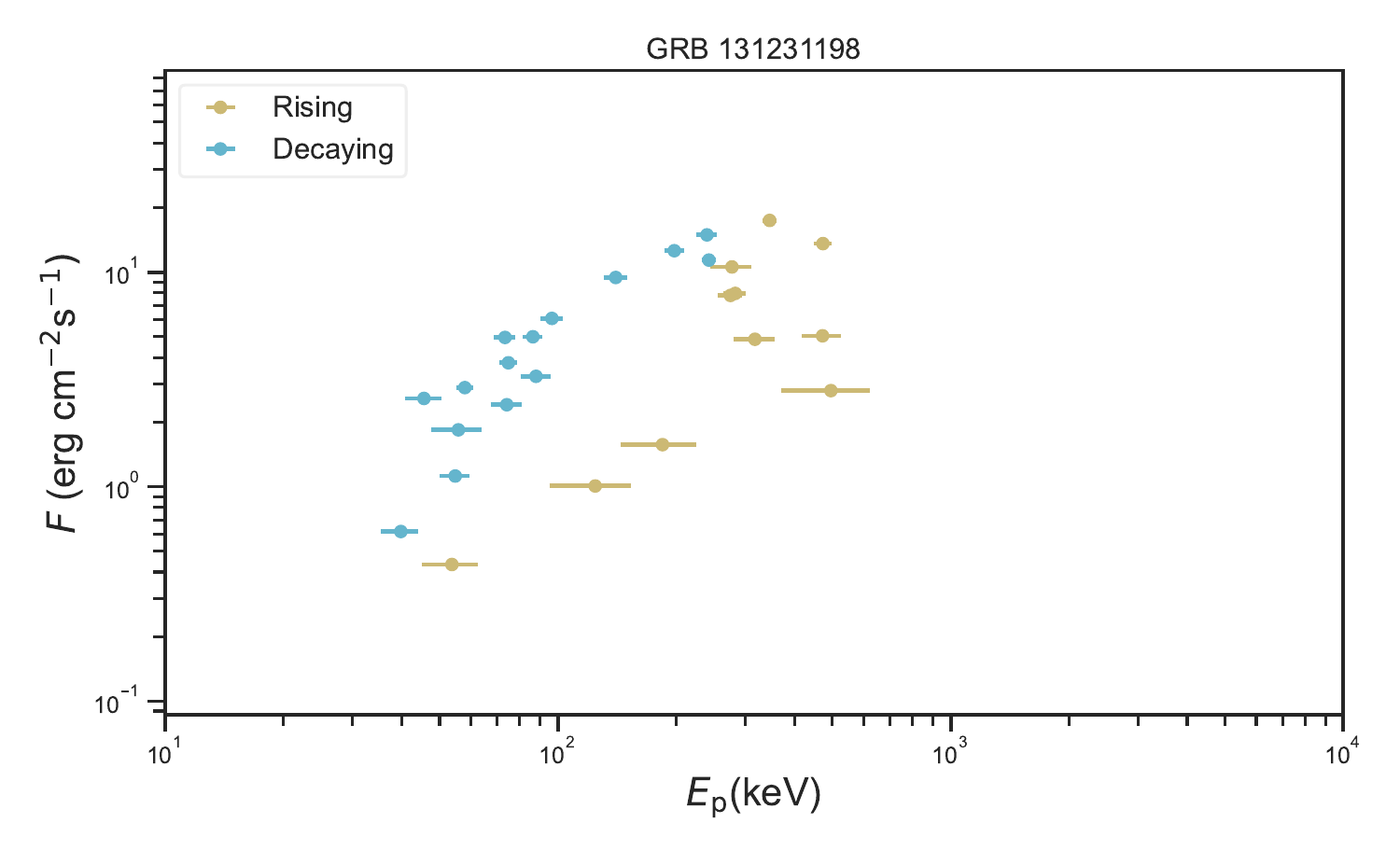}
\includegraphics[width=0.49\textwidth]{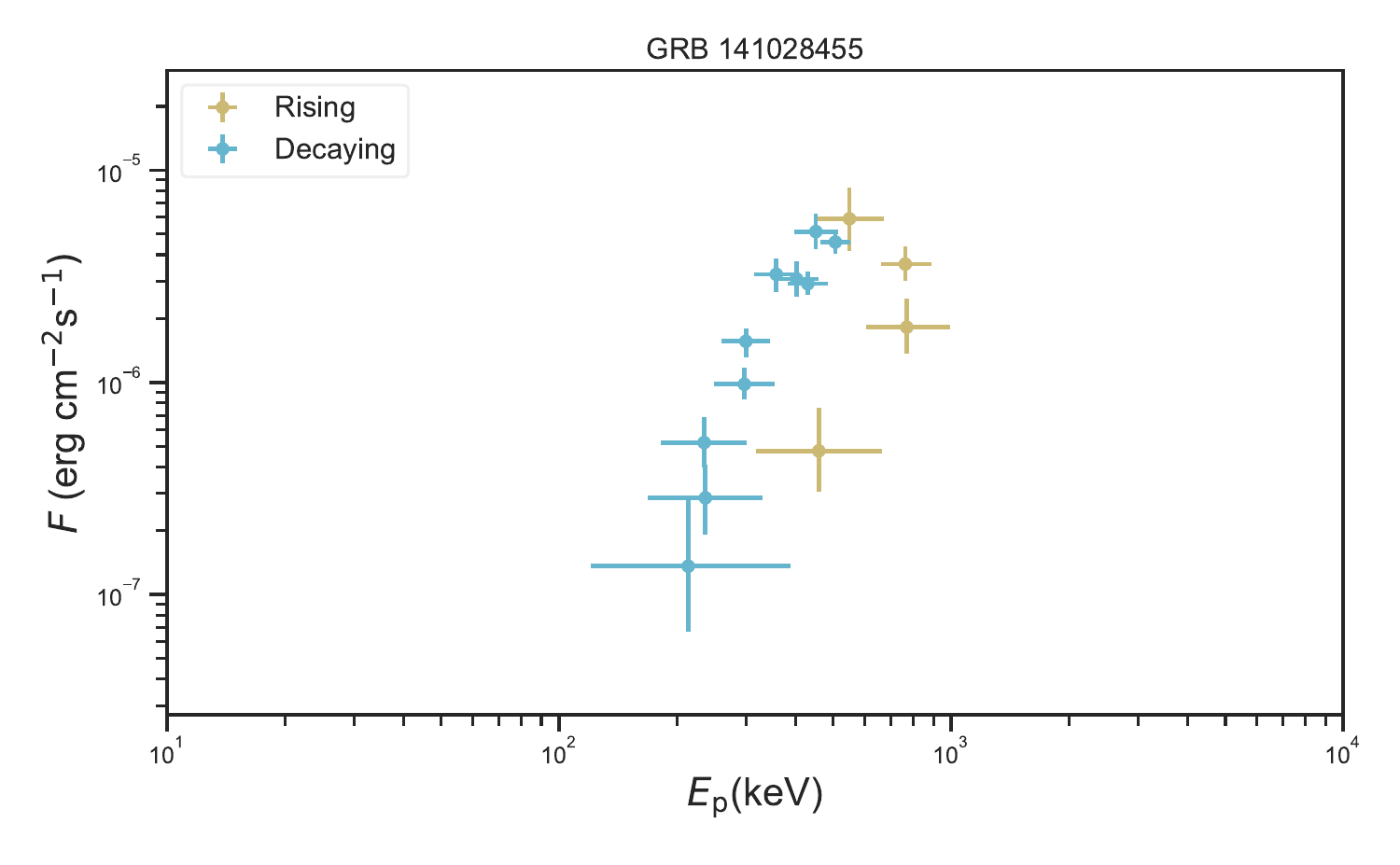}
\includegraphics[width=0.49\textwidth]{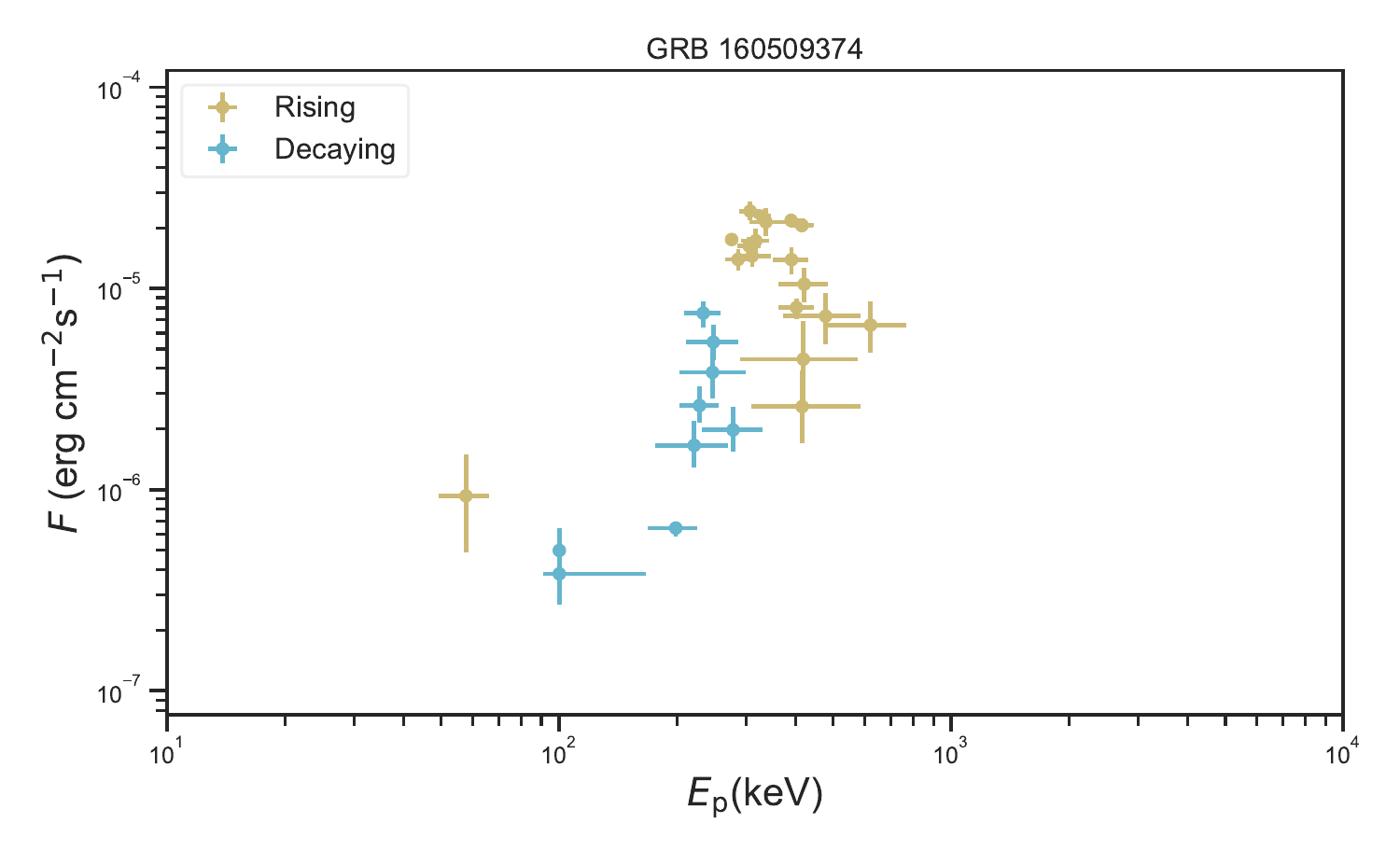}
\includegraphics[width=0.49\textwidth]{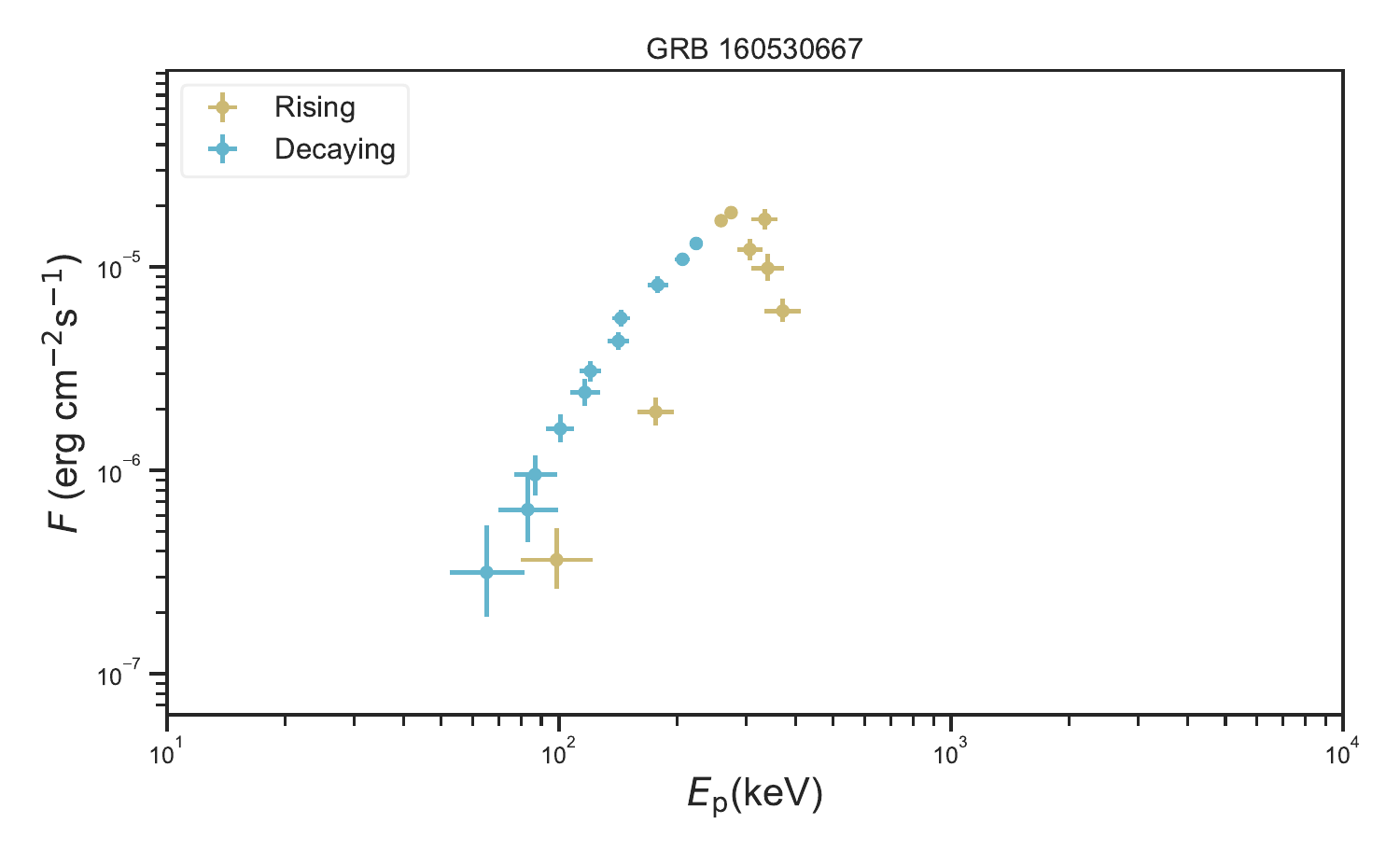}
\includegraphics[width=0.49\textwidth]{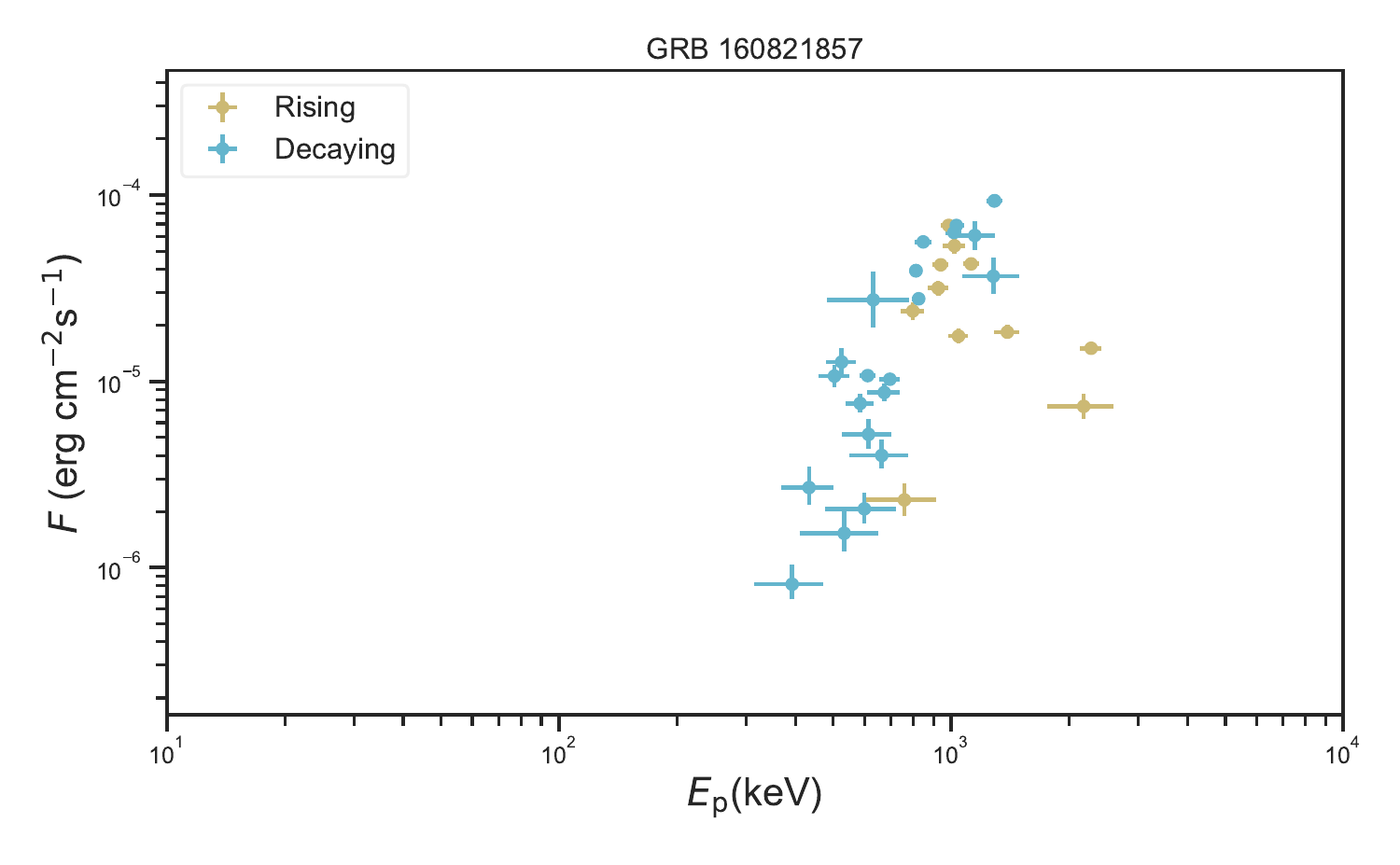}
\includegraphics[width=0.49\textwidth]{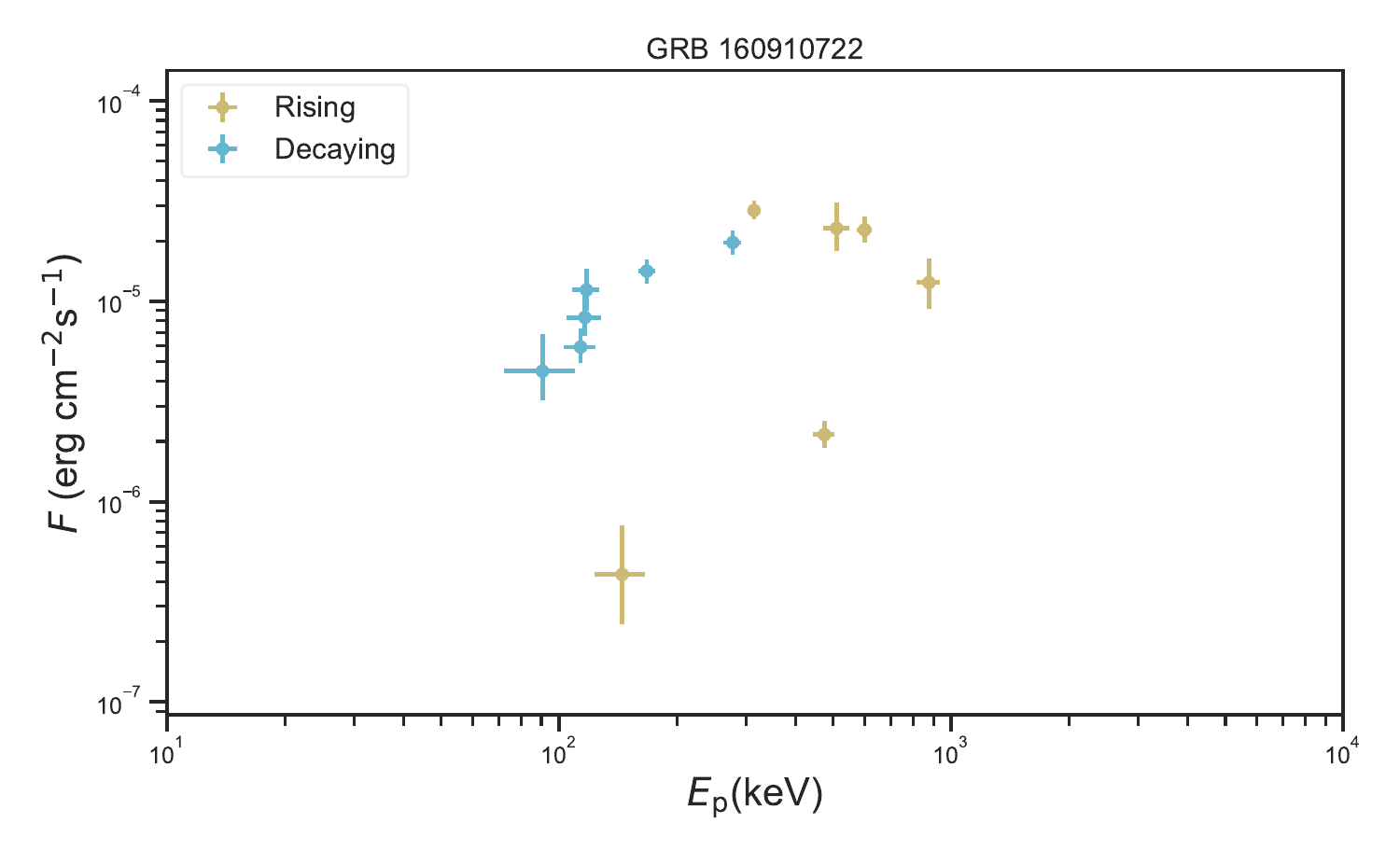}
\includegraphics[width=0.49\textwidth]{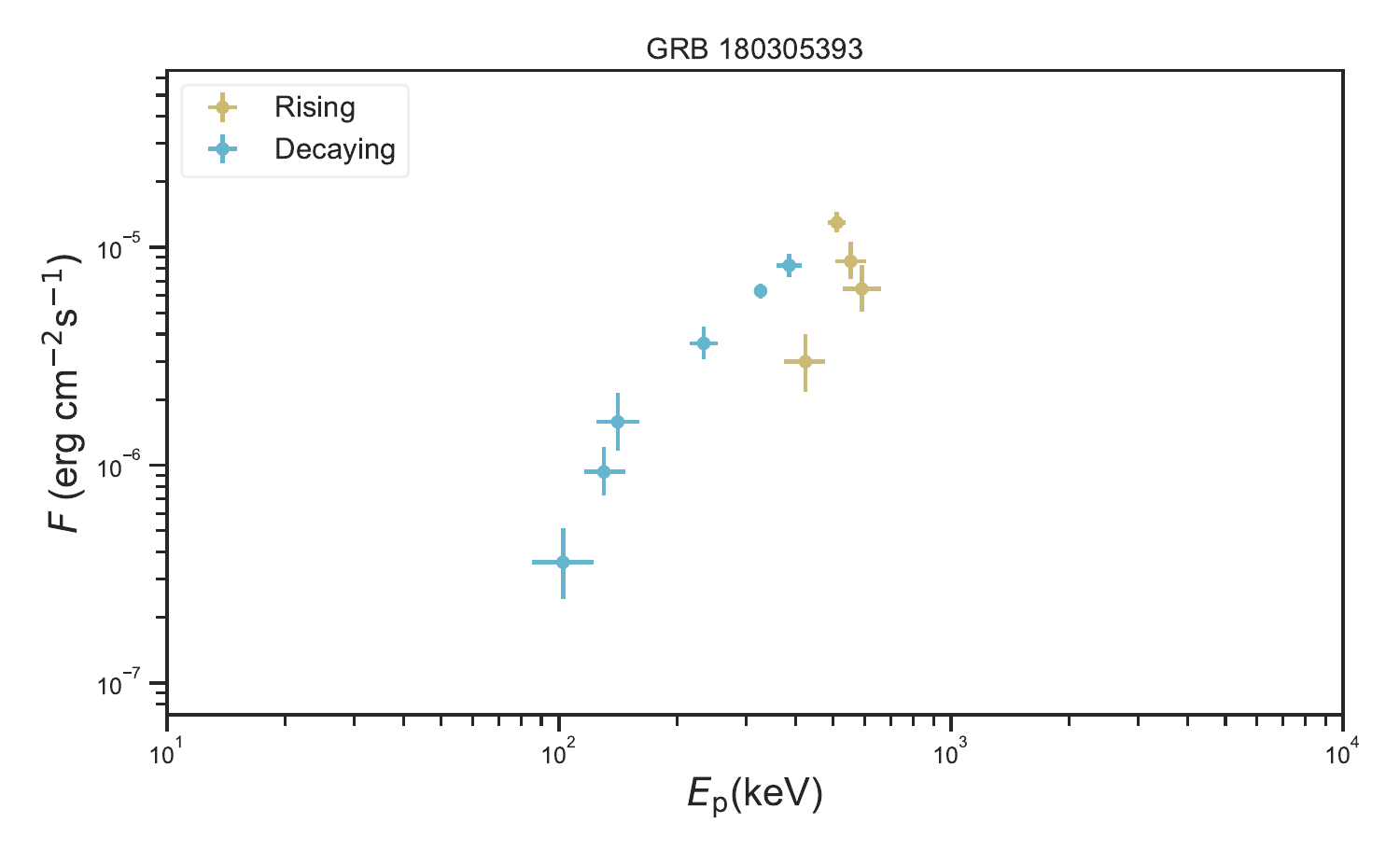}
\center{Fig. \ref{fig:Ep-Flux-TypeII}- Continued}
\end{figure*}

\subsection{Connections with Flux-pulse Morphology}

\begin{figure*}
\centering
\includegraphics[width=0.88\textwidth]{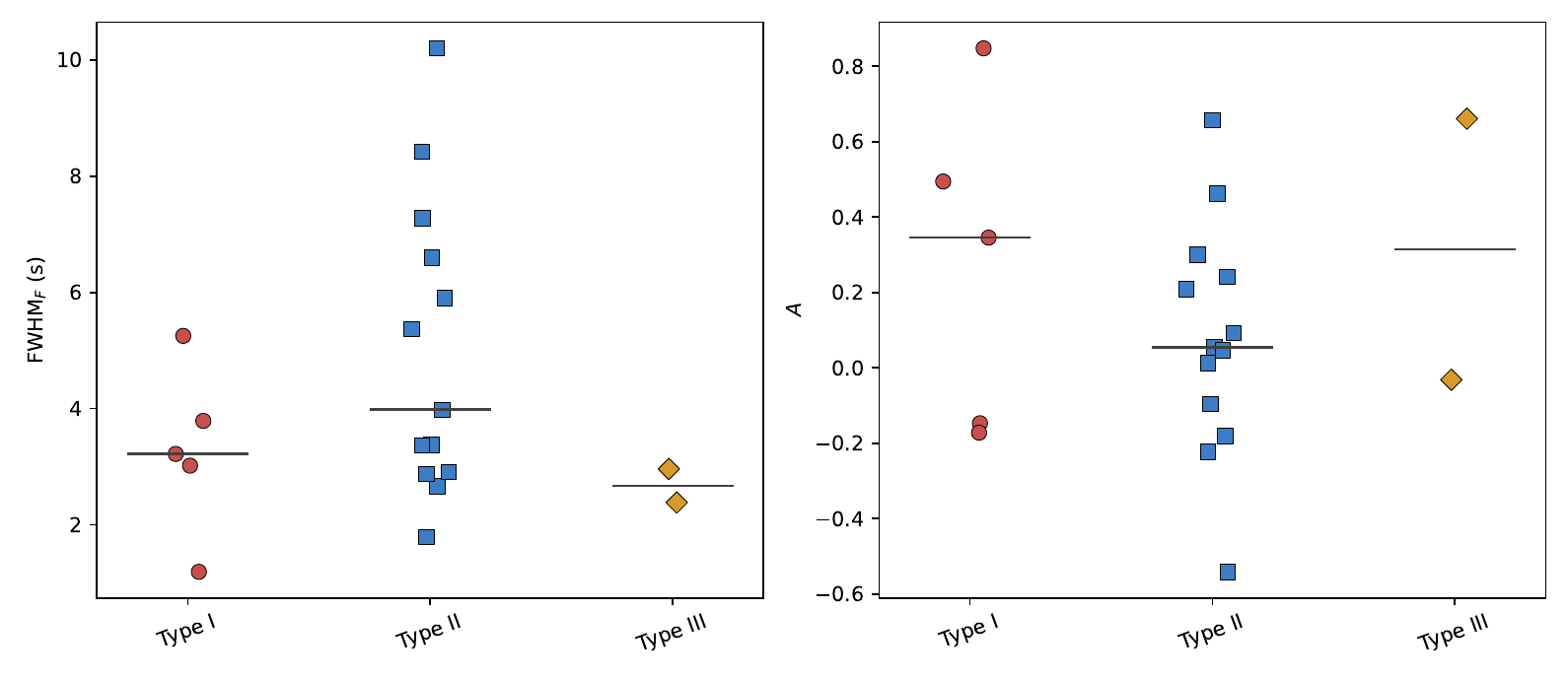}
\caption{Flux-pulse morphology by subclass. Left: ${\rm FWHM}_{F}$. Right: asymmetry parameter $A=(t_{d}-t_{r})/(t_{d}+t_{r})$. The horizontal bars mark the median value in each subclass. Type~II bursts are, on average, broader than Type~I bursts, while the Type~III sample remains small.}
\label{fig:morphology-summary}
\end{figure*}

We next examine whether the lag-based subclasses are linked to differences in the morphology of the energy-flux pulse itself. The relevant parameters are listed in Table~\ref{tab:morph-summary}, including ${\rm FWHM}_{F}$, the rise and decay times ($t_{r}$ and $t_{d}$), the pulse-asymmetry measures $\kappa=t_{d}/t_{r}$ and $A=(t_{d}-t_{r})/(t_{d}+t_{r})$, the branch-dependent $E_{\rm p}$-$F$ slopes, and the peak-prominence statistics $S_{\rm pk}(F)$ and $S_{\rm pk}(E_{\rm p})$. Here $S_{\rm pk}(X)$ is defined as the difference between the highest and second-highest bins in the quantity $X$, divided by the quadrature sum of their $1\sigma$ uncertainties, so that $S_{\rm pk}$ by construction. The subclass comparison is shown in Figure~\ref{fig:morphology-summary}. Type~II bursts have a broader median flux width (${\rm FWHM}_{F}=4.94$~s) than Type~I bursts (${\rm FWHM}_{F}=3.22$~s), while the Type~III bursts are narrower (${\rm FWHM}_{F}=2.67$~s). Type~I and Type~III bursts also tend to have larger asymmetry parameters than Type~II bursts, with relatively longer decay phases.

The peak-prominence results indicate that a strict peak-separation criterion is conservative for the present sample. Applying the threshold ${\rm min}[S_{\rm pk}(F),S_{\rm pk}(E_{\rm p})]<2$, only GRB~090820027 constitutes a clearly non-marginal detection, while the remaining bursts are formally marginal, in the sense that either the flux peak or the $E_{\rm p}$ peak is only weakly distinguished from the second-highest bin once bin-level uncertainties are accounted for. We therefore treat $S_{\rm pk}$ as a useful cautionary diagnostic rather than as the primary classification metric. The more robust discriminator in the present sample is the sign and significance of $t_{\rm lag}^{\rm F}$ measured on the matched spectral bins.

\begin{deluxetable*}{cccccccccccc}
\tablewidth{0pt}
\tabletypesize{\scriptsize}
\tablecaption{Flux-pulse Morphology and Correlation Summary\label{tab:morph-summary}}
\tablehead{
\colhead{GRB} & \colhead{${\rm FWHM}_{F}$} & \colhead{$t_{r}$} & \colhead{$t_{d}$} & \colhead{$\kappa$} & \colhead{$A$} & \colhead{$b_{\rm rise}$} & \colhead{$b_{\rm decay}$} & \colhead{$\rho_{E_{\rm p},F}$} & \colhead{$S_{\rm pk}(F)$} & \colhead{$S_{\rm pk}(E_{\rm p})$} & \colhead{$f_{\rm stable,model}$} \\
 & (s) & (s) & (s) & & & & & & & &
}
\startdata
090620400 & 2.66 & 1.46 & 1.20 & 0.82 & -0.10 & 0.44 & 0.41 & 0.86 & 1.57 & 0.07 & not tested \\
090719063 & 7.28 & 5.61 & 1.67 & 0.30 & -0.54 & 0.23 & 0.46 & 0.93 & 0.13 & 1.65 & not tested \\
090804940 & 3.22 & 0.81 & 2.41 & 2.96 & 0.49 & 0.16 & 0.45 & 0.74 & 1.44 & 0.94 & not tested \\
090820027 & 5.25 & 3.01 & 2.24 & 0.74 & -0.15 & 0.19 & 0.48 & 0.92 & 3.62 & 2.37 & not tested \\
100122616 & 3.79 & 1.24 & 2.55 & 2.06 & 0.35 & 0.58 & 0.28 & 0.90 & 0.41 & 3.12 & 0 \\
100528075 & 10.20 & 6.02 & 4.18 & 0.69 & -0.18 & 0.39 & 0.52 & 0.69 & 2.21 & 0.07 & not tested \\
100707032 & 2.87 & 0.77 & 2.10 & 2.72 & 0.46 & -1.25 & 0.51 & 0.94 & 0.64 & 0.89 & not tested \\
120426090 & 1.79 & 0.31 & 1.48 & 4.83 & 0.66 & -0.12 & 0.49 & 0.71 & 0.48 & 0.45 & not tested \\
120919309 & 3.37 & 2.06 & 1.31 & 0.64 & -0.22 & 0.45 & 0.64 & 0.89 & 0.46 & 2.28 & not tested \\
130614997 & 3.02 & 0.23 & 2.79 & 12.12 & 0.85 & \nodata & 0.56 & 1.00 & 0.72 & 0.10 & not tested \\
131231198 & 5.37 & 1.88 & 3.49 & 1.86 & 0.30 & 0.17 & 0.48 & 0.59 & 0.36 & 0.80 & 0 \\
141028455 & 5.90 & 2.24 & 3.66 & 1.64 & 0.24 & -0.27 & 0.33 & 0.65 & 0.33 & 0.03 & 1 \\
150213001 & 1.19 & 0.70 & 0.49 & 0.71 & -0.17 & 1.10 & 0.28 & 0.78 & 1.61 & 0.90 & not tested \\
150902733 & 2.96 & 0.50 & 2.46 & 4.90 & 0.66 & -0.11 & 0.38 & 0.55 & 1.33 & 2.68 & not tested \\
160509374 & 8.42 & 3.33 & 5.09 & 1.53 & 0.21 & 0.37 & 0.17 & 0.38 & 0.33 & 0.00 & 1 \\
160530667 & 3.98 & 1.96 & 2.02 & 1.03 & 0.01 & 0.30 & 0.28 & 0.87 & 0.21 & 0.19 & not tested \\
160821857 & 6.60 & 3.12 & 3.48 & 1.12 & 0.05 & -0.04 & 0.21 & 0.66 & 0.04 & 0.22 & not tested \\
160910722 & 2.91 & 1.39 & 1.52 & 1.10 & 0.05 & -0.04 & 0.69 & 0.53 & 0.71 & 4.23 & not tested \\
180305393 & 3.37 & 1.53 & 1.84 & 1.21 & 0.09 & 0.14 & 0.46 & 0.83 & 1.96 & 0.42 & not tested \\
180728728 & 2.38 & 1.23 & 1.15 & 0.94 & -0.03 & 0.62 & 0.63 & 0.88 & 2.18 & 0.82 & 1 \\
\enddata
\tablecomments{The final column gives the outcome of the model-dependence check based on the available alternative Band/CPL resolved fits, where 1 indicates that the lag-based class is unchanged, 0 indicates that it changes under the forced-model comparison, and ``not tested'' means that no suitable alternative forced-model fit is available for that burst.}
\end{deluxetable*}

\section{Discussion}\label{sec:Dis}

\subsection{Observational Results}

Our analysis reveals that the classical flux-tracking behavior does not constitute a single, homogeneous pattern. 

Based on the relative ordering of peaks in the matched time series $E_{\rm p}(t)$ and $F(t)$, the sample divides naturally into three subclasses: Type I, in which the spectral peak and the energy-flux peak are aligned; Type II, in which the $E_{\rm p}$ peak precedes the $F$ peak; and Type III, in which the $E_{\rm p}$ peak follows the $F$ peak. Among the 20 single-pulse bursts in our sample, 5 events belong to Type I, 13 to Type II, and 2 to Type III. These numbers establish that early-peaking spectral tracking is not a rare exception, but rather the dominant mode of flux-tracking behavior among single-pulse GRBs.

This conclusion is supported by both the absolute peak-time lag and the width-normalized lag distributions. The median values of $t_{\rm lag}^{\rm F}$ are 0.00 s, $-1.54$ s, and 1.27 s for Types I, II, and III, respectively. This ordering is preserved after normalizing by the flux-pulse width, with median $\hat{t}_{\rm lag}$ values of 0.00, $-0.46$, and 0.45. The predominance of Type II bursts therefore cannot be attributed solely to a small number of anomalously broad pulses. A direct comparison between $t_{\rm p}(E_{\rm p})$ and $t_{\rm p}(F)$ leads to a consistent picture: Type I bursts cluster around the one-to-one relation, Type II bursts lie systematically below it, and the two Type III bursts lie above it.

The subclasses also differ in their spectral properties. Burst-level diagnostics extracted from low-energy photon index evolution demonstrate that Type II bursts are, systematically, spectrally harder than Type I. This holds for the photon index measured at the flux peak, $\alpha_{\rm peak}^{F}$, for the inverse-variance weighted mean $\bar{\alpha}_{\rm w}$, and for the hard-bin fraction $f_{-2/3}=N(\alpha>-2/3)/N_{\rm bin}$. Across all three diagnostics, the Type III events lie intermediate between the aligned and early-peaking subclasses although the small Type III sample precludes a statistically robust characterization of this group. The Anderson-Darling tests applied to these burst-level quantities indicate that the subclass distributions are unlikely to be drawn from the same parent population, with the caveat that the limited Type III sample warrants caution in this interpretation.

\begin{figure}
\centering
\includegraphics[width=0.92\columnwidth]{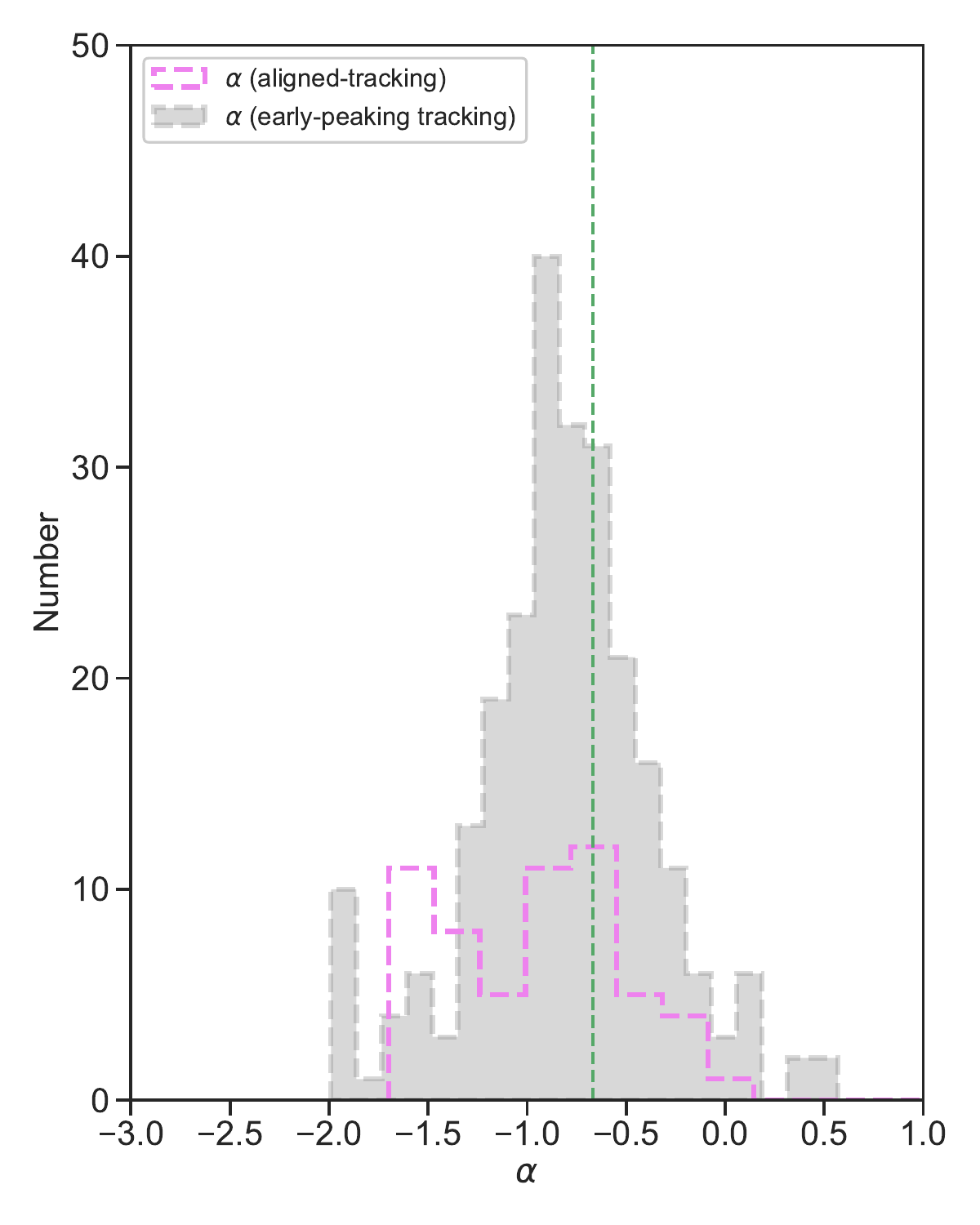}
\caption{Distribution of the time-resolved low-energy photon index $\alpha$ for the Type~I and Type~II bursts. The vertical dashed line marks $\alpha=-2/3$. Counts are shown without normalization by burst or by bin number. The Type~III class is omitted because it contains only two bursts.}
\label{fig:alpha-dist}
\end{figure}

The same tendency is apparent in the time-resolved view. Figure~\ref{fig:alpha-dist} shows that the raw distribution of $\alpha$ values for the Type II subclass is shifted toward harder slopes relative to Type I, with a larger fraction of bins extending toward and beyond the synchrotron line $\alpha=-2/3$. Because the histogram is plotted in raw counts rather than normalized by burst or by bin number, it should be interpreted as a visual diagnostic rather than as a formal population test. Nonetheless, it provides a direct empirical evidence that the harder low-energy states occur more frequently in the early-peaking subclass than in the aligned subclass. A complementary view is provided by Figure~\ref{fig:epmax-alpha-max}. In the $E_{\rm p}(\alpha_{\max})$-$\alpha_{\max}$ plane, the Type II bursts occupy a broader region and include the hardest states in the current sample, whereas the Type I bursts are concentrated in a softer, narrower locus. The comparison curves in that figure should be regarded as qualitative guides only, but the subclass contrast itself is clear.

Additional subclass differences emerge in the spectral-flux and temporal domains. The median Spearman coefficient of the $E_{\rm p}$-$F$ relation for the full sample is $\rho_{E_{\rm p},F}=0.81$, confirming that a positive $E_{\rm p}$-$F$ correlation is common to all three subclasses. The key discriminator is therefore not the sign of the correlation itself, but the ordering of the maxima and the asymmetry between the rising and decaying branches. Type I bursts tend to remain closer to $b_{\rm rise}=b_{\rm decay}$, whereas many Type II bursts exhibit $b_{\rm rise}<b_{\rm decay}$, indicating a stronger branch asymmetry. In the time domain, Type II bursts also have broader median flux-pulse widths than Type I bursts, while the two Type III bursts are notably narrower.

The robustness tests indicate that the most reliable result of the current analysis is the sign and population-level separation of $t_{\rm lag}^{\rm F}$, rather than the precise numerical value of the lag in individual burst. This is partly because the flux-based lag is measured from spectrally matched bins and therefore avoids the direct time-grid mismatch that arises when spectral and temporal quantities are defined independently. At the same time, a strict peak-prominence requirement is very conservative for the present sample, so that the prominence statistics are better treated as cautionary diagnostics than as the primary classification metric. Similarly, the limited model-dependence checks suggest that the lag ordering is more robust than the exact lag amplitude. In practice, the clearest separation in the current data comes from the sign and significance of the flux-based lag itself.

Taken together, these results demonstrate that the traditional flux-tracking category is not observationally uniform. It contains at least one tightly coupled aligned mode, one dominant early-peaking mode, and a rare positive-lag population. The physical question is therefore no longer whether flux-tracking exists, but rather what dynamical and radiative conditions are required to produce each of each mode of peak-ordering behavior.

\subsection{Type I: Aligned Tracking and Tightly Coupled Spectral and Power Evolution}

The defining feature of the Type I subclass is that the maxima of $E_{\rm p}$ and $F$ occur in the same bin, or in bins whose centers are consistent within the timing uncertainty. In this respect, Type I represents the most tightly coupled form of flux-tracking identified in the present sample. The simplest phenomenological interpretation is that the physical quantities controlling the characteristic photon energy and the total radiative power evolve nearly in concert. This picture is supported by the small values of $|t_{\rm lag}^{\rm F}|$, the compact locus of the Type I bursts in the $t_{\rm p}(E_{\rm p})$-$t_{\rm p}(F)$ plane, and their comparatively symmetric branch-dependent $E_{\rm p}$-$F$ slopes.

A natural physical scenario is that this behavior is arises preferentially when the prompt emission is dominated by a photosphere emission component, or at least when the spectral peak and the radiative power are set by closely linked thermodynamic variables. In photosphere emission models, a higher luminosity is often associated with a higher effective temperature, and hence a higher observed $E_{\rm p}$. In that case, near-exact tracking between the $E_{\rm p}$ peak and the flux peak is therefore qualitatively compatible with a compact emission region and a relatively tight luminosity-temperature coupling, as expected in photosphere emission scenarios \citep[e.g.,][]{Rees2005,Peer2006,Peer2017IJMPD}.

\begin{figure}
\centering
\includegraphics[width=\columnwidth]{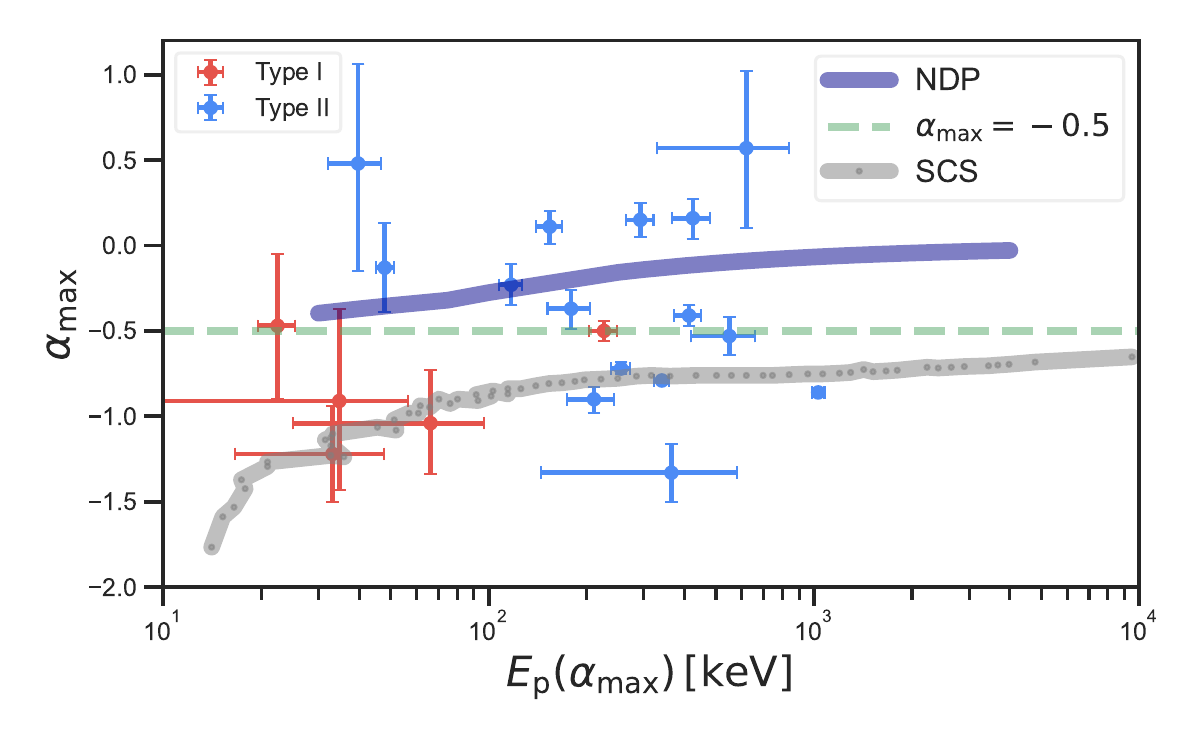}
\caption{Relation between the maximum low-energy photon index $\alpha_{\rm max}$ and the corresponding $E_{\rm p}$ measured in the same time bin for the Type~I and Type~II bursts. The comparison curves and the horizontal reference line are kept as in the original exploratory figure and are intended only as qualitative guides. The Type~III class is omitted because of its small sample size.}
\label{fig:epmax-alpha-max}
\end{figure}

The present data do not, however, support the expectation that the aligned subclass should also be the spectrally hardest subclass. In fact, the opposite trend is seen. Both the burst-level diagnostics and the raw time-resolved $\alpha$ distribution indicate that Type II, rather than Type I, is systematically harder. This is also seen in Figure~\ref{fig:epmax-alpha-max}: the Type I bursts remain concentrated in a relatively soft and narrow region of the $E_{\rm p}(\alpha_{\max})$-$\alpha_{\max}$ plane, while the harder states are more commonly found among Type II events. Therefore, the current sample does not support a one-to-one identification of Type I with a purely photosphere channel.

This distinction merits emphasis. What the data support at present is not that Type I bursts are uniquely photospheric, but rather that they represent a state of tightly coupled spectral and power evolution. Such a state may indeed occur in a photosphere-dominated outflow, but it may also arise in a hybrid jet or in a nonthermal dissipation scenario. If the characteristic electron energy, magnetic-field strength, and injection power evolve together over the pulse. The observational signature of Type I is therefore primarily temporal, not mechanistic.

The qualitative reference curves shown in Figure~\ref{fig:slope-summary} should be interpreted in this spirit. They provide useful heuristic loci for slow-cooled synchrotron and non-dissipative photosphere emission expectations, but are not by themselves sufficient to assign a unique radiation mechanism to any individual burst or subclass. In particular, the fact that the Type I points do not preferentially occupy the hardest part of the diagram argues against an overly simple ``Type I = photosphere'' interpretation. A more conservative reading is that Type I may preferentially trace conditions under which the relevant emission variables remain well synchronized, regardless of whether the underlying radiation is quasi-thermal, nonthermal, or mixed.

We therefore interpret Type I as the observational subclass most consistent with a regime of prompt emission in which spectral hardening and radiative power are closely locked in time. Photosphere emission models remain attractive candidates for producing such behavior, owing to their naturally small emission radii and the expected luminosity-temperature coupling, but the current $\alpha$-based evidence is not sufficient to claim that they dominate the Type I sample. For now, Type I is best regarded as a phenomenological class characterized by near-synchronous evolution, whose underlying radiation mechanism still requires direct model-based tests, in particular, targeted searches for thermal components near the aligned peaks.

\subsection{Type II: Early-peaking Tracking as the Dominant and Harder Subclass}

The Type II subclass constitutes the dominant observational mode in the present sample. In these bursts, the spectral peak energy reaches its maximum while the energy flux is still rising, so that the hardest spectrum appears before the flux peaks. This behavior is not a minor perturbation around the aligned case. It is the most common configuration in the current sample, and remains dominant even after the lag is normalized by the flux-pulse width. The median lag of the Type II sample is negative both in absolute time and in normalized form, showing that the early-peaking behavior does not arise simply from a few unusually broad pulses.

The Type II subclass is also spectrally harder than the Type I. This is already evident from the burst-level diagnostics. The median values of $\alpha_{\rm peak}^{F}$, $\bar{\alpha}_{\rm w}$, and $f_{-2/3}$ are all shifted toward harder spectra in Type II relative to Type I. The same trend is seen in Figure~\ref{fig:alpha-dist}, where the time-resolved $\alpha$ distribution for the Type II sample extends more frequently toward hard low-energy slopes, particularly near the synchrotron line $\alpha=-2/3$. Figure~\ref{fig:epmax-alpha-max} provides a complementary view: in the $E_{\rm p}(\alpha_{\max})$-$\alpha_{\max}$ plane, the Type II bursts occupy a broader region and include the hardest spectral states in the current sample, whereas the Type I bursts remain more concentrated. Collectively, these results indicate that the early-peaking subclass is not only temporally distinct, but also preferentially associated with harder low-energy spectral states.

The most direct physical implication of Type II is that the quantity controlling the characteristic photon energy reaches its maximum earlier than the quantity controlling the total radiative power. In optically thin synchrotron or magnetic-dissipation scenarios, this outcome is physically natural. The spectral peak energy is directly sensitive to the magnetic-field strength, the characteristic electron Lorentz factor, and the local acceleration and cooling conditions, whereas the energy flux depends on the total dissipated power, the number of emitting particles, the emitting area, and geometric effects. Consequently, $E_{\rm p}$ may peak before $F$ even when the two quantities still broadly track each other over the pulse. This interpretation is qualitatively consistent with magnetically dissipative or ICMART-like models, in which spectral hardening and power release need not be perfectly synchronized \citep[e.g.,][]{ZhangYan2011,Uhm2014NP,Uhm2018}. The broader flux-pulse widths of Type II bursts and their tendency to show $b_{\rm rise}<b_{\rm decay}$ support the idea that this subclass involves a stronger temporal decoupling between spectral evolution and radiative power. The representative $E_{\rm p}$-$F$ tracks themselves offers a useful physical visualization of this decoupling. In the Type~I events, the rising and decaying branches largely retrace each other, suggesting that spectral hardening and radiative output respond to a common control parameter. In contrast, many Type II events show a clearer branch separation, with the rise tracing a flatter path and the decay occupying a broader dynamic range in the $E_{\rm p}$-$F$ plane. This behavior is qualitatively consistent with a scenario in which the quantity governing the characteristic photon energy reaches its maximum before the quantity governing the total radiative power. The Type~III events still preserve an overall positive $E_{\rm p}$-$F$ relation, but their small number prevents identification of a comparably stable branch morphology. The branch geometry seen in Figures~\ref{fig:Ep-Flux-TypeI}-\ref{fig:Ep-Flux-TypeIII} therefore supports the view that peak ordering and branch asymmetry are two linked manifestations of the same underlying timing structure.

The present results do not support a simple one-to-one mapping between the subclasses and the standard radiation mechanisms. In particular, one might have expected the aligned Type I bursts to show systematically harder $\alpha$ values if exact tracking were preferentially associated with a photosphere emission origin. The data show the opposite trend: Type II, not Type I, is the harder subclass. This creates a direct tension with the simplest photosphere emission expectation and suggests that the Type II bursts cannot be dismissed as merely a softened or perturbed version of Type I. Rather, they likely represent a genuinely different dynamical regime.

This does not imply that all Type II bursts must be purely synchrotron, nor does it exclude dissipative photospheres or hybrid jets. A hybrid outflow, in which thermal and nonthermal contributions coexist and evolve differently, may naturally produce hard low-energy slopes together with a measurable offset between $t_{\rm p}(E_{\rm p})$ and $t_{\rm p}(F)$ \citep[e.g.,][]{Gao2015,Li2020}. Similarly, hidden thermal components and model dependence can affect the inferred $\alpha$ values \citep[e.g.,][]{Burgess2014,Li2019a,Li2019b,Li2020}. Nevertheless, once the harder $\alpha$ distribution is considered together with the early peak ordering, the broader pulses, and the branch-dependent $E_{\rm p}$-$F$ asymmetry, the Type II subclass is more naturally compatible with nonthermal or hybrid prompt-emission scenarios than with a simple non-dissipative photosphere.

We therefore interpret Type II as the dominant flux-tracking regime in which spectral hardening and radiative power are correlated but temporally no longer locked to the same peak. As such, the Type II subclass may be tracing a prompt-emission zone in which the characteristic particle-acceleration or magnetic-field evolution precedes the maximal radiative output. This makes Type II particularly important for future model discrimination, because any viable prompt-emission model must explain not only the existence of flux-tracking, but also why its most common manifestation in the current sample is an early-peaking, spectrally harder mode.

\subsection{Type III: A Rare Late-peaking Tail and Its Possible Physical Origin}

The Type III subclass is currently represented by only two bursts, but it is nevertheless important because it demonstrates that the peak ordering can also occur in the opposite direction. In these events, the energy flux reaches its maximum first, and subsequent the latest $E_{\rm p}$ maximum appears afterward. The median lag of the Type III sample is positive both in absolute units and after normalization by the flux-pulse width. Despite the limited sample size, the existence of these two events indicates that the internal timing structure of flux-tracking is more diverse than a simple aligned-versus-early dichotomy.

Nevertheless, Type III should be treated with caution. The two bursts do not by themselves define a statistically mature population on their own. In the burst-level $\alpha$ summaries, they lie between the Type I and Type II groups rather than forming a clearly isolated spectral population. The same caution applies to the two-dimensional diagnostic plane shown in Figure~\ref{fig:slope-summary}. The contrast in that figure is dominated by the broader Type II distribution and the more concentrated Type I points, while the two Type III events are too few to occupy a distinct locus of their own. Accordingly, Type III is best described at present as a rare positive-lag tail, rather than a fully established third class on an equal footing with Types I and II.

Several physical interpretations are conceivable. One possibility is that the flux can peak while the spectrum is still hardening, for example, if a later phase of dissipation preferentially increases the characteristic particle energy or magnetic-field strength without increasing the total radiative power by the same factor. In magnetically dissipative outflows, this could happen if reconnection or turbulent acceleration becomes more efficient slightly after the main power release. In a hybrid outflow, it could also reflect an evolving balance between thermal and nonthermal components, so that the flux maximum is set by one component while the latest spectral hardening is driven by another. A delayed reheating episode or a progressive hardening of the dominant nonthermal component would qualitatively lead to the same observational signature.

However, the current data do not allow a decisive mechanism to be identified. Because the Type III sample is so small, and because one of the two events is much less significant than the other, unresolved substructure, binning choices, or model dependence may still play a non-negligible role. The safest conclusion is therefore that Type III identifies a phenomenologically real positive-lag tail in the current sample, but its physical origin remains open. Although the existence of the Type III is of considerable interest, a conservative approach is warranted here to avoid unsubstantiated over-interpretation.

Future progress on this subclass will most likely require targeted studies of the positive-lag events themselves. In particular, it will be important to search for additional thermal components in the bins around both the flux peak and the delayed $E_{\rm p}$ peak, in order to test whether the positive lag is associated with a changing thermal-to-nonthermal balance. A larger single-pulse sample, together with a fully joint-posterior classification of $t_{\rm p}(E_{\rm p})$ and $t_{\rm p}(F)$, will also be needed to determine whether Type III is a genuine rare channel of prompt emission or simply the extreme tail of a broader distribution. Until such tests are carried out, Type III is best interpreted as a rare late-peaking extension of the flux-tracking phenomenology, one that reveals the temporal ordering between spectral hardening and radiative power to be bidirectional.

\section{Conclusion}\label{sec:Con}

In this study, we have revisited the classical flux-tracking behavior of GRB prompt spectral evolution using a sample of 20 well-defined single-pulse \textit{Fermi}/GBM bursts. Rather than comparing the peak time of $E_{\rm p}$ with an independently binned counts light curve, we employed a flux-based definition, $t_{\rm lag}^{\rm F}$, where both $E_{\rm p}$ and the energy flux $F$ are derived from the same BBlocks time bins through time-resolved spectroscopy. This matched-bin approach eliminates the direct temporal mismatch between spectral and intensity measures and allows a self-consistent classification of the internal timing structure of the flux-tracking pattern.

With this method, we find that the flux-tracking pattern is not observationally uniform. The sample separates into three subclasses: Type I, in which the peaks of $E_{\rm p}$ and $F$ are aligned; Type II, in which $E_{\rm p}$ peaks before $F$; and Type III, in which $E_{\rm p}$ peaks after $F$. In the present sample, 5 bursts belong to Type I, 13 to Type II, and 2 to Type III. Type II remains the dominant subclass in both the absolute lag and the width-normalized lag distributions, confirming that early-peaking tracking is the most common realization of flux-tracking in this sample.

The subclasses also differ spectrally. The Type II bursts are systematically harder than the aligned Type I bursts, as indicated by the burst-level summaries of the low-energy spectral index, including $\alpha_{\rm peak}^{F}$, $\bar{\alpha}_{\rm w}$, and the hard-bin fraction $f_{-2/3}$. The same tendency is seen in the raw time-resolved $\alpha$ distributions and in the $E_{\rm p}(\alpha_{\max})$-$\alpha_{\max}$ plane, where Type II occupies a broader region and includes the hardest states in the sample. Furthermore, Type II bursts tend to have broader flux pulses and a stronger asymmetry between the rising and decaying $E_{\rm p}$-$F$ branches than Type I. These results indicate that the distinction among the subclasses is not limited to peak ordering alone, but reflects broader differences in both spectral and temporal evolution.

Physically, Type I is most naturally interpreted as a regime in which spectral hardening and radiative power remain tightly coupled in time. Such behavior is qualitatively compatible with photosphere emission models, but the present data do not support a simple one-to-one association between Type I and a purely photosphere emission origin. In particular, the harder $\alpha$ distribution of Type II directly challenges the simplest expectation that the aligned subclass should also be the hardest if it were preferentially photospheric. The Type II subclass appears more naturally compatible with nonthermal or hybrid prompt-emission scenarios, in which the quantity controlling the characteristic photon energy reaches its maximum before the quantity controlling the total radiative output. The Type III subclass, by contrast, is currently best regarded as a rare positive-lag tail whose physical origin remains open.

Overall, our results show that the traditional flux-tracking picture contains a richer internal structure than previously recognized. Any viable prompt-emission model must therefore account for not only the existence of a positive $E_{\rm p}$-$F$ correlation, but also the peak ordering differs from burst to burst and the dominant early-peaking subclass is spectrally harder than the aligned one. A larger single-pulse sample, together with joint-posterior classification and targeted searches for thermal components near the $E_{\rm p}$ and $F$ peaks, will be essential for clarifying the origin of these subclasses and for establishing peak ordering as a practical diagnostic of GRB prompt-emission physics.

\acknowledgments

This work is supported by the National Natural Science Foundation of China (Grant Nos. 11874033 and 12588101), the KC Wong Magna Foundation at Ningbo University, and made use of the High Energy Astrophysics Science Archive Research Center (HEASARC) Online Service at the NASA/Goddard Space Flight Center (GSFC). The computations were supported by the high performance computing center at Ningbo University. I thank Bing~Zhang, En-Wei Liang, Asaf~Pe{\textquoteright}er, and Felix Ryde for many useful discussions.

\vspace{30mm}
\facilities{{\it Fermi}/GBM}
\software{
{\tt 3ML} \citep{Vianello2015}, 
{\tt matplotlib}, 
{\tt NumPy}, 
{\tt SciPy}, 
{\tt $lmfit$}, 
{\tt astropy},
{\tt pandas},
{\tt seaborn}
}

\clearpage
\appendix
\setcounter{figure}{0}    
\setcounter{section}{0}
\setcounter{table}{0}
\renewcommand{\thesection}{A\arabic{section}}
\renewcommand{\thefigure}{A\arabic{figure}}
\renewcommand{\thetable}{A\arabic{table}}
\renewcommand{\theequation}{A\arabic{equation}}
\renewcommand{\theHsection}{appendix.\arabic{section}}
\renewcommand{\theHfigure}{appendix.\arabic{figure}}
\renewcommand{\theHtable}{appendix.\arabic{table}}
\renewcommand{\theHequation}{appendix.\arabic{equation}}

In this appendix, we present below the counts-based comparison figures.

\section{Counts-based Comparison Figures}

For comparison with the conventional observational light curve, we define an auxiliary counts-based lag,
\begin{equation}
t_{\rm lag}^{\rm cnt} \equiv t_{\rm p}(E_{\rm p}) - t_{\rm p}(C),
\end{equation}
where $t_{\rm p}(C)$ is the peak time of the counts light curve. This quantity is included in the figures to provide context relative to the traditional counts-based timing reference, but does not serve as the formal classification criterion in this work. This choice reflects the fact that the counts light curve and the time-resolved spectral sequence do not necessarily share the same temporal binning, whereas $E_{\rm p}$ and $F$ do. We therefore adopt the flux-based lag as the more self-consistent diagnostic for characterizing the internal timing structure of the flux-tracking behavior.

For completeness, we also present the figures comparing $E_{\rm p}(t)$ with the counts light curve. These plots are not employed in the formal classification scheme presented in the main text and are retained solely as an auxiliary visual reference for comparison with the conventional counts-based timing picture.

\clearpage
\begin{figure*}
\includegraphics[width=0.5\hsize,clip]{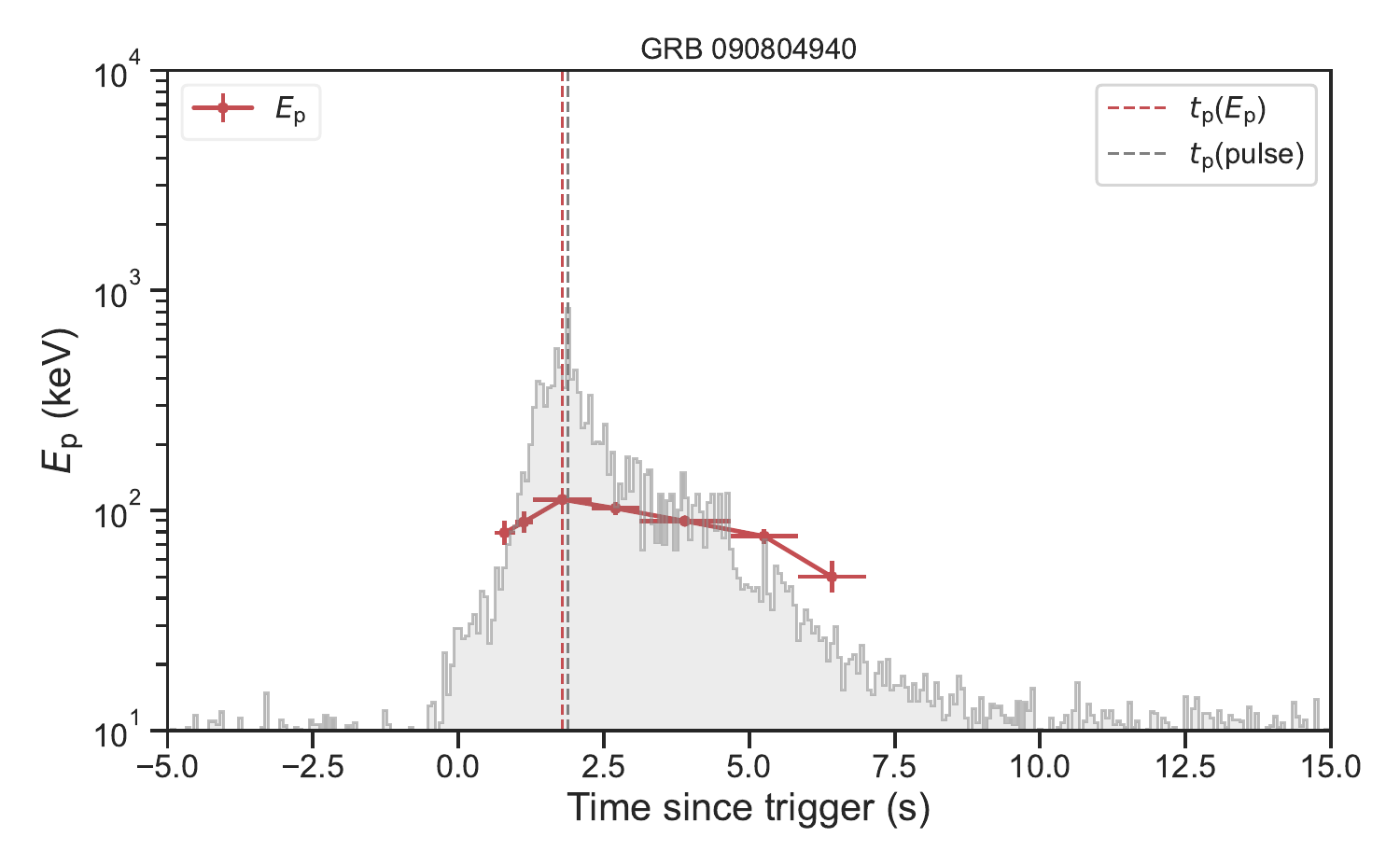}
\includegraphics[width=0.5\hsize,clip]{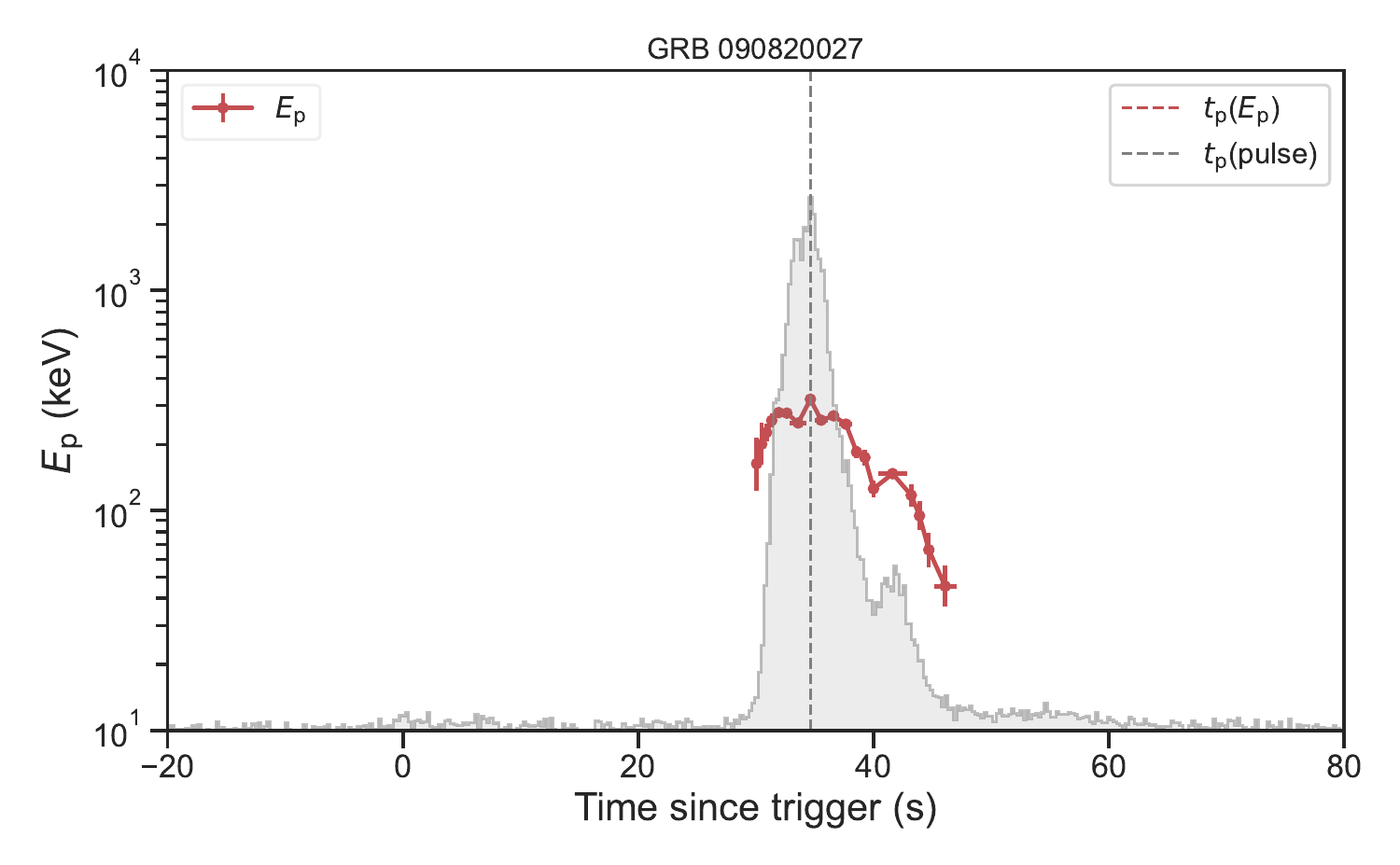}
\includegraphics[width=0.5\hsize,clip]{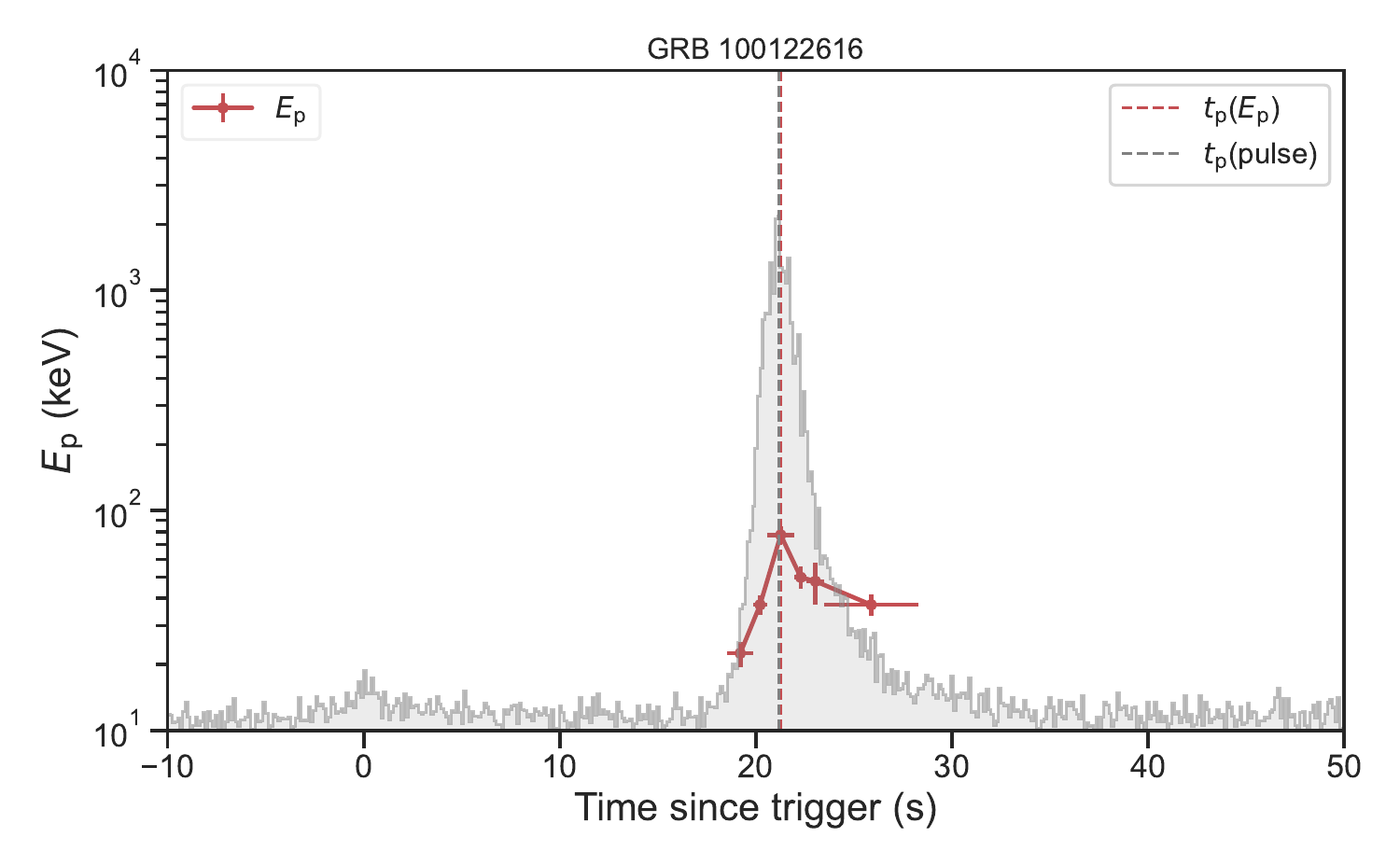}
\includegraphics[width=0.5\hsize,clip]{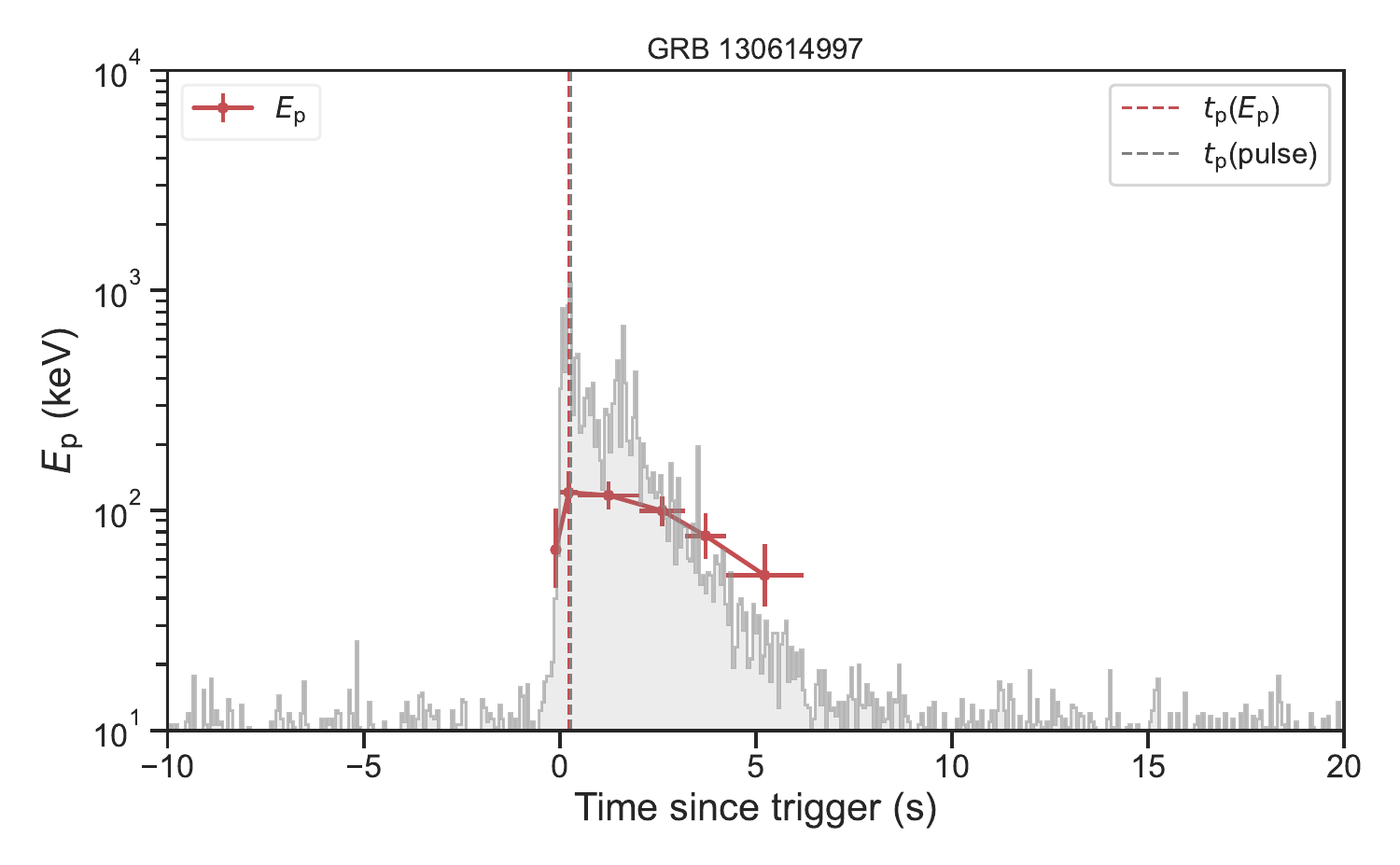}
\includegraphics[width=0.5\hsize,clip]{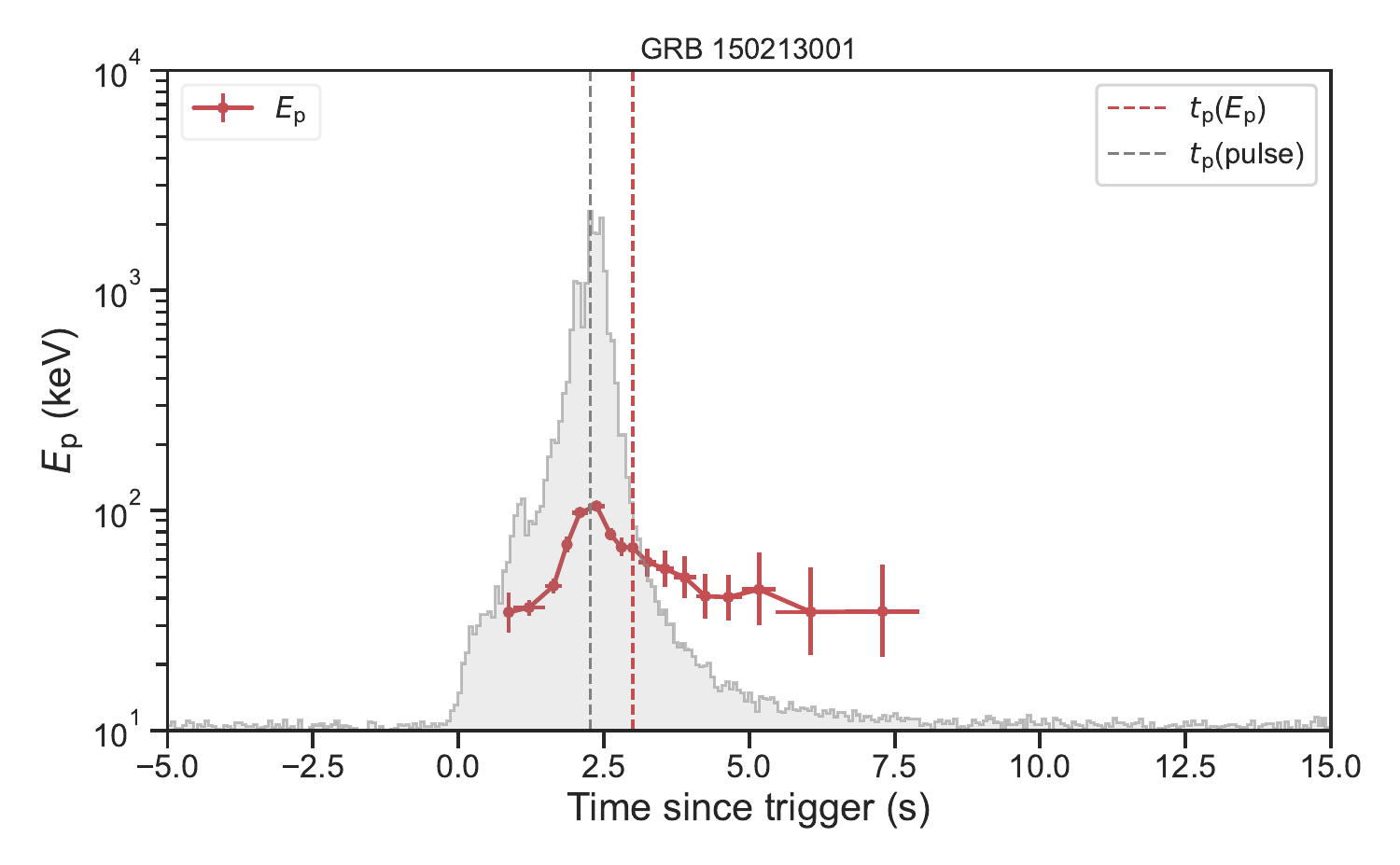}
\caption{Counts-based comparison plots for all Type~I bursts. The red points show $E_{\rm p}(t)$ and the gray background shows the counts light curve.}
\label{fig:appendix-counts-typeI}
\end{figure*}

\clearpage
\begin{figure*}
\includegraphics[width=0.5\hsize,clip]{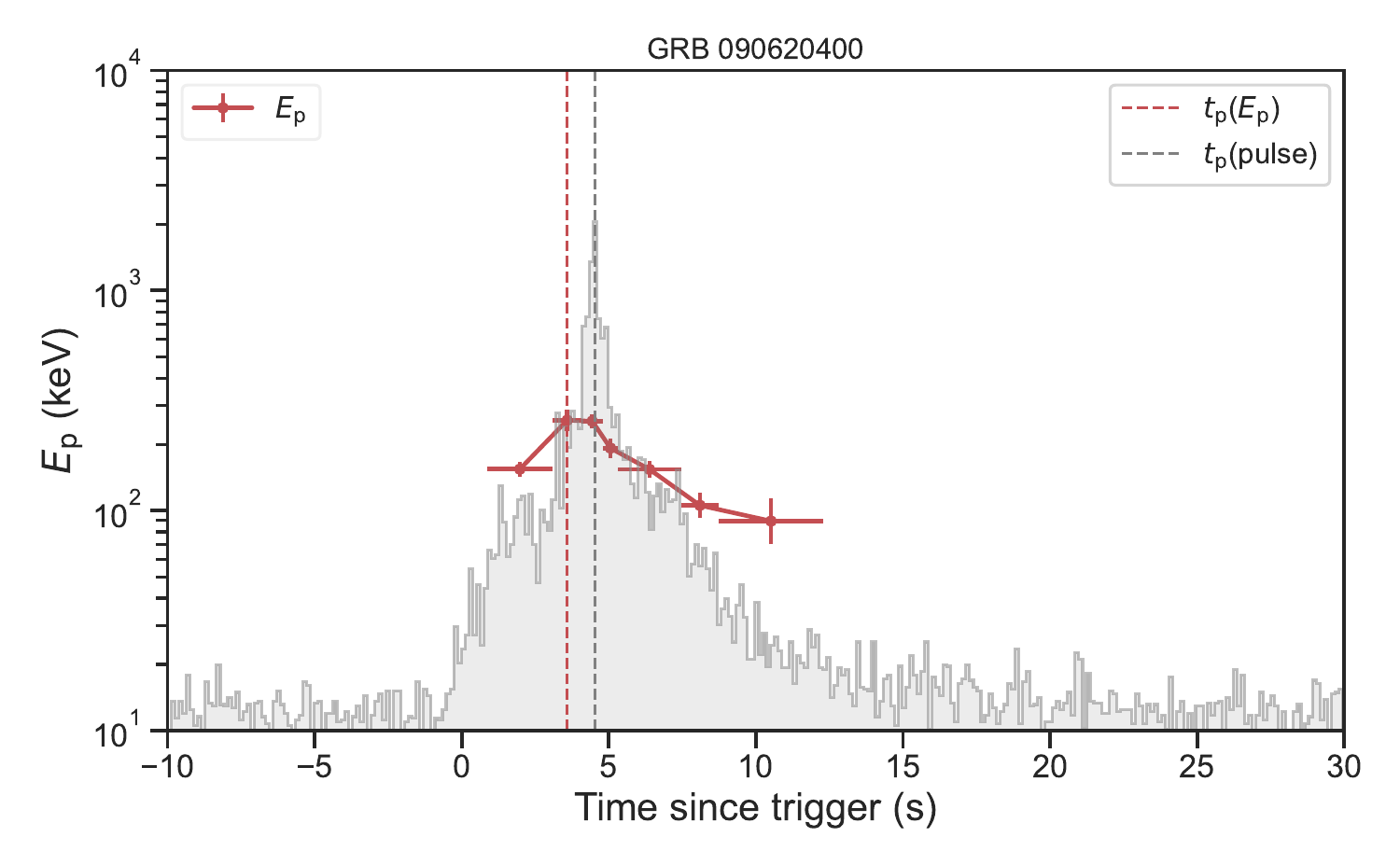}
\includegraphics[width=0.5\hsize,clip]{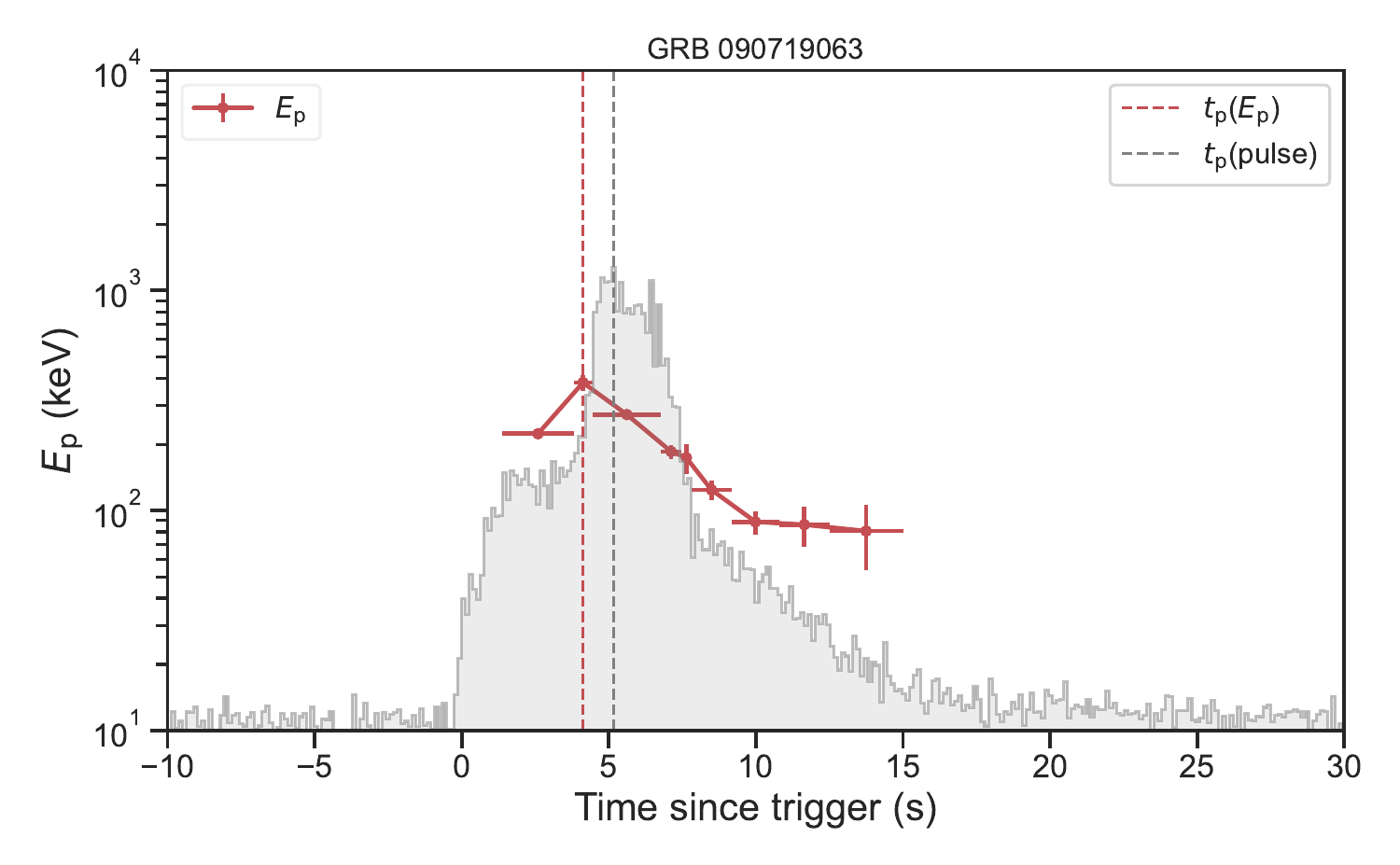}
\includegraphics[width=0.5\hsize,clip]{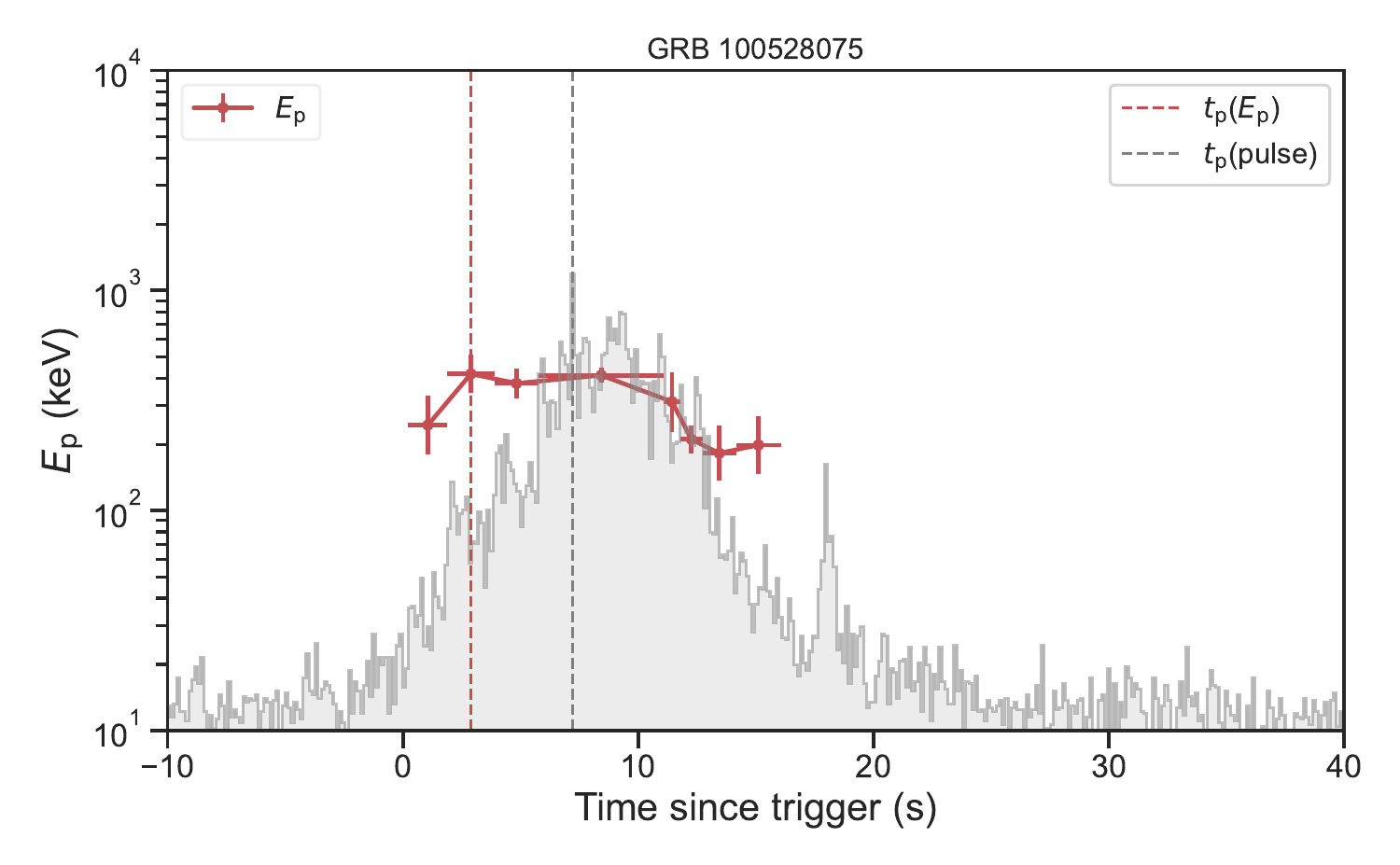}
\includegraphics[width=0.5\hsize,clip]{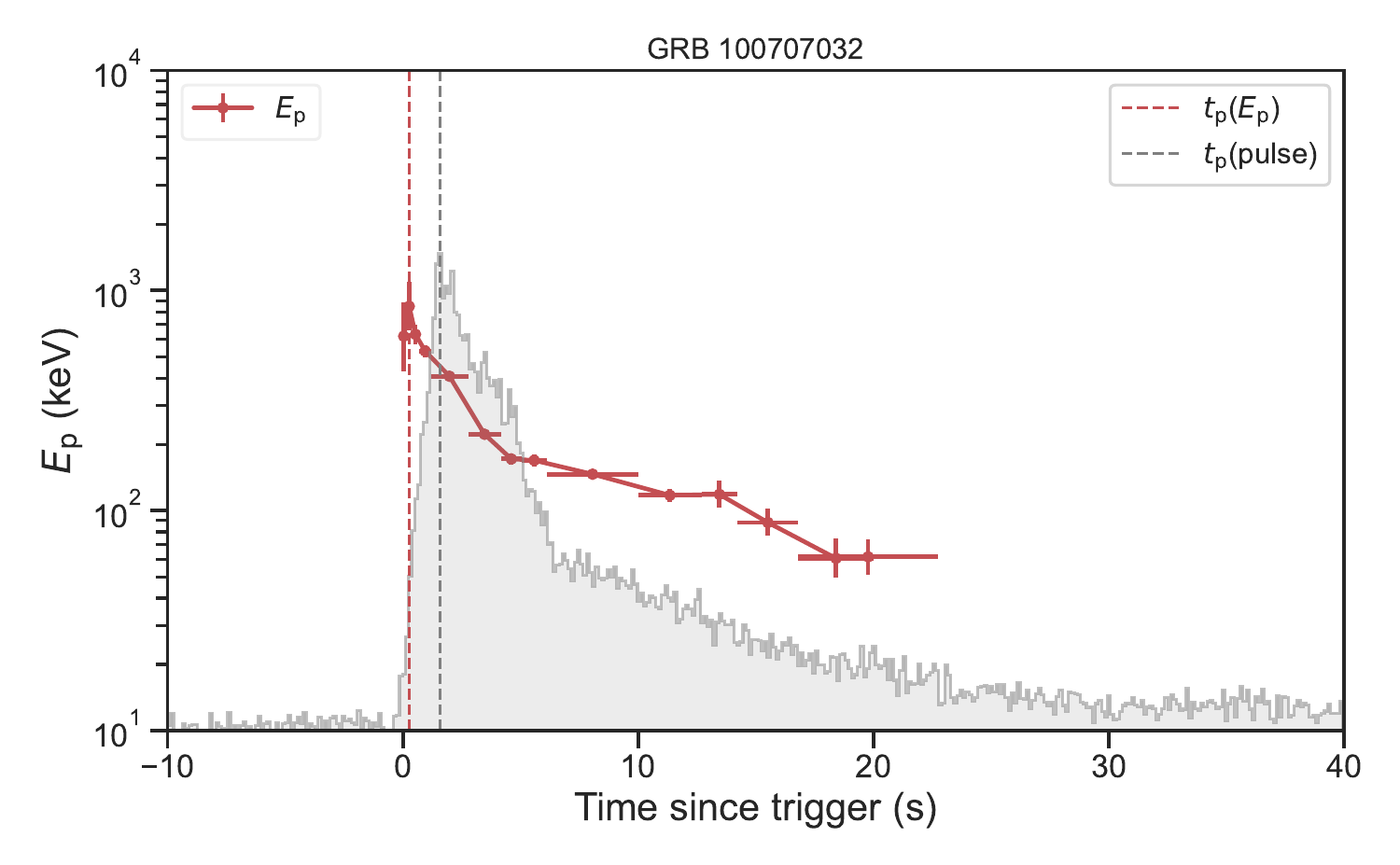}
\includegraphics[width=0.5\hsize,clip]{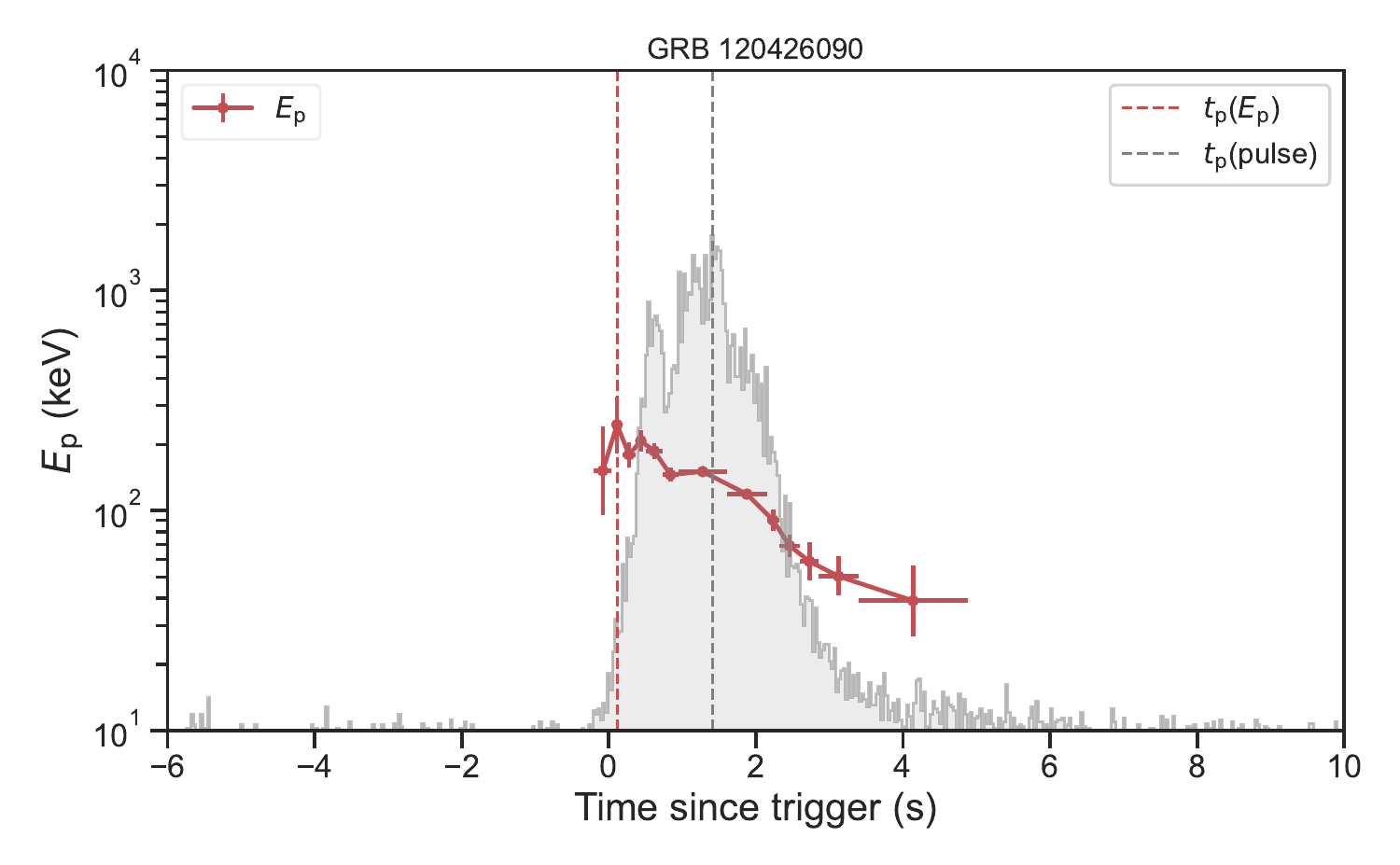}
\includegraphics[width=0.5\hsize,clip]{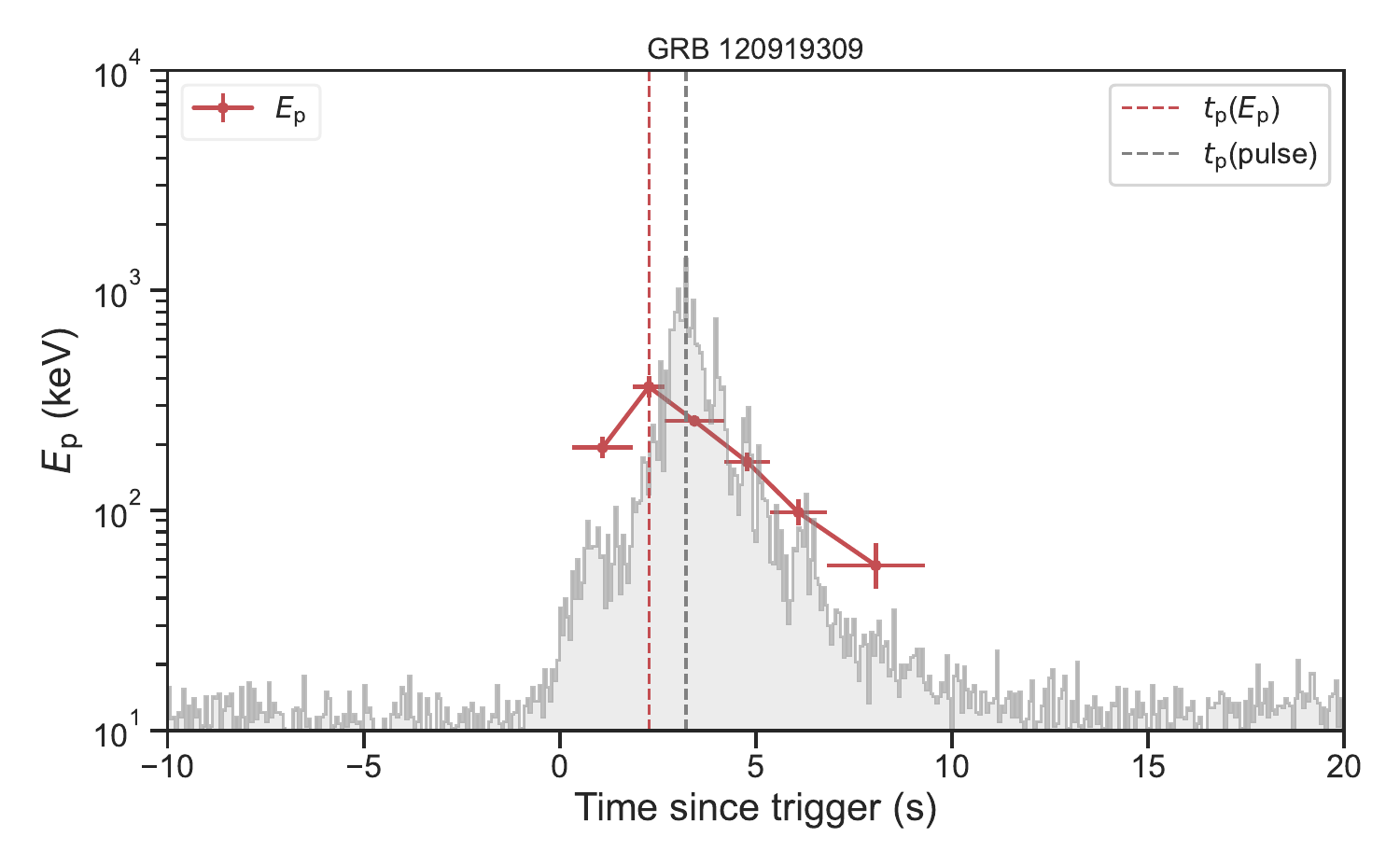}
\caption{Sample as Fig.\ref{fig:appendix-counts-typeI} but for all Type~II bursts.}
\label{fig:appendix-counts-typeII}
\end{figure*}
\begin{figure*}
\includegraphics[width=0.5\hsize,clip]{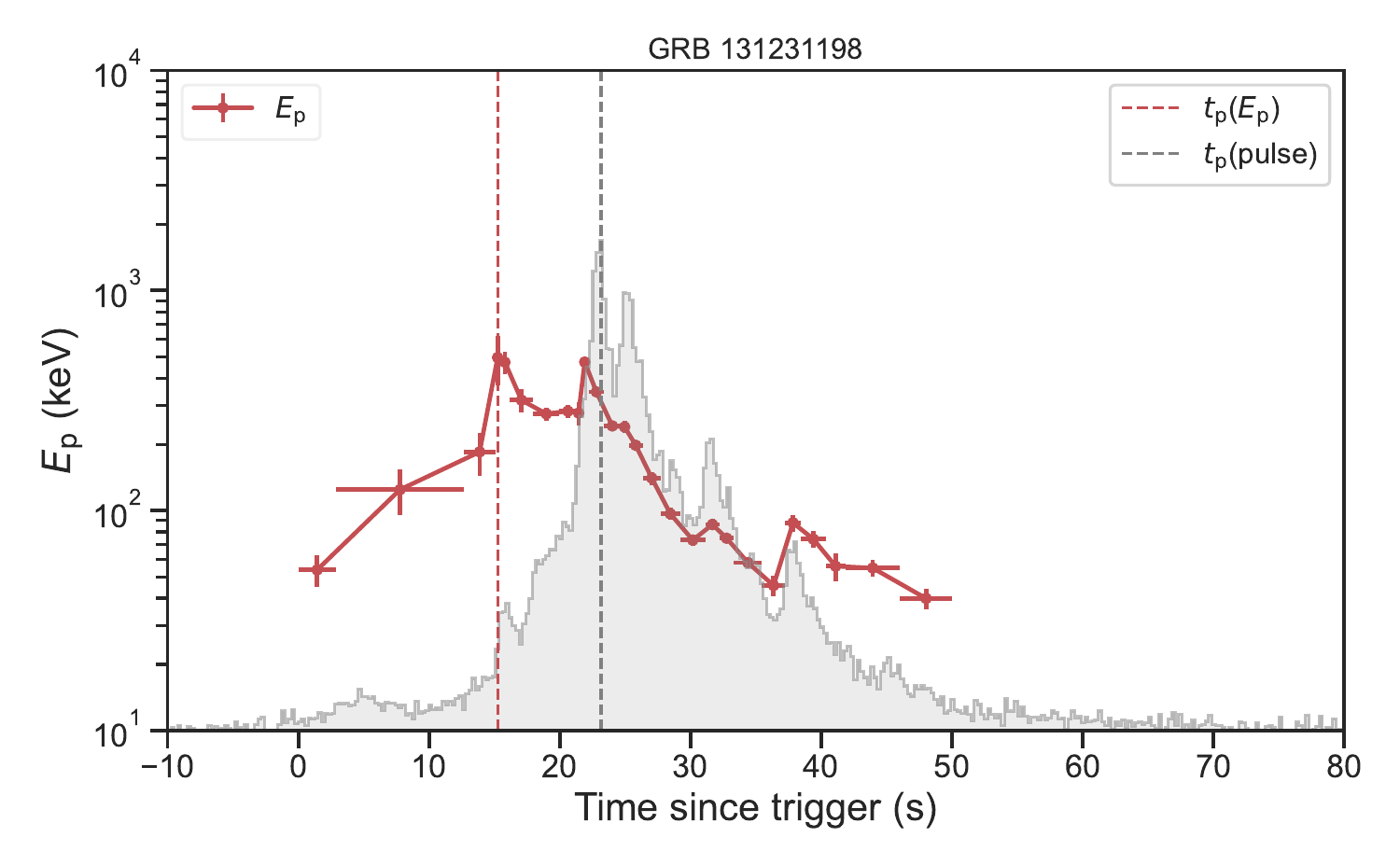}
\includegraphics[width=0.5\hsize,clip]{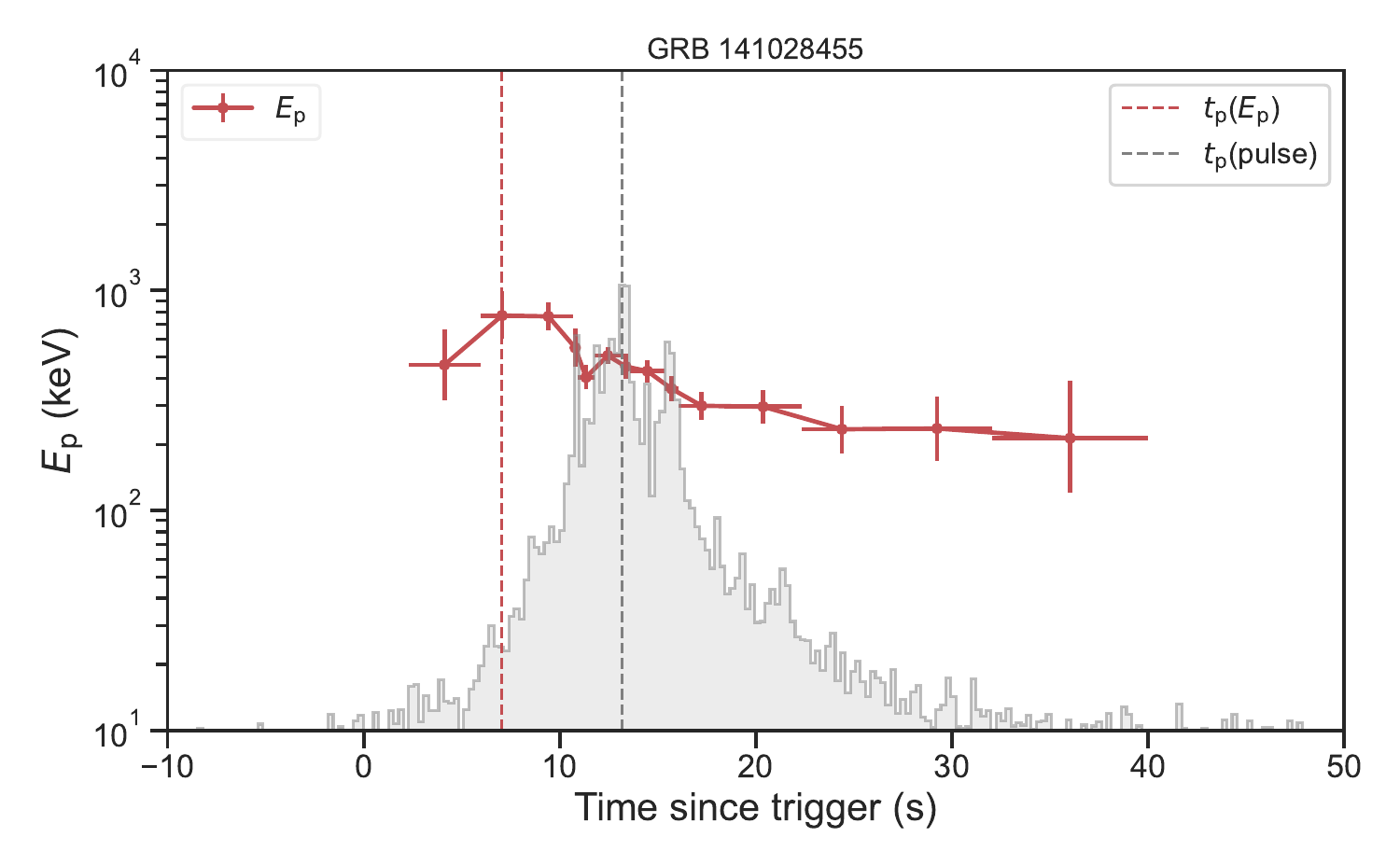}
\includegraphics[width=0.5\hsize,clip]{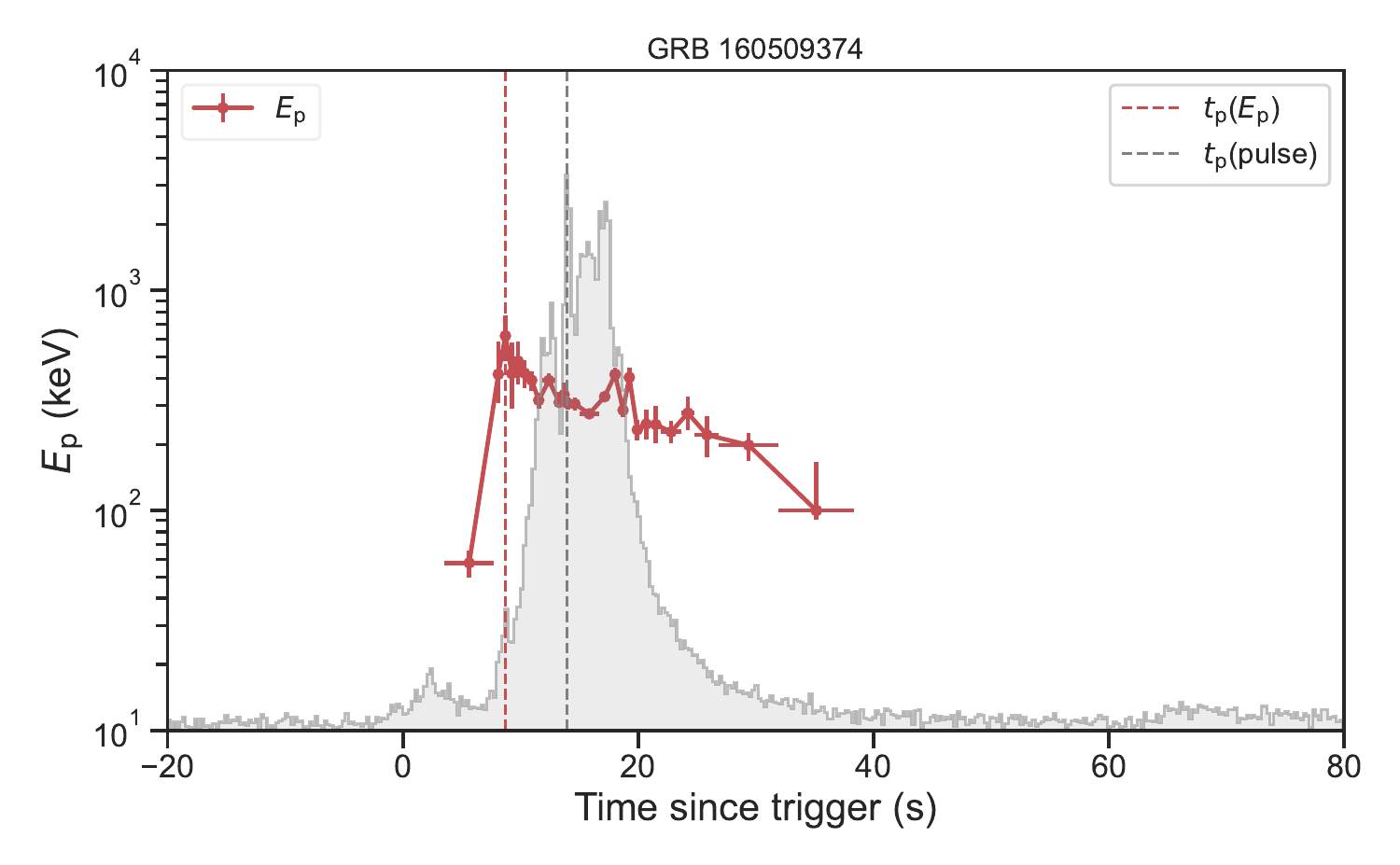}
\includegraphics[width=0.5\hsize,clip]{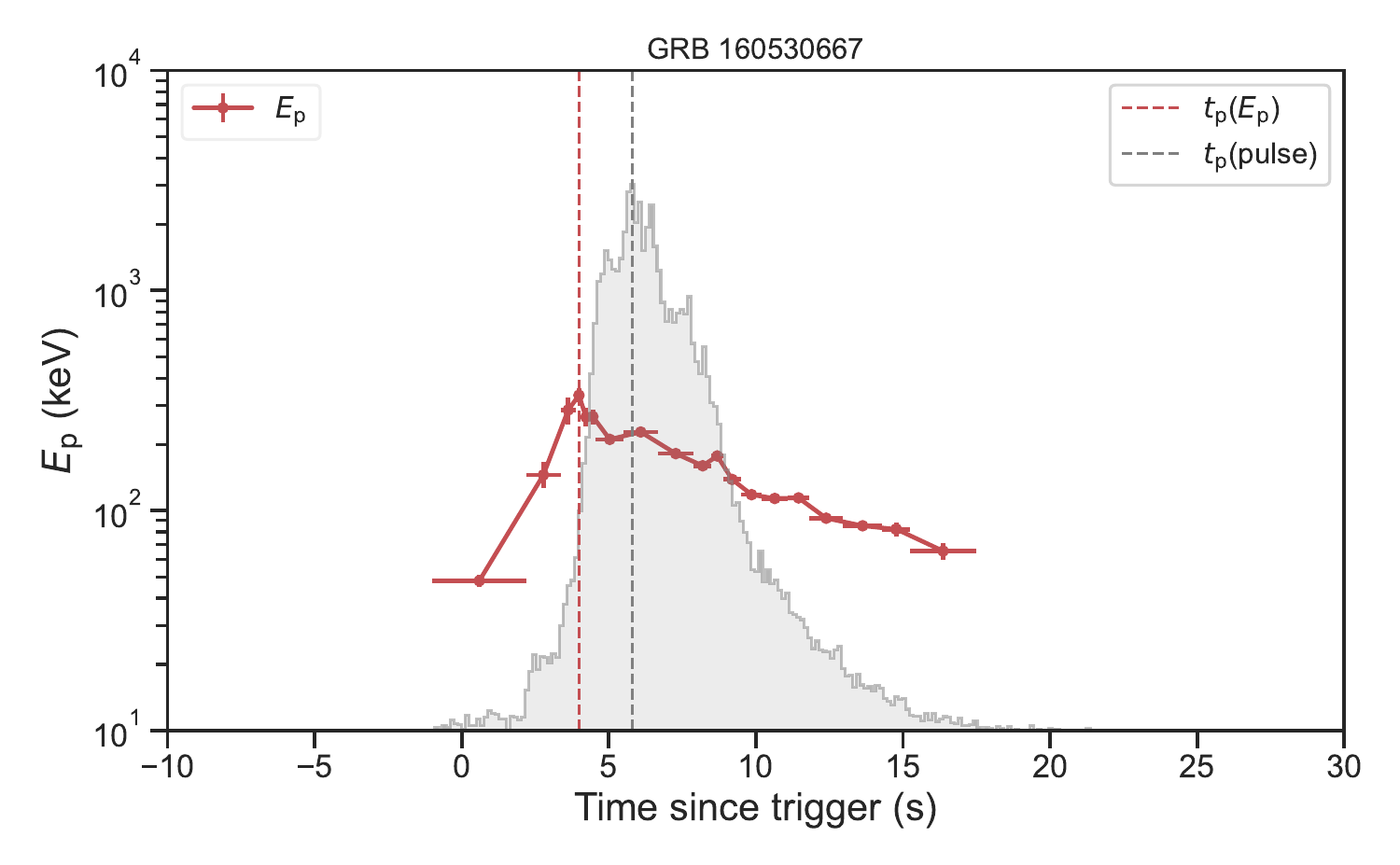}
\includegraphics[width=0.5\hsize,clip]{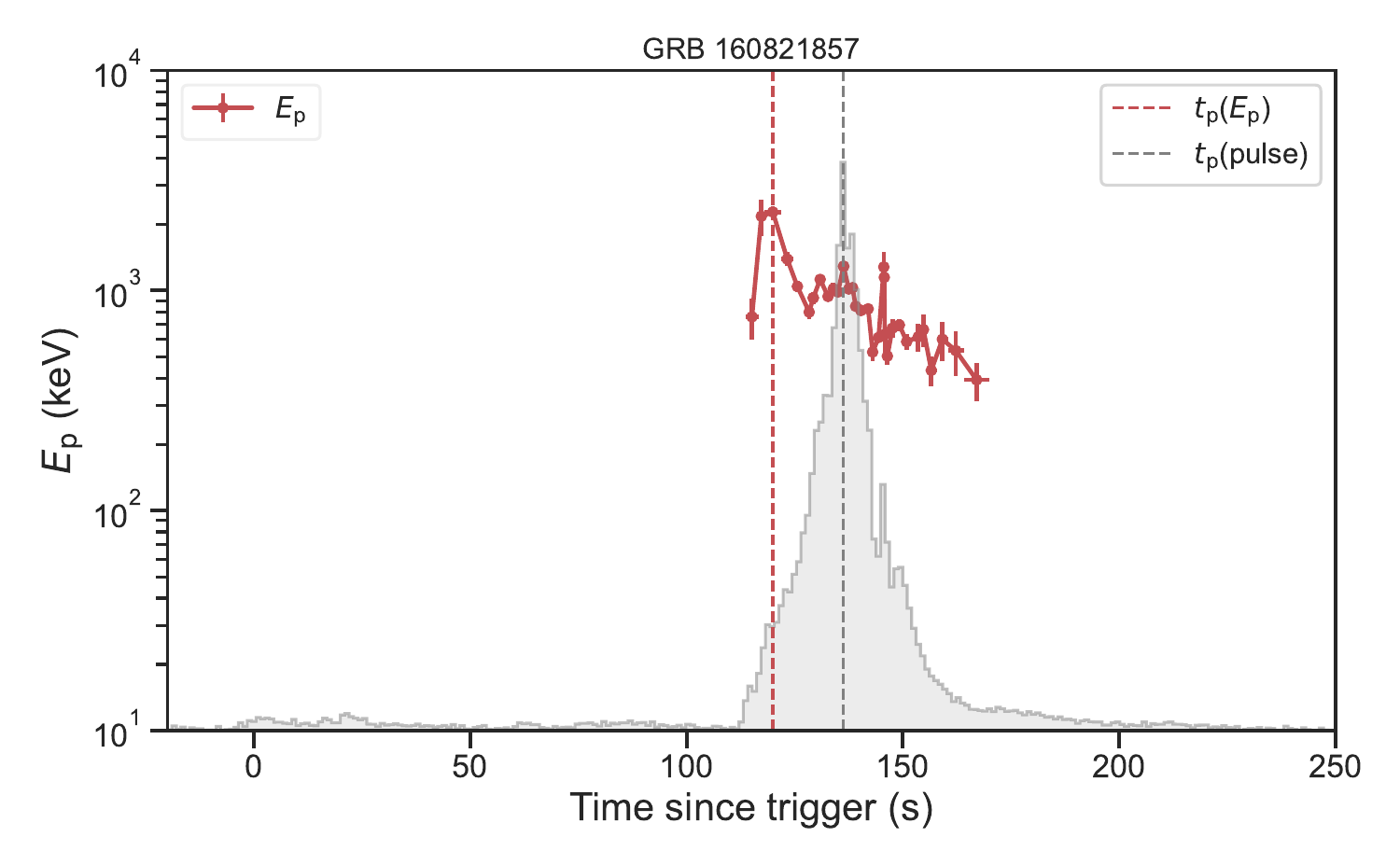}
\includegraphics[width=0.5\hsize,clip]{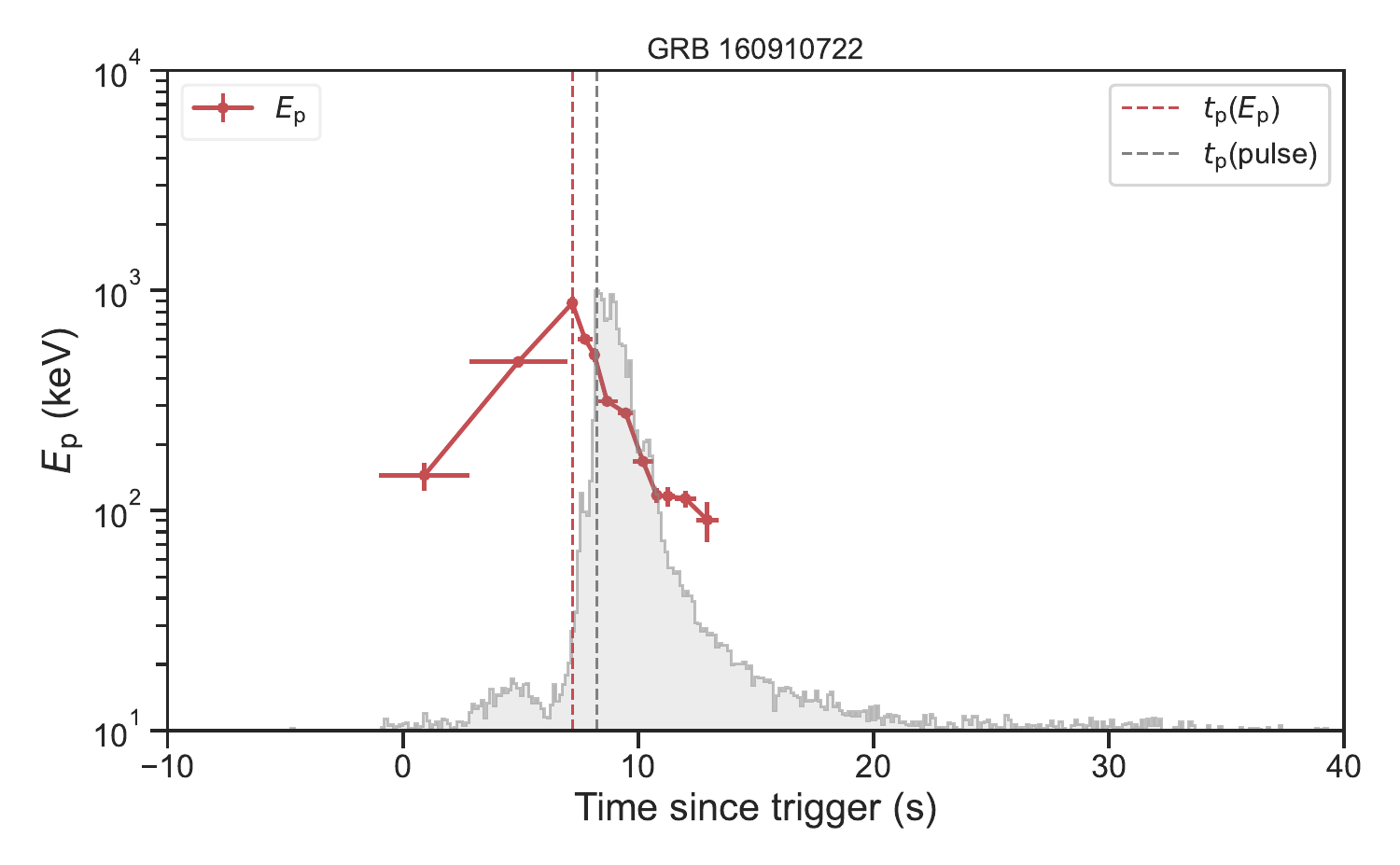}
\includegraphics[width=0.5\hsize,clip]{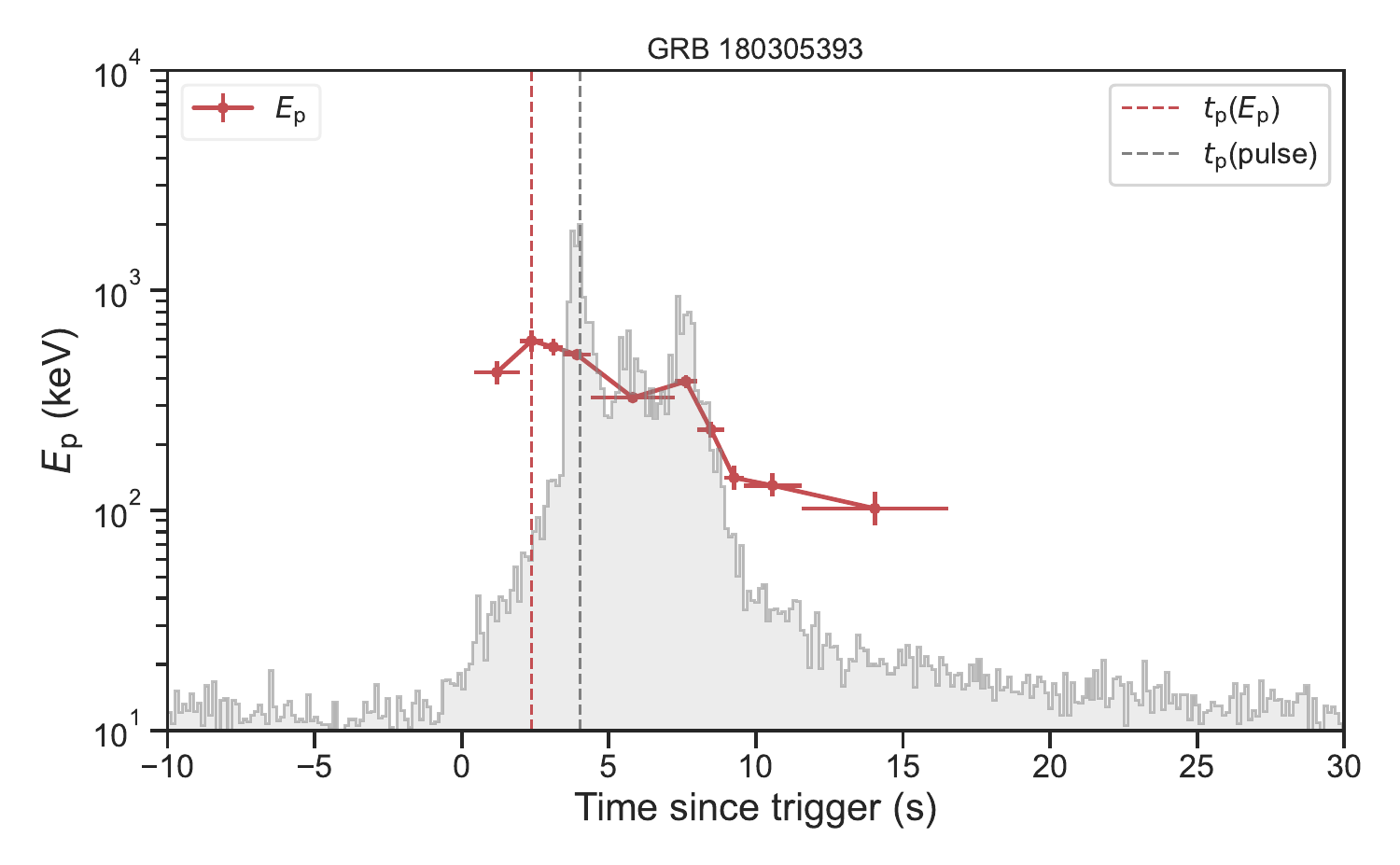}
\center{Fig. \ref{fig:appendix-counts-typeII}- Continued}
\end{figure*}

\clearpage
\begin{figure*}
\includegraphics[width=0.5\hsize,clip]{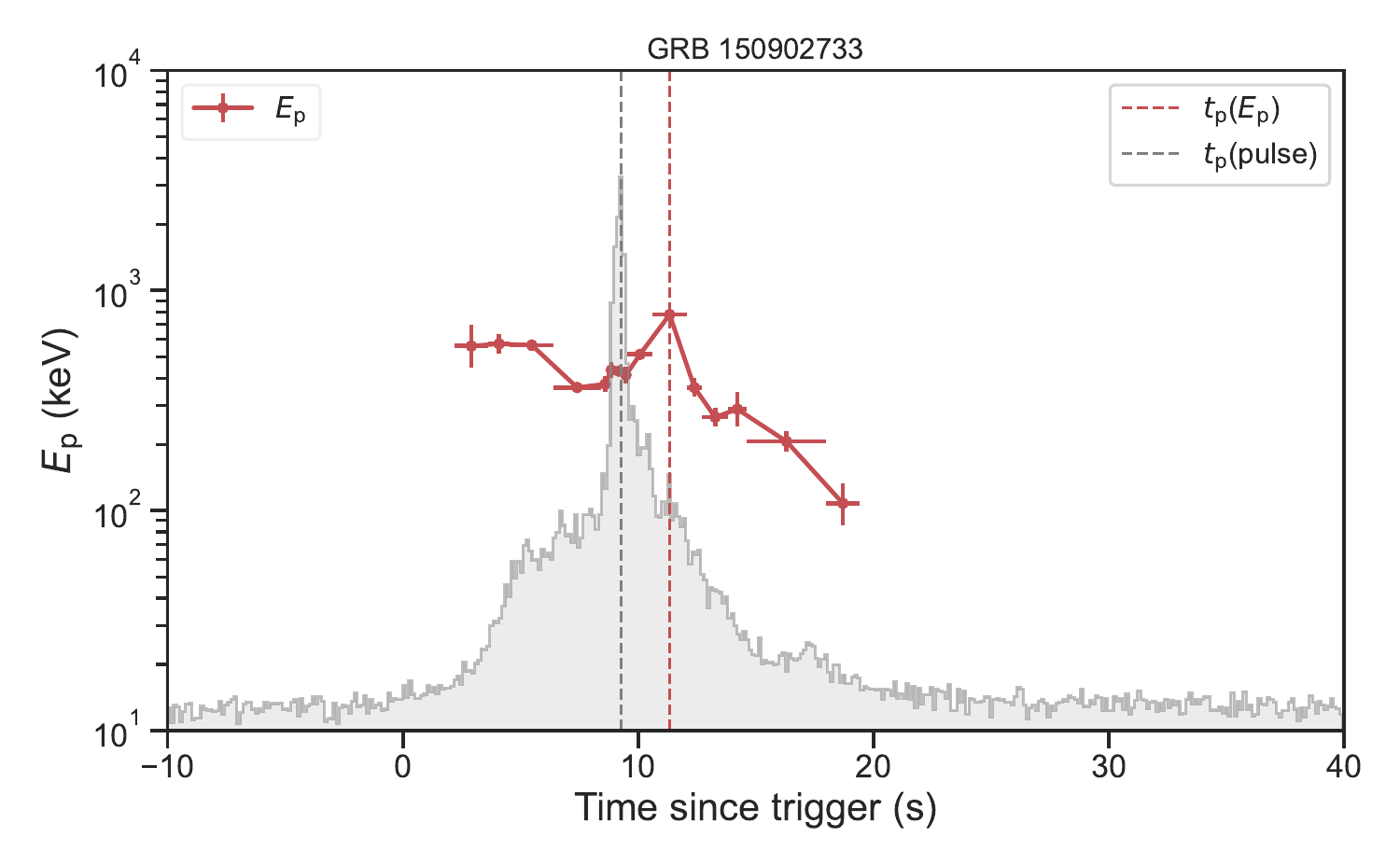}
\includegraphics[width=0.5\hsize,clip]{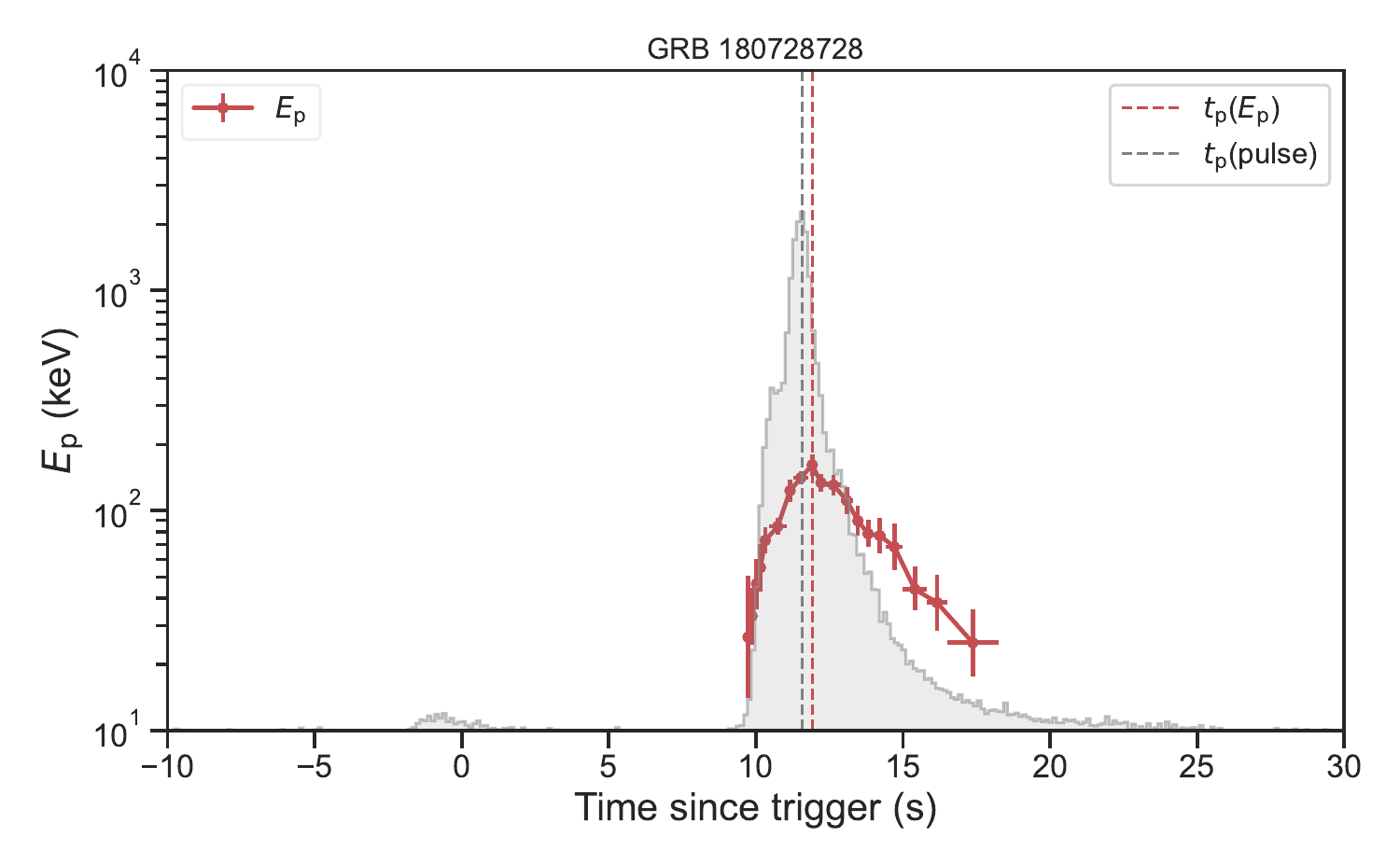}
\caption{Sample as Fig.\ref{fig:appendix-counts-typeI} but for all Type~III bursts.}
\label{fig:appendix-counts-typeIII}
\end{figure*}

\end{document}